\documentclass[useAMS,usenatbib]{mn2e}   
\usepackage{amssymb}
\usepackage{amsmath}
\usepackage{amstext}
\usepackage{deluxetable}
\usepackage{epsfig}
\usepackage{thumbpdf}
\usepackage{epstopdf}
\usepackage{graphicx}
\usepackage{longtable}
\usepackage[dvipsnames]{xcolor}
\usepackage[citecolor=blue]{hyperref}
\hypersetup{colorlinks=true}

\newcommand{\HI}{H\,{\sc i}}
\newcommand{\HII}{H\,{\sc ii}}
\newcommand{\Hi}{H\,{\sc i}}
\newcommand{\Hii}{H\,{\sc ii}}
\newcommand{\Nii}{[N\,{\sc ii}]}
\newcommand{\Sii}{[S\,{\sc ii}]}
\newcommand{\Oii}{[O\,{\sc ii}]}
\newcommand{\Oiii}{[O\,{\sc iii}]}
\newcommand{\NII}{[N\,{\sc ii}]}

\newcommand{\OIII}{[O\,{\sc iii}]}

\newcommand{\Ha}{H$\alpha$}
\newcommand{\Hb}{H$\beta$}
\newcommand{\Hg}{H$\gamma$}

\newcommand{\abox}{12+log(O/H)}

\newcommand{\Te}{$T_{\rm e}$}

\newcommand{\Ne}{$n_{\rm e}$}

\DeclareRobustCommand{\ion}[2]{
\relax\ifmmode
\ifx\testbx\f@series
{\mathbf{#1\,\mathsc{#2}}}\else
{\mathrm{#1\,\mathsc{#2}}}\fi
\else\textup{#1\,{\mdseries\textsc{#2}}}
\fi}

\title[SN~Ia luminosity dependence on oxygen abundance]
{Using the local gas-phase oxygen abundances to explore a metallicity-dependence in SNe~Ia luminosities}

\author[M.E. Moreno-Raya et al.]
{M.E. Moreno-Raya$^{1}$\thanks{E-mail: manuelemilio.moreno@ciemat.es},
{\'A}.R. L{\'o}pez-S{\'a}nchez$^{2,3}$, 
M. Moll{\'a}$^{1}$, 
L. Galbany$^{4,5}$,
\newauthor
J.M. V{\'\i}lchez$^{6}$, 
\& A. Carnero$^{7,8}$\\
$^1$Departamento de Investigaci{\'o}n B{\'a}sica, CIEMAT,Avda. Complutense 40, 28040, Madrid, Spain\\
$^2$Australian Astronomical Observatory, PO Box 915, North Ryde, NSW 1670, Australia\\ 
$^3$Department of Physics and Astronomy, Macquarie University, NSW 2109, Australia\\ 
$^4$Millennium Institute of Astrophysics, Chile\\ 
$^5$Departamento de Astronom{\'\i}a, Universidad de Chile, Casilla 36-D, Santiago, Chile\\ 
$^6$Instituto de Astrof{\'\i}sica de Andaluc{\'\i}a-CSIC, Apdo. 3004, 18008, Granada, Spain \\ 
$^7$ Laborat{\'o}rio Interinstitucional de e-Astronomia - LIneA, Rua Gal. Jos\'e Cristino 77, Rio de Janeiro, RJ 20921-400, Brazil \\ 
$^8$ Observat{\'o}rio Nacional, Rua Gal. Jos{\'e} Cristino 77, Rio de Janeiro, RJ 20921-400, Brazil \\
}

\date{Received date: 9 March 2016; accepted date: 12 July 2016}

\begin{document}

\maketitle


\begin{abstract}

We present an analysis of the gas-phase oxygen abundances of a sample of 28  galaxies in the local Universe ($z<0.02$) hosting Type Ia Supernovae (SNe~Ia). The data were obtained with the 4.2m William Herschel Telescope (WHT). We derive local oxygen abundances for the regions where the SNe~Ia exploded by calculating oxygen gradients through each galaxy (when possible) or assuming the oxygen abundance of the closest \Hii region. The sample selection only considered galaxies for which distances not based on the the SN~Ia method are available. Then, we use a principal component analysis to study the dependence of the absolute magnitudes on the color of the SN~Ia, the oxygen abundances of the region where they exploded, and the stretch of the SN light curve. We demonstrate that our previous result suggesting a metallicity-dependence on the SN~Ia luminosity for not-reddened SNe~Ia \citep{2016ApJ...818L..19M} can be extended to our whole sample. These results reinforce the need of including a metallicity proxy, such as the oxygen abundance of the host galaxy, to minimize the systematic effect induced by the metallicity-dependence of the SN~Ia luminosity in future studies of SNe~Ia at cosmological distances.

\end{abstract}

 
 
\section{Introduction} \label{Section1}

Type Ia Supernovae (SNe~Ia) are claimed to be thermonuclear explosions of carbon-oxygen white dwarfs (CO WD) \citep{1960ApJ...132..565H}. Their origin is not well-established, since there is still an open discussion about the different possibles progenitor scenarios. The single-degenerate scenario \citep[SD;][]{1982ApJ...253..798N, 1973ApJ...186.1007W} occurs when a WD in a binary system accretes mass from its non-degenerate companion until the Chandrasekhar mass limit ($\sim$~1.44 $M_{\odot}$) is reached. At that moment, the degenerate-electron pressure is not longer supported, and the thermonuclear explosion occurs. On the other hand, the double-degenerate (DD) scenario \citep{1984ApJS...54..335I,1984ApJ...277..355W} consists of two CO WDs gravitationally bounded that lose angular momentum and merge \citep{1976Afz....12..521T, 1979AcA....29..665T}. At that moment, the SN explodes resulting no fossil but the SN remnant \citep{2012Natur.489..533G}. SNe~Ia are very bright (M$_{\rm B}\sim$-19.4 mag at peak), and show very low intrinsic luminosity dispersion \citep[around 0.36 mag,][]{1993ApJ...405L...5B} so they are considered extraordinary tools for measuring distances in cosmological scales. Although SNe~Ia are called \textit{standard candles}, they are not pure \textit{standard}, but \textit{standarizable}. 

\cite{1993ApJ...413L.105P}, \cite{1996AJ....112.2391H,1996AJ....112.2438H}, and \cite{1999AJ....118.1766P} proved that a correlation between the absolute magnitude at maximum brightness and the luminosity decline after maximum, lately parametrized as the light-curve (LC) width.
\citet{1996ApJ...473...88R} also found a relation between the peak magnitude and the SN color.

In this way, the distance to these objects can be estimated from their distance modulus $\mu = m_{B} - M_{B}$ (where $m_{B}$ is the apparent magnitude and $M_{B}$ the absolute magnitude, both in band $B$) by just studying SNe~Ia multiwavelength LCs. These calibrations allowed to reduce the scatter of distances in the \textit{Hubble Diagram} (HD), in which $\mu$ is represented as function of redshift, $z$. 

In fact, diverse standardization techniques have been developed to standardize SN~Ia LCs and obtain the absolute magnitudes at maximum, and simultaneously the parameters that better reduce the scatter in the HD. Modern techniques, such as SALT2 \citep{2007A&A...466...11G} adjust SN LC templates to the observed LC and determine the SN~Ia color at maximum brightness ($C$), the stretch applied to the LC template ($s$), the apparent magnitude at maximum brightness in the $B$ band ($m_{B}$), and the epoch of the maximum brightness ($t_{max}$). Then, the distance modulus $\mu_{SALT}$ can be calculated using the equation
\begin{equation}
\mu_{SALT}=m_{B}-(M_{B}-\alpha\,(s-1)+\beta\,C), 
\label{eq:salt}
\end{equation}
where $\alpha$, $\beta$ and $M_{B}$ are obtained by minimizing the HD residuals. With similar techniques the SNe~Ia-based cosmology projects discovered that the Universe is in accelerated expansion \citep{1999ApJ...517..565P,1998AJ....116.1009R}.

However after this method, there still exists a certain inhomogeneity in SNe~Ia at peak. A plausible source of inhomogeneity is a dependence of the properties of the SN~Ia  on the characteristics of its environment. Since
the average properties of host galaxies evolve with redshift, any such dependence not included in the standardization techniques will impact on the cosmological parameter determination. Many recent studies have indeed analyzed the dependence of SNe~Ia properties on global characteristics of their hosts \citep{2006ApJ...648..868S,2008ApJ...685..752G, 2009ApJ...691..661H, 2009ApJ...700..331H,2010ApJ...715..743K, 2010MNRAS.406..782S, 2010ApJ...722..566L, 2011ApJ...743..172D, 2011ApJ...740...92G, 2011ApJ...734...42N,2010MNRAS.406..782S, 2012ApJ...755..125G, 2013MNRAS.435.1680J,2013ApJ...770..108C, 2014A&A...568A..22B, 2014MNRAS.438.1391P}. 
In summary, all found that SNe~Ia are systematically brighter in more massive galaxies than in less massive ones {\sl after LC shape  and color corrections}. Through the mass-metallicity relation \citep{2010MNRAS.406..782S} this would lead to a correlation between SNe~Ia magnitudes and the metallicities of their host galaxies: more metal-rich galaxies would host brighter SNe~Ia {\sl after corrections}. However the cause of these correlations is not well-understood.  
In addition, due to the metal enrichment of galaxies with time, a change in chemical abundances with redshift \citep{2006ApJ...644..813E,2009A&A...505..529L} is expected. All these SNe~Ia calibrations are based on local objects mostly having around solar abundances\footnote{Here we use the terms metallicity, total abundance in metals, Z, (being X+Y+Z=1 in mass), and oxygen abundances indistinctly, assuming that $\log(\rm Z/Z_{\odot})=\log({\rm O/H})-\log({\rm O/H})_{\odot}$, $12+\log({\rm O/H})_{\odot}=8.69$, and $Z_{\odot}=0.019$ being the solar values \citep{2009ARA&A..47..481A}.}. Therefore, a standard calibration between the LC shape and the M$_{B}$ of SNe~Ia might not be completely valid for objects with chemical abundances which are different to those for which the calibration was made. Therefore, the metallicity may be one source of systematic errors when using these techniques.

The dependence of SNe~Ia luminosity on metallicity was studied by \citet{2005ApJ...634..210G}, who estimated elemental abundances using emission lines from host-galaxy spectra following the \cite{2002ApJS..142...35K} method. They found that most metal-rich galaxies have the faintest SNe~Ia. \citet{2008ApJ...685..752G} analyzed the spectral absorption indices in early-type galaxies, also finding a correlation between magnitudes and the metal abundance of their galaxies, in agreement with the above trend observed for late-type galaxies reported. These results are however not precise enough: \citet{2005ApJ...634..210G} based their conclusion on the analysis of the  Hubble residuals (see their Figure 15a), which implies the use of the own SN~Ia LC to extract the information, while  \citet{2008ApJ...685..752G} used theoretical evolutive synthesis models which still have many caveats, since predictions are very dependent on the code used technique and input spectra, with the extra bias included by the well-known age-metallicity degeneracy in theoretical evolutive synthesis models.

Theoretically, there is a predicted dependence between the maximum luminosity of the SN~Ia and the metallicity of the binary system: assuming the progenitor mass (WD) is constant, the parameter which leads the relation between the light curve width and its maximum magnitude is the opacity of the outer part of the ejected material \citep{1996ApJ...457..500H,2001ApJ...547..988M}, which depends on temperature and, thus, on the heating due to the radioactive $\rm ^{56}Ni$ decay. Then the luminosity of the supernova depends basically on the $\rm ^{56}Ni$ mass ejected from the explosion \citep{1982ApJ...253..785A}:
\begin{equation}
L \propto M(^{56}\rm Ni)\hspace{0.13cm} erg\,s^{-1}.
\label{timmes}
\end{equation} 
\citet{2003ApJ...590L..83T} showed that the neutron excess, which controls the radioactive ($\rm ^{56}Ni$) to non-radioactive (Fe-peak elements) abundance ratio, in the exploding WD is a direct function of the initial metallicity of the WD progenitor. This acts upon the maximum luminosity of the explosion \citep[see][for detailed calculations]{2005A&A...443.1007T,2006astro.ph..8324P}. The maximum luminosity of the SN~Ia depends thus on the initial abundances of C, N, O, and Fe of the progenitor WD. Models by \citet{2003ApJ...590L..83T} predicted this dependence, suggesting that a variation of a factor 3 in the metallicity may cause a variation up to
$\sim 25\%$ in the mass of $^{56}$Ni synthesized during the explosion for initial metallicities higher than solar. 

More recently, \cite{2010ApJ...711L..66B} computed a series of SNe~Ia scenarios, finding an even stronger dependence on metallicity (see their Figure 1 and Eq.~2) than that estimated using \cite{2003ApJ...590L..83T},
\begin{equation} \label{bravo2}
M(^{56}{\rm Ni}) \sim f({\rm Z})=1-0.075\frac{\rm Z}{\rm Z_{\odot}}.
\end{equation}
Following \citet{2007ApJ...655L..93C} prescriptions, \cite{2010ApJ...711L..66B} also explored the dependence of the explosion parameters on the local chemical composition, C mass fraction, and neutronization. These authors found a non-linear relation between the synthesized mass of $^{56}$Ni and the metallicity of the progenitor binary system (see their Figure 1 and Eq.~3):
\begin{equation} M(^{56}Ni) \sim
f({\rm Z})=1-0.18\frac{\rm Z}{\rm Z_{\odot}}\Bigg(1-0.10\frac{\rm Z}{\rm Z_{\odot}}\Bigg).
\label{bravo3}
\end{equation}
This dependence on Z translates into different bolometric LC luminosity-width relationships for different metallicities. Their Figure 3, which plots the LC luminosity-width relationship for three initial metallicities (Z/Z$_{\odot}$=0.1, 1, and 3) and the same LC width, clearly shows this effect: the luminosity is {\it smaller} at higher Z than at lower Z. \cite{2010ApJ...711L..66B} results imply that SNe~Ia located in galaxies with metallicity higher than solar {\it might be dimmer} than expected as compared to those with solar and subsolar abundances.

Since the number of SNe~Ia detections will extraordinarily increase in the forthcoming surveys, statistical errors will decrease while systematic errors will dominate, limiting the precision of SNe~Ia as indicators of extragalactic distances. Hence the importance of  characterizing a possible dependence of the SN~Ia luminosity on the metallicity.

The final purpose of this project is to seek if a dependence between the SNe~Ia maximum luminosity and the metallicity of its host galaxy (provided by the gas-phase oxygen abundance) does exist. For this, we perform a careful analysis of a sample of galaxies of the local Universe hosting SNe~Ia to estimate the oxygen abundances in the regions where those SNe~Ia exploded. Our aim is to perform this analysis in a very basic way, just searching for a simple dependence of the magnitude $M_{B}$ of the SNe~Ia in their LC maximum with the oxygen abundance without using any standardization technique. Therefore,  we build our sample considering local SNe~Ia host galaxies that have distances well determined by methods which are different of those following  SNe~Ia techniques. This way we estimate the absolute peak magnitude for each SN~Ia using the classic equation: $M_{B}=m_{B}-5\log{D} +5$,  eliminating possible problems coming from the use of cosmological techniques.
On the other hand, since galaxies are nearby enough, gas-phase abundances may be estimated in several \Hii~regions across the galaxies, and at different galactocentric distances (GCDs), thus allowing us to derive, in many cases, metallicity gradients. We then use the corresponding value of the oxygen abundance at the same GCD the SN~Ia exploded as a proxy of its metallicity. This method has been already used in \citet{2012A&A...545A..58S, 2016arXiv160307808G}, who also studied galaxies hosting SNe Ia and estimated the oxygen abundances in the regions where the explosions took place, using radial gradients. The technique is different, though. They use the PPAK/PPMAS Integral Field Spectrograph (IFS) mounted in the 3.5m telescope at the Calar Alto Observatory, as part of the CALIFA collaboration project. They obtained oxygen abundances at every position along the galaxy disk of each galaxy. It also allows to obtain azimuthal averaged values and estimate the oxygen abundance at each SN Ia location. We have three galaxies in our sample that are in common with (NGC\,0105, UGC\,04195 and  NGC\,3982), and we use these data to check and improve the accuracy of our results.
In \citet{2016ApJ...818L..19M} we presented our results obtained for non-reddened SNe~Ia ($z\le 0.02$), thus eliminating possible dependences of luminosities on the color of the objects, and found that this dependence on metallicity seems to exist. Our data show a trend, with an 80\% of chance not being due to random fluctuation, between SNe~Ia absolute magnitudes and the oxygen abundances of the host galaxies, in the sense that luminosities tend to be higher for galaxies with lower metallicities. Our result agrees with the theoretical expectations and with other findings suggested in previous works. 

In this paper we present all the details about the \mbox{analysis} of the oxygen abundances derived for our low-redshift SN~Ia host galaxies. Section~2 discusses the sample selection and the data reduction process. The analysis of the spectra and the determination of oxygen abundances and absolute magnitudes for the SN~Ia are described in Sect~3. Section~4 presents our results, which are discussed in Sect.~5. For this we are taking into consideration the SNe~Ia color and stretch parameters used in the classic Supernova Cosmology, performing a principal component analysis to seek dependences among observed parameters, including the oxygen abundance. Our conclusions are given in Sect.~6.

\begin{figure} 
\centering
\includegraphics[width=\linewidth]{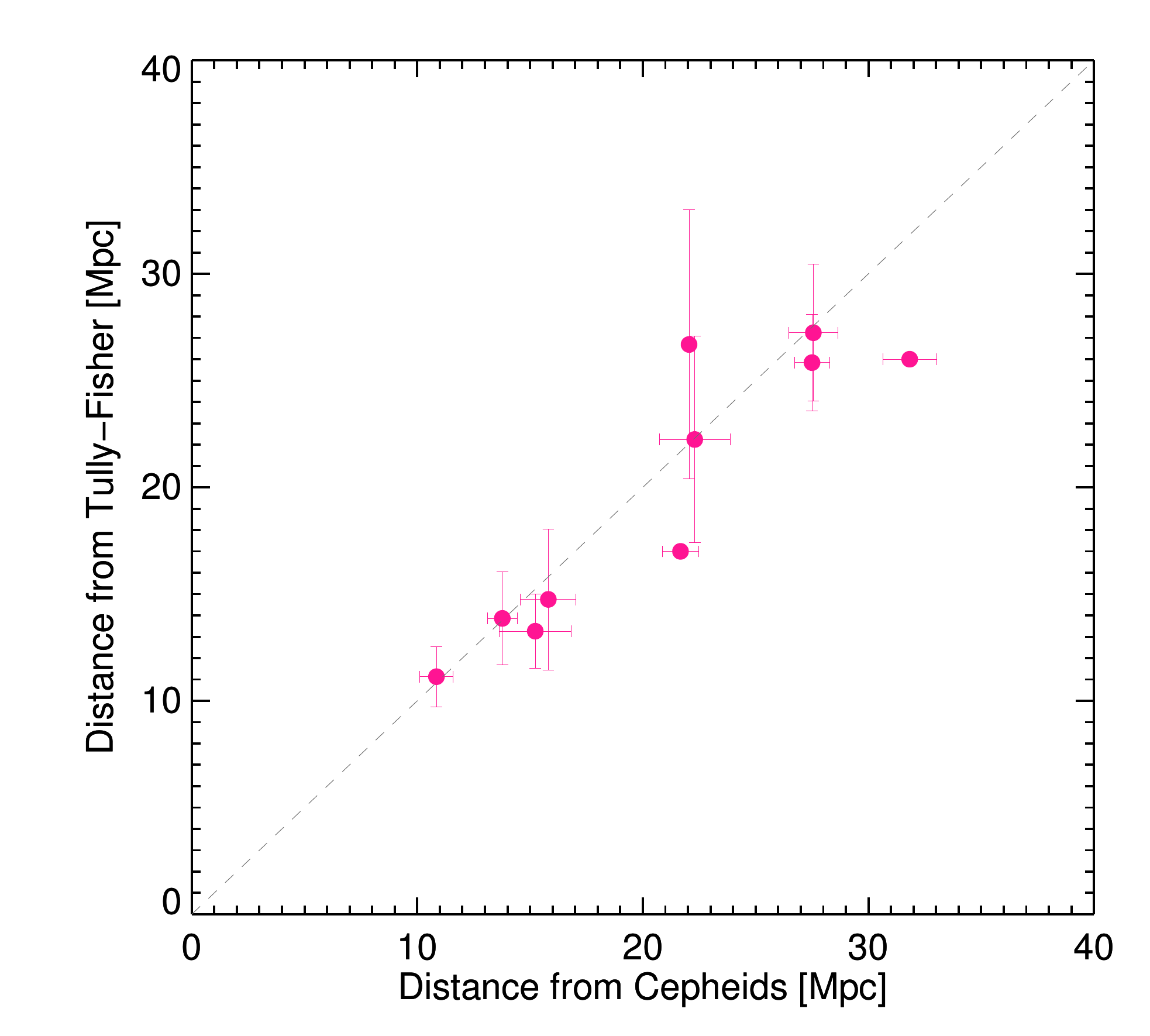}    
\caption{Distances to galaxies having both Tully-Fisher an Cepheids distances. Grey dashed line represents identity. When both measurements are available, a mean distance is calculated considering as many values as possible.}
\label{distance_TF_Ceph}
\end{figure}


\section{Galaxy sample, data reduction}  \label{Section2}

\subsection{Sample selection and distance measurement}\label{sec:sample_sel}

Our sample is selected from \cite{2009ApJ...707.1449N} who lists a large sample of 168 galaxies hosting SNe~Ia. We chose those with $z \le$0.02 and observable from the Observatorio del Roque de los Muchachos (La Palma, Spain), and for which accurate distances {\bf not based on SN~Ia methods} are available.
We obtained intermediate long-slit spectroscopy of a total of 28 galaxies which follow our requested criteria.
The key issue here is that we choose galaxies with distances that are well measured using methods {\it which are different} from those using SNe~Ia. That is, {\it the distances to our sample galaxies are totally independent from SNe~Ia}, because they are not assuming a fixed absolute magnitude. We have exhaustive searched in the NASA Extragalactic Database (NED\footnote{http://ned.ipac.caltech.edu}) and adopted one distance for each galaxy considering the following criteria:
\begin{enumerate}
\item The galaxy have as many independent measurements as possible, including Tully-Fisher, Cepheids, Planetary Nebula Luminosity Functions, etc.
\item If one distance measurement does not agree with the others significantly, we neglect this value  (e.g., early 70s, 80s measurements which have been improved in the present).
\item When possible, we take into consideration measurements from different distance indicators.\\
\end{enumerate}

%
%
%
%
%
%
\begin{table*}
\caption{Galaxies observed at 4.2m-WHT, with their morphology in column 2. Redshifts in columns 3. Columns 4 and 5 show parallactic anges and ratios between major an minor axis. SNe Ia names are shown in column 6; and their positions, in terms of RA, DEC offsets and distance in arcsecs from galactic centers, are in columns 7, 8 and 9. PA, b/a and offsets from Asiago SN catalogue. Distance indicators, number of measurements and distance values are shown in columns 10, 11 and 12. Henceforth, data are shown this same order.}
\begin{tabular}{c@{\hspace{6pt}} c@{\hspace{8pt}} c@{\hspace{8pt}} c@{\hspace{8pt}} c@{\hspace{8pt}} c@{\hspace{8pt}} c@{\hspace{8pt}} c@{\hspace{8pt}} c@{\hspace{8pt}} c@{\hspace{8pt}} c@{\hspace{8pt}} c}\noalign{\smallskip}
\hline
\hline
{Host Galaxy} & {Morphology} &{$z$}&{PA}&{b/a}&{SN Ia}&{RA offset}&{DEC offset}&{Separation} &{Distance} &{Number of} &{Distance}\\
{}&{}&{}&{[deg]}&{}&{}&{[arcsec]}&{[arcsec]}&{[arcsec]}&{indicator}&{measures}&{[Mpc]} \\
 \hline 
\noalign{\smallskip}

M\,82			&	I0		&		0.000677	&	155	&	0.37	&	2014J  	&	$-$54.0	&	$-$21.0	&	57.9		&	PNLF 		&	6	&	3.8		 $\pm$ 	0.7	 \\
MCG-02-16-02		&	Sb?		&		0.007388	&	15	&	0.19	&	2003kf 	&	+9.2    	&	$-$14.3	&	17.0		&	T-F			&	2	&	22.6		 $\pm$ 	0.8	 \\
NGC\,0105		&	Sab:		& 		0.017646	&	77	&	0.64	&	1997cw 	&	+7.6	          &	+4.2   	&	8.7		&	SNIa			&	6	&	64.2		 $\pm$ 	5.9	 \\
NGC\,1275		&	S0		&		0.017559	&	20	&	0.64	&	2005mz 	&	+19.2	&	$-$23.6	&	30.4		&	T-F			&	2	&	61.4		 $\pm$ 	7.5	 \\
NGC\,1309		&	Sbc:		&		0.007125	&	135	&	0.93	&	2002fk 	&	$-$12.0	&	$-$3.5	&	12.5		&	Ceph \& T-F	&	5	&	29.3		 $\pm$ 	0.9	 \\
NGC\,2935		&	SBb		&		0.007575	&	90	&	0.73	&	1996Z  	&	+0.0	          &	$-$70.0	&	70.0		&	T-F			&	12	&	28.2		 $\pm$ 	3.2	 \\
NGC\,3021		&	Sbc		&		0.005140	&	20	&	0.56	&	1995al 	&	$-$15.0	&	+2.9   	&	15.3		&	Ceph \& T-F	&	11	&	26.3		 $\pm$ 	1.9	 \\
NGC\,3147		&	Sbc		&		0.009346	&	65	&	0.86	&	1997bq 	&	+50.0	&	$-$60.0	&	78.1		&	T-F		         &	1	&	40.9		 $\pm$ 	4.1	 \\
NGC\,3169		&	Sa		&		0.004130	&	135	&	0.50	&	2003cg 	&	+14.0	&	+5.0   	&	14.9		&	T-F			&	3	&	17.1		 $\pm$ 	2.9	 \\
NGC\,3368		&	Sab		&		0.002992	&	95	&	0.69	&	1998bu 	&	+4.3	         &	+55.3	&	55.5		&	Ceph \& T-F	&	27	&	11.0		 $\pm$ 	1.0	 \\
NGC\,3370		&	Sc		&		0.004266	&	58	&	0.56	&	1994ae 	&	$-$30.3	&	+6.1    	&	30.9		&	Ceph \& T-F	&	24	&	27.4		 $\pm$ 	1.9	 \\
NGC\,3672		&	Sc		&		0.006211	&	98	&	0.47	&	2007bm 	&	$-$2.4	&	$-$10.8	&	11.1		&	T-F			&	3	&	22.8		 $\pm$ 	1.8	 \\
NGC\,3982		&	Sb:		&		0.003699	&	90	&	0.87	&	1998aq 	&	$-$18.0	&	+7.0	         &	19.3		&	Ceph \& T-F	&	25	&	21.5		 $\pm$ 	0.8	 \\
NGC\,4321		&	SBbc	&		0.005240	&	120	&	0.81	&	2006X  	&	$-$12.0	&	$-$48.0	&	49.5		&	Ceph \& T-F	&	35	&	15.5		 $\pm$ 	1.9	 \\
NGC\,4501		&	Sb		&		0.007609	&	50	&	0.53	&	1999cl 	&	$-$46.0	&	+23.0	&	51.4		&	T-F			&	12	&	20.7		 $\pm$ 	3.2	 \\
NGC\,4527		&	SBbc	&		0.005791	&	157	&	0.33	&	1991T  	&	+26.0	&	+45.0	&	52.0		&	Ceph \& T-F	&	21	&	13.8		 $\pm$ 	1.4	 \\
NGC\,4536		&	SBbc	&		0.006031	&	40	&	0.39	&	1981B  	&	+41.0	&	+41.0	&	58.0		&	Ceph \& T-F	&	53	&	14.8		 $\pm$ 	1.6	 \\
NGC\,4639		&	SBbc	&		0.003395	&	33	&	0.66	&	1990N  	&	+63.2	&	$-$1.8	&	63.2		&	Ceph \& T-F	&	44	&	22.3		 $\pm$ 	2.1	 \\
NGC\,5005		&	Sbc		&		0.003156	&	155	&	0.48	&	1996ai 	&	+24.0	&	+4.0   	&	24.3		&	T-F			&	4	&	23.2		 $\pm$ 	2.1	 \\
NGC\,5468		&	Scd		&		0.009480	&	15	&	0.91	&	1999cp 	&	$-$52.0	&	+23.0	&	56.9		&	T-F			&	1	&	41.5		 $\pm$ 	4.2	 \\
NGC\,5584		&	Scd		&		0.005464	&	50	&	0.72	&	2007af 	&	$-$40.0	&	$-$22.0	&	45.7		&	Ceph \& T-F	&	10	&	24.3		 $\pm$ 	1.3	 \\
UGC\,00272		&	Sd		&		0.012993	&	40	&	0.42	&	2005hk 	&	+17.2	&	+6.9   	&	18.5		&	T-F			&	2	&	60.5		 $\pm$ 	3.7	 \\
UGC\,03218		&	Sb		&		0.017432	&	55	&	0.37	&	2006le 	&	$-$12.4	&	+40.1	&	42.0		&	T-F			&	4	&	59.0		 $\pm$ 	6.0	 \\
UGC\,03576		&	SBb		&		0.019900	&	38	&	0.54	&	1998ec 	&	$-$8.7	&	+19.5	&	21.4		&	T-F			&	5	&	87.4		 $\pm$ 	8.2	 \\
UGC\,03845		&	SBbc	&		0.010120	&	86	&	0.67	&	1997do 	&	$-$2.6	&	$-$3.8	&	4.6		&	T-F			&	1	&	38.5		 $\pm$ 	3.9	 \\
UGC\,04195		&	SBb		&		0.016305	&	110	&	0.50	&	2000ce 	&	+15.1	&	+17.3	&	23.0		&	T-F			&	4	&	78.8		 $\pm$ 	2.2	 \\
UGC\,09391		&	SBdm	&		0.006384	&	110	&	0.59	&	2003du 	&	$-$8.8	&	$-$13.5	&	16.1		&	T-F			&	1	&	31.8		 $\pm$ 	3.2	 \\
UGCA\,017		&	Scd:		&		0.006535	&	111	&	0.16	&	1998dm 	&	$-$13.8	&	$-$37.0	&	39.5		&	T-F			&	13	&	26.3		 $\pm$ 	4.4	 \\

\hline
\hline
\end{tabular}
\label{tab:paper_table_1}
\end{table*}

The distance we finally adopt for each galaxy is given by the mean of all the considered measurements, within its standard deviation. For those galaxies having distance measurements from Tully-Fisher and from Cepheids we have checked there is not a bias depending on the chosen distance indicator. Figure \ref{distance_TF_Ceph} shows that it is equivalent to choose either Tully-Fisher or Cepheids distances.

It is crucial that we do not select distances obtained from the SNe~Ia studies. As it is shown in \citet{2016ApJ...818L..19M} a discrepancy appears when distances obtained using independent and dependent indicators from SNe~Ia are compared. SN~Ia techniques provide overestimated distances compared with other methods, and the effect becomes more important at greater distances. 

Table \ref{tab:paper_table_1} shows the sample, sorted by the name of the host galaxies. This table lists the galaxies morphology, their redshifts, parallactic angle and semi-axis ratio. Introduces as well the SN~Ia they host and shows their positions within the galaxies. Distance indicators and distances are also provided.

\begin{figure} 
\centering
\includegraphics[width=\linewidth]{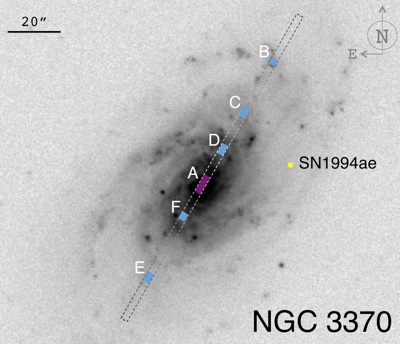}
\caption{$R-$Band~images for NGC\,3370 hosting SN1994ae. Slit size is not to scale. Blue zones correspond to regions in which we have obtained spectra, whereas purple zones represent regions without emission lines. SNe~Ia positions are marked with a yellow dot. All figures can be seen in Appendix~\ref{App:B}.}
\label{rendijas}
\end{figure}

\subsection{Observations}

We obtained intermediate-resolution long slit spectroscopy for all our sample of local SN~Ia host galaxies. For this we used the 4.2m \textit{William Herschel Telescope} (WHT), located at the Roque de los Muchachos Observatory (ORM, La Palma, Canary Islands, Spain). We completed two observation runs on December 2011 and January 2014. In both cases, the double-arm ISIS (Intermediate dispersion Spectrograph and Imaging System) instrument located at the Cassegrain focus was used. The dichroic used to separate the blue and red beams was set at 5400 \AA. The slit was 3.7' long and 1" wide. As an example, Figure \ref{rendijas} shows the slit positions over three galaxies. Blue zones correspond to regions with extracted useful spectra, while the location of the SN~Ia is represented with a yellow circle. Appendix~\ref{App:B} provides the same figures for all our sample galaxies.

The observational details of each observing run are:
\begin{enumerate}
\item {\bf December 2011}. We were granted with two observation nights, $\rm 22^{nd}$ and $\rm 23^{rd}$ of December, in which we observed 12 galaxies following this set-up:
\begin{itemize}
\item {\bf Blue arm.} An EVV CCD with a $4096 \times 2048$ pixels array and 13.5 $\mu$m size was used with a spatial scale of 0.20" pix$^{-1}$. The grating used was the R600B, giving a dispersion of 33 \AA\,mm$^{-1}$ (0.45 \AA~pix$^{-1}$).
\item {\bf Red arm}. We used a REDPLUS CCD with a configuration of $4096 \times 2048$ pixels of 24 $\mu$m pixel size, having a spatial scale of 0.22" pix$^{-1}$. We used the grating R600R, which has a dispersion of 33 \AA\,mm$^{-1}$ (0.49 \AA\,pix$^{-1}$).
\end{itemize}

Table~\ref{longitudes} compiles the different central wavelengths used among these two nights due to the redshift range of the sample, in order to cover the emission lines for the subsequent analysis.

\item{\bf January 2014:} 16 galaxies were observed during the four nights of this run: from $\rm 23^{rd}$ to $\rm 26^{th}$ January. We used the same instrumental setup that for the December 2011 run. However, in this case only one pair of central wavelengths was enough to cover the whole spectral range. The central wavelengths of this configuration and other details  are also compiled in Table~\ref{longitudes}.
\end{enumerate}

\begin{figure*}
\centering
\includegraphics[width=0.99\linewidth]{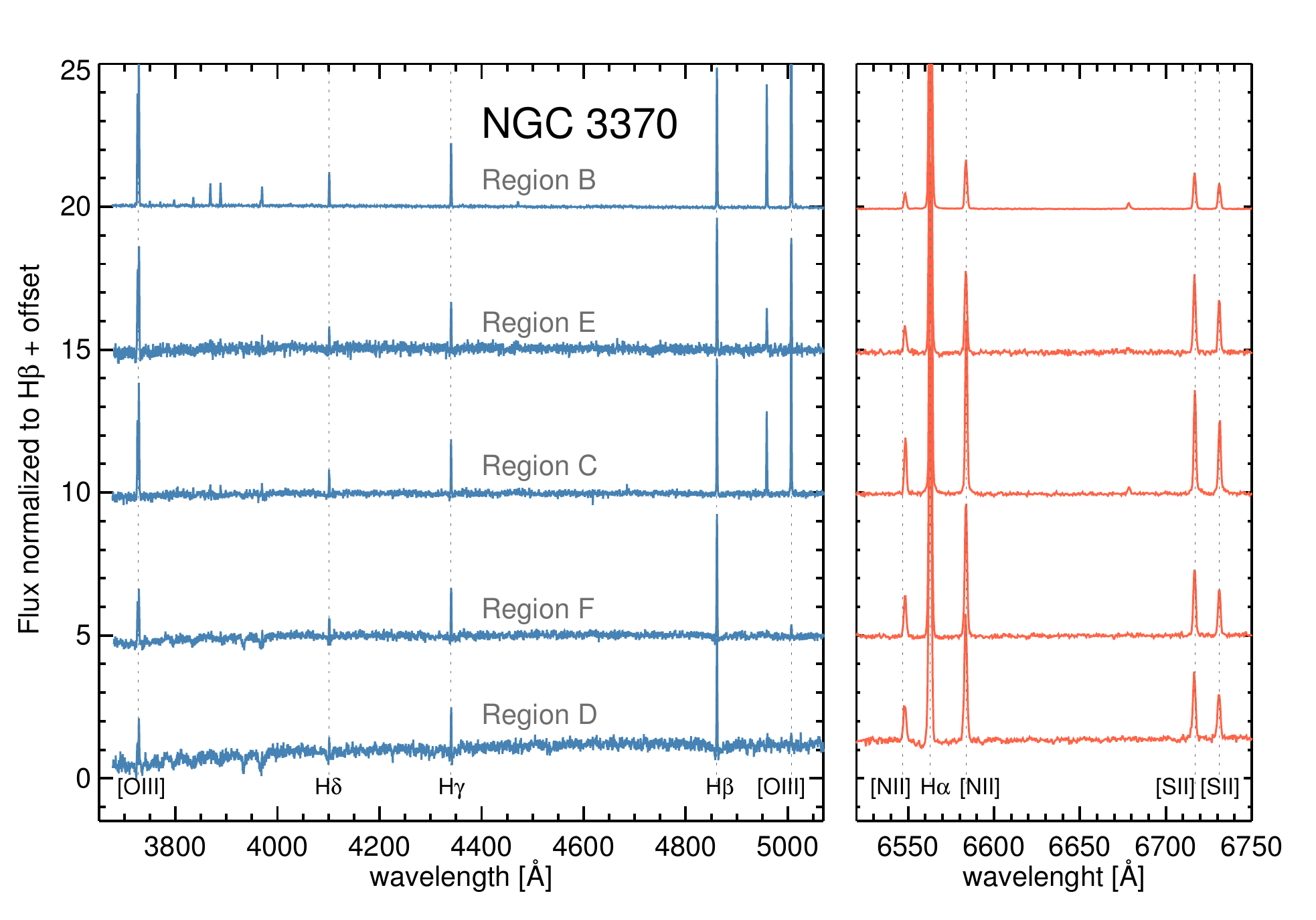}  
\caption{Spectra obtained for the five \Hii~regions analyzed in NGC\,3370, as labelled in top panel of Figure \ref{rendijas}. The main emission lines are identified. Note that the GCD of the  \Hii~regions decreases from top (region~B) to bottom (region~D).}
\label{espectro_prueba}
\end{figure*}

\begin{table}
\centering
\scriptsize
\caption{Central wavelengths for both the blue and red ISIS arms for the four configurations used in our observations. Identifications $z_0$, $z_1$ and $z_2$ refer to the December 2011 run. $z_{2014}$ refers to those used for the January 2014 run.}
\begin{tabular}{c@{\hspace{10pt}} cc@{\hspace{10pt}}cc@{\hspace{10pt}}  }\noalign{\smallskip}
\hline
\hline
{Configuration} & \multicolumn{2}{c}{ Central $\lambda$ [\AA] } & \multicolumn{2}{c}{ Spectral range $\lambda$ [\AA] }   \\
{}&{Blue arm}&{Red arm} & {Blue arm}&{Red arm} \\
 \hline 
\noalign{\smallskip}

z$_{0}$ & 4368 & 6720 & 3560 - 5186  & 5892 - 7538  \\
z$_{1}$ & 4452 & 6950 & 3652 - 5284 &  6122 - 7768 \\
z$_{2}$ & 4549 & 7023 & 3749 - 5385 & 6195 - 7841   \\
z$_{2014}$ & 4438 & 6783 & 3630 - 5266 &  5995 - 7601  \\

\hline
\hline
\end{tabular}
\label{longitudes}
\end{table}

Arc lamps were used for the wavelength calibration of the spectra. Specifically, the lamps used were `CuAr' for the blue arm, and `CuAr+CuNe' for the red arm.
The absolute flux calibration was achieved by observing the standard stars Hiltner 600, HD 19445 and HD 84937 \citep{1977ApJ...218..767S, 1983ApJ...266..713O}. Between two and four individual exposures were obtained for each slit position in order to get  a good signal-to-noise ratio ($SNR$) and achieve a proper cosmic rays removal. Table~\ref{tab:setup_table} compiles all the intermediate-resolution long-slit spectroscopy observations performed for the 28 SN~Ia host galaxies included in this paper.

%
%
%
%
%

\begin{table*}

\caption{Instrument specifications for the observations taken at the 4.2m WTH in both runs. Galaxies are sorted as in Table \ref{tab:paper_table_1}.}
\begin{tabular}{c@{\hspace{8pt}} c@{\hspace{8pt}}c@{\hspace{8pt}} c@{\hspace{8pt}} c@{\hspace{8pt}} c@{\hspace{8pt}} c@{\hspace{8pt}} c}\noalign{\smallskip}
\hline
\hline
{Host Galaxy} &{SN Ia} &{Date}&{Exp. Time}&{Spatial R.}&{Grism}&{P.A.}&{Airmass} \\
{}&{}&{}&{[s]}&{[$" {\rm pix}^{-1}$]}&{}&{$\circ$}&{}\\
 \hline 
\noalign{\smallskip}		

M\,82		&	2014J	&	24-Jan-2014	&	3x600	&	0.20	&	R600B	&	248		&	1.699358	\\
			&			&	24-Jan-2014	&	4x300	&	0.22	&	R600R	&	248		&	1.703212	\\
MCG-02-16-02	&	2003kf	&	24-Jan-2014	&	2x1800	&	0.20	&	R600B	&	290		&	1.415056	\\
			&			&	24-Jan-2014	&	2x1800	&	0.22	&	R600R	&	290		&	1.414584	\\
NGC\,0105	&	1997cw	&	26-Jan-2014	&	2x1800	&	0.20	&	R600B	&	-6		&	1.358633	\\
			&			&	26-Jan-2014	&	2x1800	&	0.22	&	R600R	&	-6		&	1.433069	\\
NGC\,1275	&	2005mz	&	26-Jan-2014	&	2x1800	&	0.20	&	R600B	&	62		&	1.196434	\\
			&			&	26-Jan-2014	&	2x1800	&	0.22	&	R600R	&	62		&	1.196645	\\
NGC\,1309	&	2002fk	&	26-Jan-2014	&	2x1800	&	0.20	&	R600B	&	28		&	1.487035	\\
			&			&	26-Jan-2014	&	2x1800	&	0.22	&	R600R	&	28		&	1.487479	\\
NGC\,2935	&	1996Z	&	25-Jan-2014	&	2x1800	&	0.20	&	R600B	&	329		&	1.753246	\\
			&			&	25-Jan-2014	&	2x1800	&	0.22	&	R600R	&	329		&	1.689944	\\
NGC\,3021	&	1995al	&	22-Dec-2011	&	2x1200	&	0.20	&	R600B	&	93		&	1.285784	\\
			&			&	22-Dec-2011	&	2x1200	&	0.22	&	R600R	&	93		&	1.285847	\\
NGC\,3147	&	1997bq	&	22-Dec-2011	&	2x1200	&	0.20	&	R600B	&	320		&	1.509696	\\
			&			&	22-Dec-2011	&	2x1200	&	0.22	&	R600R	&	320		&	1.542244	\\
NGC\,3169	&	2003cg	&	23-Dec-2011	&	2x1200	&	0.20	&	R600B	&	172		&	1.262523	\\
			&			&	23-Dec-2011	&	2x1200	&	0.22	&	R600R	&	172		&	1.295006	\\
NGC\,3368	&	1998bu	&	24-Jan-2014	&	2x1800	&	0.20	&	R600B	&	298		&	1.306252	\\
			&			&	24-Jan-2014	&	2x1800	&	0.22	&	R600R	&	298		&	1.307326	\\
NGC\,3370	&	1994ae	&	22-Dec-2011	&	2x1800	&	0.20	&	R600B	&	330		&	1.160587	\\
			&			&	22-Dec-2011	&	2x1800	&	0.22	&	R600R	&	330		&	1.125012	\\
NGC\,3672	&	2007bm	&	25-Jan-2014	&	2x1800	&	0.20	&	R600B	&	364		&	1.497680	\\
			&			&	25-Jan-2014	&	2x1800	&	0.22	&	R600R	&	364		&	1.498453	\\
NGC\,3982	&	1998aq	&	24-Jan-2014	&	2x1800	&	0.20	&	R600B	&	335		&	1.154017	\\
			&			&	24-Jan-2014	&	2x1800	&	0.22	&	R600R	&	335		&	1.140907	\\
NGC\,4321	&	2006X	&	22-Dec-2011	&	2x1800	&	0.20	&	R600B	&	340		&	1.134375	\\
			&			&	22-Dec-2011	&	2x1800	&	0.22	&	R600R	&	340		&	1.171063	\\
NGC\,4501	&	1999cl	&	25-Jan-2014	&	2x1800	&	0.20	&	R600B	&	322		&	1.145303	\\
			&			&	25-Jan-2014	&	3x1800	&	0.22	&	R600R	&	322		&	1.145677	\\
NGC\,5005	&	1996ai	&	22-Dec-2011	&	2x1200	&	0.20	&	R600B	&	250		&	1.080511	\\
			&			&	22-Dec-2011	&	2x1200	&	0.22	&	R600R	&	250		&	1.096719	\\
NGC\,4527	&	1991T	&	24-Jan-2014	&	2x1800	&	0.20	&	R600B	&	64		&	1.266297	\\
			&			&	24-Jan-2014	&	2x1800	&	0.22	&	R600R	&	64		&	1.317050	\\
NGC\,4536	&	1981B	&	24-Jan-2014	&	2x1800	&	0.20	&	R600B	&	265		&	1.663660	\\
			&			&	24-Jan-2014	&	2x1800	&	0.22	&	R600R	&	265		&	1.666170	\\
NGC\,4639	&	1990N	&	24-Jan-2014	&	2x1800	&	0.20	&	R600B	&	335		&	1.060286	\\
			&			&	24-Jan-2014	&	2x1800	&	0.22	&	R600R	&	335		&	1.077132	\\
NGC\,5468	&	1999cp	&	26-Jan-2014	&	2x1800	&	0.20	&	R600B	&	7		&	2.028415	\\
			&			&	26-Jan-2014	&	2x1800	&	0.22	&	R600R	&	7		&	2.030882	\\
NGC\,5584	&	2007af	&	25-Jan-2014	&	2x1800	&	0.20	&	R600B	&	308		&	1.439699	\\
			&			&	25-Jan-2014	&	2x1800	&	0.22	&	R600R	&	308		&	1.438900	\\
UGC\,00272	&	2005hk	&	22-Dec-2011	&	4x600	&	0.20	&	R600B	&	126		&	1.160984	\\
			&			&	22-Dec-2011	&	4x600	&	0.22	&	R600R	&	126		&	1.158374	\\
UGC\,03218	&	2006le	&	25-Jan-2014	&	2x1800	&	0.20	&	R600B	&	142		&	1.200891	\\
			&			&	25-Jan-2014	&	2x1800	&	0.22	&	R600R	&	142		&	1.200936	\\
UGC\,03576	&	1998ec	&	25-Jan-2014	&	2x1800	&	0.20	&	R600B	&	130		&	1.073760	\\
			&			&	25-Jan-2014	&	2x1800	&	0.22	&	R600R	&	130		&	1.073590	\\
UGC\,03845	&	1997do	&	26-Jan-2014	&	2x1800	&	0.20	&	R600B	&	214		&	1.055408	\\
			&			&	26-Jan-2014	&	2x1800	&	0.22	&	R600R	&	214		&	1.055376	\\
UGC\,04195	&	2000ce	&	26-Jan-2014	&	2x1800	&	0.20	&	R600B	&	188		&	1.268553	\\
			&			&	26-Jan-2014	&	2x1800	&	0.22	&	R600R	&	188		&	1.268342	\\
UGC\,09391	&	2003du	&	23-Dec-2011	&	3x900	&	0.20	&	R600B	&	190		&	1.765348	\\
			&			&	23-Dec-2011	&	3x900	&	0.22	&	R600R	&	190		&	1.810054	\\
UGCA\,017	&	1998dm	&	23-Dec-2011	&	3x1200	&	0.20	&	R600B	&	202		&	1.584673	\\
			&			&	23-Dec-2011	&	3x1200	&	0.22	&	R600R	&	202		&	1.529884	\\

\hline
\hline
\end{tabular}
\label{tab:setup_table}
\end{table*}

The slit position was fixed independently for each galaxy by looking at the acquisition image. Usually the slit was orientated following the direction that allowed to observe the highest number of \HII~regions. This orientation does not coincide with the SNe~Ia positions, as the SNe~Ia neighborhood usually lacked of measurable star-formation activity. Hence, we prioritized that position angle (P.A.) that provides the largest number of \HII~regions in order that we could use their data to derive a metallicity gradient for the galaxy.

\subsection{Reduction of the spectra}

{\sc Iraf}\footnote{Image Reduction Analysis Facility, distributed by the National Optical Astronomy Observatories (NOAO), which is operated by AURA Inc., under cooperative agreement with NSF}  software was used to reduce the CCD frames (debiasing, flat-fielding, cosmic-ray rejection, wavelength and flux calibration, sky subtraction) and to extract the one-dimensional spectra. Correction for atmospheric extinction was performed using an average curve for the continuous atmospheric extinction at Roque de los Muchachos Observatory. For each two-dimensional spectrum several apertures were defined along the spatial direction to extract the final one-dimensional spectra of each galaxy or emission knot. The apertures were centered at the brightest point of each aperture and the width was fixed to obtain a good $SNR$ in each spectrum. In this case, we have the optical spectrum separated in two different wavelength intervals with different spatial resolutions, so it was essential to be precise in order to get identical apertures in both spectral ranges. For the blue arm, \Hb~was the reference line to extract these apertures,while in the red arm, was \Ha~the used line.

Figure \ref{espectro_prueba} shows a set of spectra for NGC\,3370. The main emission line features are identified with labels. The 1D spectra usually have a high $SNR$ for all the lines (e.g. $\sim$22, and always over 6 for \Hb).


\section{Analysis} \label{Section3}

\subsection{Line measurements}

For the 28 galaxies, a total of 102 apertures have been extracted. 13 of these apertures lack of measurable emission lines, and hence these have not been included in the subsequent analysis. For the remaining 89 regions, we  tried to measure all the emission lines we are interested in. In most of the spectra we have measured eight emission lines: three hydrogen Balmer lines (\Ha, \Hb, \Hg) and the brightest collisional excited lines of metallic elements: \Oii~$\lambda\lambda$3726,29 (blended), \Oiii~$\lambda$5007, \Nii $\lambda$6583, \Sii~$\lambda$6716 and \Sii~$\lambda$6731. If a line is measured and its $SNR \le 3$, we do not consider it. Line intensities and equivalent widths were measured integrating all the emission between the limits of the line and over a local adjacent continuum. All these measurements were made with the {\sc splot} routine of {\sc Iraf}. However, due to the faintness of some of the detected emission lines, a detailed inspection of the spectra was needed to get a proper estimation of the adjacent continuum and the line flux in these cases. Uncertainties in the line fluxes were estimated for each line considering both the $rms$ of the continuum and the width of each emission line.

\subsection{Correction for reddening}

Interstellar medium redden spectra as dust blocks more efficiently the blue wavelengths than the red wavelengths. Hence, emission lines fluxes should be reddening-corrected to get appropriate flux ratios. This correction is usually made by using the Balmer decrement between the $H\alpha$ and $H\beta$ line fluxes according to: 
\begin{equation}
\frac{I(\lambda)}{I({\rm H\beta})} = \frac{F(\lambda)}{F({\rm H\beta})}10^{c({\rm H\beta})[f(\lambda)-f({\rm H\beta})]},
\label{redeq}
\end{equation}
where $I(\lambda)/I({\rm H\beta})$ is the line intensity flux unaffected by reddening or absorption, $F(\lambda)/F({\rm H\beta})$ is the observed line measured flux, $c({\rm H\beta})$ is the reddening coefficient and $f(\lambda)$ is the reddening curve normalized to \Hb\ using the \citet{1989ApJ...345..245C} law.

To calculate the reddening coefficient, $c({\rm H\beta})$, both H${\rm \alpha}$ and H${\rm \beta}$ are usually used in Eq.~\ref{redeq}. Sometimes we can measure other pairs of \Hi\ Balmer lines --e.g., \Hg/\Hb\ or H$\delta$/\Hb--  and the reddening coefficient can be determined with higher accuracy \citep[e.g., see][]{2009A&A...508..615L} and agrees with that derived using the \HI\ Paschen lines \citep{2007ApJ...656..168L}. However, in extragalactic objects the fluxes of nebular Balmer lines are affected by absorption produced by the underlying stellar population (mainly B and A stars). We have considered in our analysis that the Balmer lines are indeed affected by underlying stellar absorptions. Including the absorption in the hydrogen lines, $W_{abs}$ , which we assume are the same for all the Balmer lines within the same object, the reddening coefficient can be derived applying
\begin{equation}
c({\rm H\beta})=\frac{1}{f(\lambda)} \log\Bigg[\frac{\frac{I(\lambda)}{I({\rm H\beta})}\times
\Big(1+\frac{W_{abs}}{W_{{\rm H\beta}}}\Big)} {\frac{F(\lambda)}{F({\rm H\beta})}\times 
\Big(1+\frac{W_{abs}}{W_{\lambda}}\Big)}\Bigg],
\end{equation}
as introduced by \citet{1993ApJS...85...27M}. In this equation, $W_{abs},\ W_{\lambda}$, and $W_{\rm H\beta}$ are the equivalent widths of the underlying stellar absorption, the considered Balmer line, and H$\beta$, respectively.

Typical analysis of star-forming regions always consider that the theoretical $I$(\Ha)/$I$(\Hb) ratio is 2.86, following the case B recombination for an electron temperature of \Te=10000\,K and electron density of \Ne=100\,cm$^{-2}$. However, the theoretical \Hii~Balmer ratios also depend \mbox{--although} weakly-- on the electron temperature \citep{1995MNRAS.272...41S}. 
Objects with \Te$\sim$15000\,K have $I$(\Ha)/$I$(\Hb)=2.86, whereas objects with \Te$\sim$5000\,K have $I$(\Ha)/$I$(\Hb)=3.01. These values are also related to the oxygen abundance of the ionized gas, in the sense that objects with higher (lower) electron temperature have lower (higher) oxygen abundances. We have used the prescriptions given by \citet{2015MNRAS.450.3381L} --see their Appendix~B-- to consider the electron temperature dependence of the theoretical \Hi~Balmer line ratios assuming the best value to the oxygen abundance provided by the empirical calibrations (see next subsection).

Appendix~\ref{App:A} provides the line intensities for the regions which make up the sample. Redden coefficient is also shown here.

\subsection{Nature of the emission\label{diagnostic}}

\begin{figure}
\centering
\includegraphics[width=0.99\linewidth]{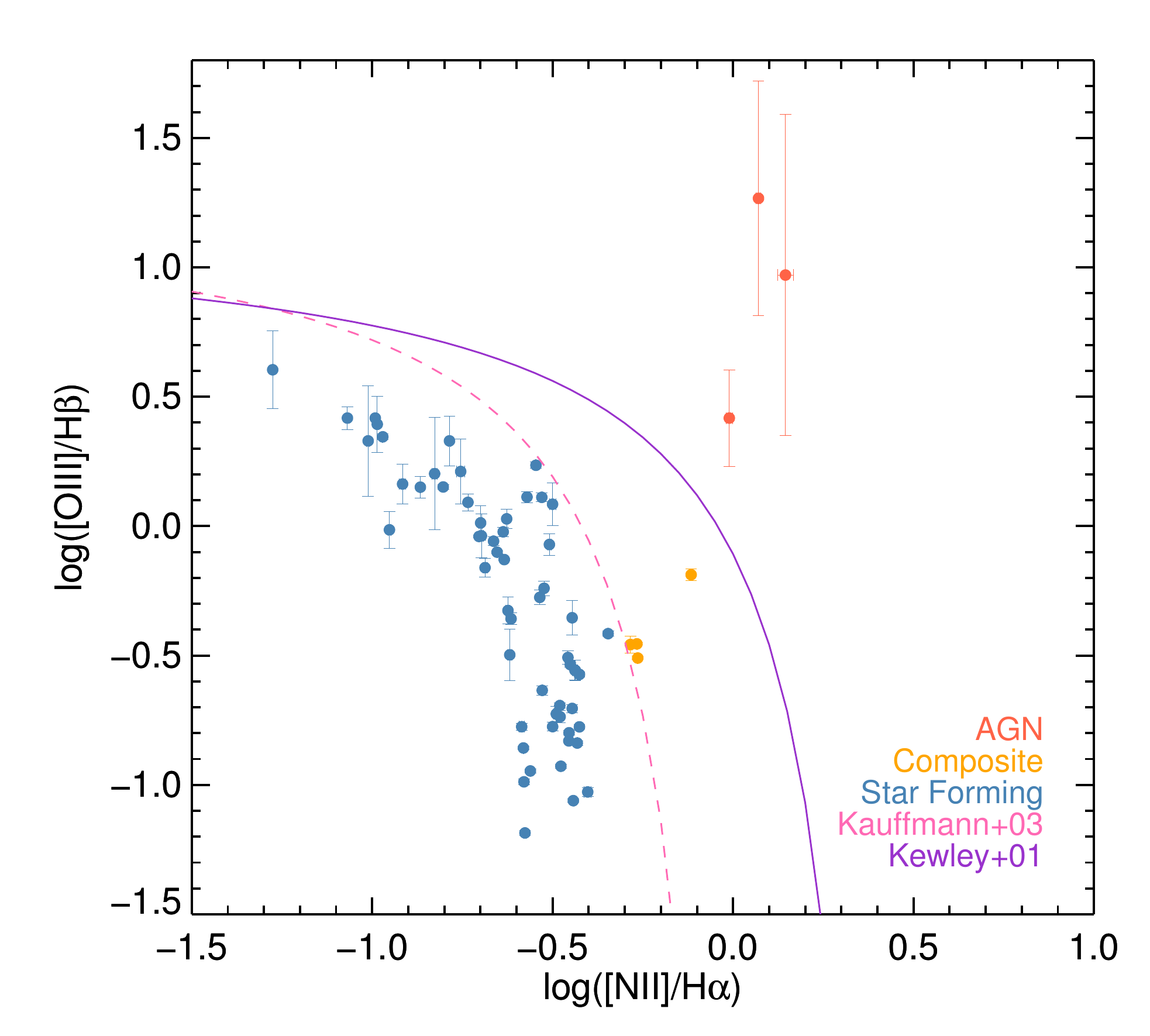}  
\includegraphics[width=0.99\linewidth]{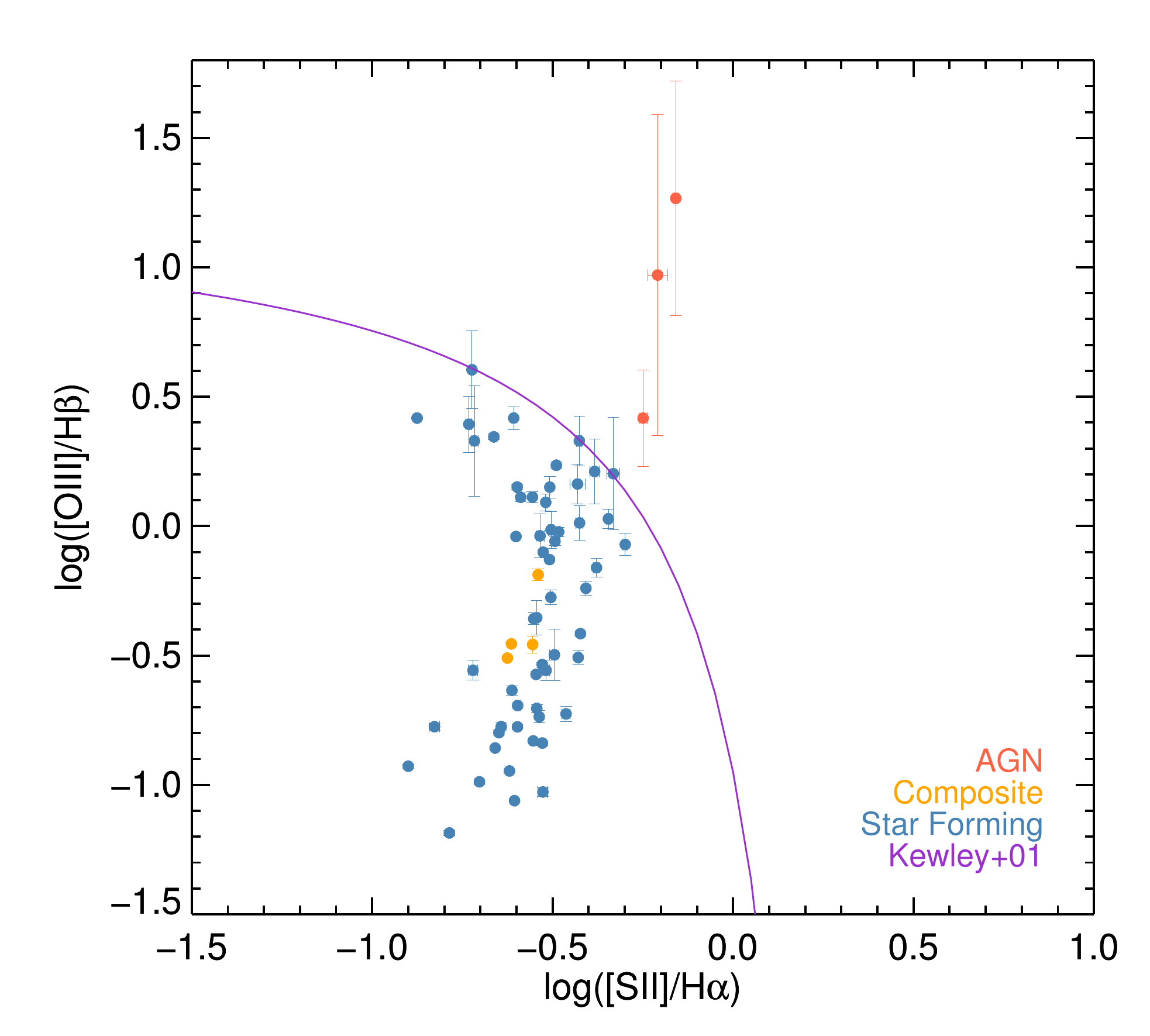} 
\caption{BPT diagrams comparing the observational flux ratio \Oiii~$\lambda$5007/\Hb~($y$-axis) with the \Nii~$\lambda$6583/\Ha~(top) and (\Sii~$\lambda$6716+$\lambda$6731)/\Ha~(bottom) flux ratios ($x$-axis) obtained for our galaxy sample. The theoretical line proposed by \citet{2001ApJ...556..121K} is plotted using a purple continuous line, while the empirical line obtained by \citet{2003MNRAS.346.1055K} is shown with a pink dashed line. Red circles indicate galaxies classified as AGN, orange circles are the galaxies with composite nature, while blue circles represent pure star-forming galaxies.}
\label{diagnosticFIG}
\end{figure}

We first check the nature of the ionization of the gas within the regions observed in our sample galaxy. For this we use the so-called diagnostic or BPT diagrams, firstly proposed by \citet{1981PASP...93....5B} and \citet{1987ApJS...63..295V}. These diagrams are excellent tools for distinguishing between Active Galactic Nucleus (AGN) or Low-Ionization Narrow-Emission Region (LINER) activity and pure star-forming regions since \HII~regions and starburst galaxies lie within a narrow band. Figure~\ref{diagnosticFIG} the typical BPT diagrams considering \Oiii~$\lambda$5007/\Hb~versus \Nii~$\lambda$6583/\Ha~(top) and \Oiii~$\lambda$5007/\Hb~versus (\Sii~$\lambda$6716+$\lambda$6731)/\Ha~(bottom).   

63 of the observed spectra have all the emission lines needed for classifying their nature following BPT diagram; their data points are included in Figure \ref{diagnosticFIG}. For the remaining 26 spectra the\Oiii~$\lambda$5007/\Hb~flux radio was not available, and hence these regions couldn't be classified following the BPT method.
Figure~\ref{diagnosticFIG} also include the analytic relations given by \citet{2001ApJ...556..121K} --continuous purple line--, which were derived for starburst galaxies and which represent an upper envelope of positions of star-forming galaxies.

For regions where the gas is excited by shocks, accretion disks, or cooling flows (as is in the case of AGNs or LINERs) their position in these diagnostic diagrams is away from the locus of \HII\ regions, and usually well above the theoretical \citet{2001ApJ...556..121K} curves. Hence, the nature of the ionization of the gas within these areas is not due to massive stars, and therefore they cannot be classified as star-forming regions. Data points not classified as \HII\ regions are colored in red in Figure \ref{diagnosticFIG} and are not further considered in our analysis.  We find only 3 of these regions, that are therefore classified as AGN. Thus we have finally classified 60 spectra as coming from star-forming regions.

The top panel of Figure \ref{diagnosticFIG} includes the empirical relation between the \Oiii~$\lambda$5007/\Hb~and the \Nii~$\lambda$6583/\Ha~provided by \cite{2003MNRAS.346.1055K} --dashed pink line- after analyzing a large data sample of star-forming galaxies from SDSS data is also drawn. All galaxies lying below this curve are considered to be pure star-forming objects, and hence are plotted using a blue color in the figure. 

\citet{2006MNRAS.372..961K} suggested that those objects between the theoretical line computed by \citet{2001ApJ...556..121K} and the empirical line found by \citet{2003MNRAS.346.1055K} may be ionized by both massive stars and shocks, i.e., to have a composite nature, although \citet{2009MNRAS.398..949P} showed that objects located in this area may also be pure star-forming galaxies with high \Nii~intensities due to a high N content. In these cases we use an orange color to distinguish these regions, which will be also analyzed in our work. All regions classified as AGN in the top panel of Figure \ref{diagnosticFIG} lie above the theoretical curve shown in the bottom panel, too. Similarly, both star-forming and composite objects lie below this curve in the bottom panel, within the errors.

\begin{figure} 
\centering
\includegraphics[width=0.99\linewidth]{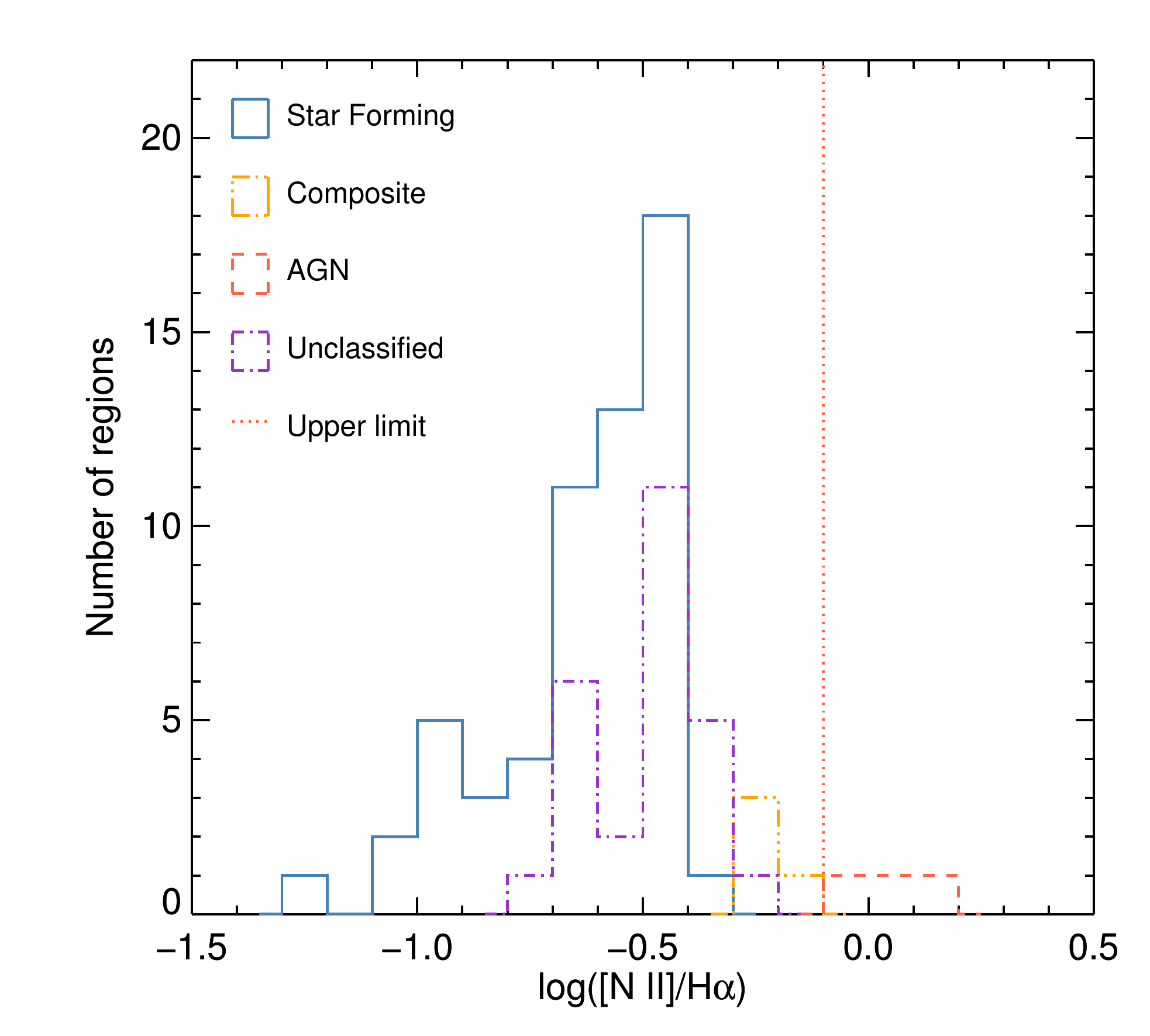}  
\includegraphics[width=0.99\linewidth]{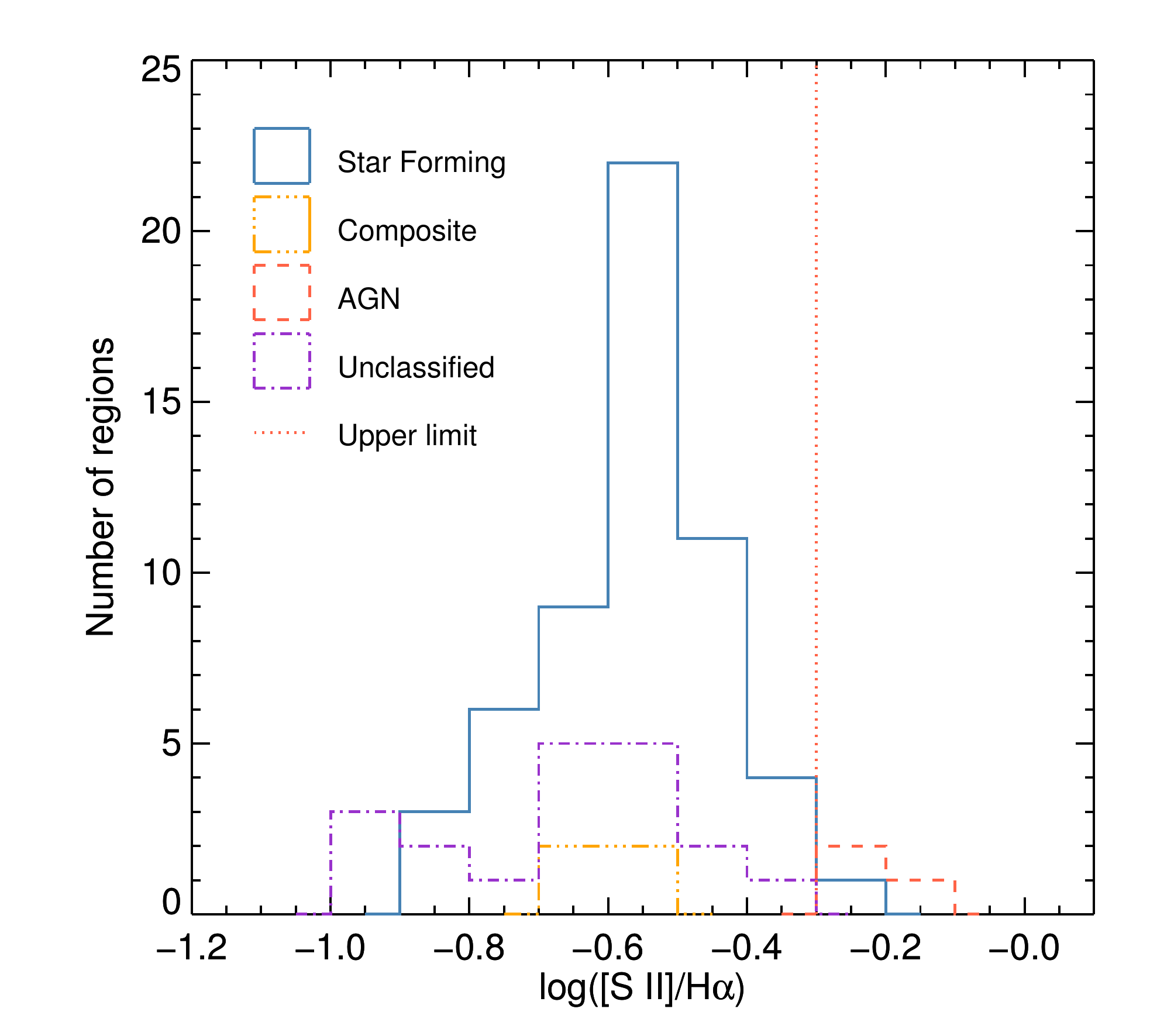} 
\caption{$N2$ and $S2$ parameter distributions in the measured regions. Dotted line represent an upper limit to differenciate between AGN and SF regions. As all unclassified regions lie above the upper limit, we consider them as SF (composite at least), and we include them in the subsequent analysis.}
\label{fig:histoN2}
\end{figure}

The 26 regions lacking of \Oiii~$\lambda$5007/\Hb~could not, obviously, be classified in the basis of these two BPT diagrams. However we may use the information coming from \Nii/\Ha. Top panel of Figure \ref{fig:histoN2} shows the distribution of \Nii/\Ha\ for the 63 regions 
classified by the diagnostic diagrams. The distribution are split into star-forming (blue), composite (orange) and AGNs (red). Figure~\ref{fig:histoN2} includes the distribution of unclassified regions (purple). All the 26 unclassified regions lie left to the limit defined by our AGNs, and they follow distribution which is equivalent to that observed for the star-forming regions. Besides that, we may  look at the spectra: if a region lacks of \Oiii~$\lambda$5007 with \Hb\ being present, then the ratio log\Oiii/\Hb\ will be always negative, so its position will be at the bottom part of the Figure \ref{diagnosticFIG} and at the left of the minimum value of \Nii/\Ha\ of our AGNs. In other words: combining the information of the distribution of log(\Nii/\Hb) seen in Figure \ref{fig:histoN2} and the experience at examining spectra we may consider that these unclassified spectra are actually coming from star forming regions. Adding these 26 regions to the 60 regions already classified as star-forming, we finally get a sample of 86 pure star-forming regions to be analyzed in our study.


\section{Results}

\subsection{Chemical Abundances using Strong Emission Lines methods}

When faint auroral lines, such as \Oiii~$\lambda$4363 or \Nii~$\lambda$5755, are not detectable in the optical spectrum of an \HII~region, the so-called strong emission line (SEL) methods can be used to estimate the chemical abundances of the ionized gas. The majority of the empirical calibrations rely on ratios between bright emission lines, which are defined by parameters involving ratios among some of the brightest  emission lines. Parameter definitions, and reviews of the most-common empirical calibrations and their limitations can be found in \citet{2008ApJ...681.1183K,2010A&A...517A..85L, 2012MNRAS.426.2630L}. Here we use:
\begin{eqnarray}
N2 & = & {\rm log}  \frac{I([\textsc{N\,ii}]) \lambda 6584}{\rm H\alpha},\\
O3N2 &= & {\rm log} \frac{I([\textsc{O\,iii}]) \lambda 5007/{\rm H\beta}}{I([\textsc{N\,ii}]) \lambda 6584/{\rm H\alpha}}.
\end{eqnarray}

These two indexes present some advantages with respect using other parameters:   
\begin{enumerate}
\item Both $N2$ and $O3N2$ are not affected for reddening, since the lines involved are so close in wavelength that this effect is cancelled. The reddening correction is important, for example, when considering the $R_{23}$ or $N2O2$ parameters \citep[see][for more details]{2010A&A...517A..85L}

\item As the intensity of oxygen lines does not monotonically increase with metallicity,  parameters involving oxygen ratios (i.e., $R_{23}$) are actually bi-valuated. Again this does not affect either the $N2$ nor $O3N2$ indexes. In case of the $R_{23}$ parameter, different calibrations must be given for the low --\mbox{\abox$ \lesssim$ 8.0}-- and the high --\abox$ \gtrsim$ 8.4-- metallicity regimes.
\end{enumerate}
Therefore, it is very convenient to rely on well-behaved parameters, such as $N2$ or $O3N2$, to derive the oxygen abundance of the ionized gas in galaxies when auroral lines are not detected, as they do not suffer either the problems of the reddening correction nor are bi-valuated. This has been extensively used in the literature in the last decade, although some precautions should still be taken into account when using these parameters \citep[see][]{2012MNRAS.426.2630L}.  For what refers to those indexes used in this work, $N2$ saturates at high metallicities --\abox$\gtrsim$8.6--, while $O3N2$ is not valid in the low-metallicity regime --\abox$\lesssim$8.1--.

\citet{2004MNRAS.348L..59P} proposed empirical calibrations for the $N2$ and $O3N2$ indexes that have been extensively used in the literature. However, these calibrations did not cover a proper range of oxygen abundances, since values in the high abundances regime were computed using photoionization models whose results are then put together with the
empirical ratios at other lower abundances. The empirical calibrations using $N2$ and $O3N2$ were
recently revisited by \citet[][hereafter MAR13]{2013A&A...559A.114M} by adding new empirical data from well-resolved \HII~regions from the CALIFA survey\footnote{http://califa.caha.es} \citep{2012A&A...538A...8S, 2015A&A...576A.135G}. These authors provided new linear calibrations:
\begin{equation}
{\rm OH}_{O3N2} = 12+\log{\rm (O/H)}  =  8.533-0.214\, O3N2,
 \label{MAR13}
\end{equation}
\begin{equation}
{\rm OH}_{N2} =  12+\log{\rm (O/H)}  =  8.743+0.462\, N2.
\label{MAR13b}
\end{equation}
In this work we use the calibrations by MAR13, which are suitable for \HII\ regions\footnote{It is important to note that a distinction has to be made when deriving oxygen abundances from resolved \HII\ regions and from galaxies (i.e. spectra from the Sloan Digital Sky Survey or the Galaxy And Mass Assembly survey). Each situation requires the proper empirical calibration, since integrated galaxy spectra do not consider only the pure \HII\ regions but the whole galaxy. In those cases some aperture corrections must be taken into account \citep{2013A&A...553L...7I}}. As all the oxygen abundance we derive are higher than \abox$\ge$8.1, we can use the $O3N2$ parameter, which is valid in all our abundances range \citep{2012MNRAS.426.2630L}.

As said before, the \Oiii~$\lambda$5007 emission line was not detected in all regions. Therefore, we cannot always provide an ${\rm OH}_{O3N2}$ oxygen abundance. But the $O3N2$ index is a more reliable parameter than the $N2$ parameter, since it does not saturate in the high-metallicity regime, besides including (at some level) a dependence on the ionizing degree of the gas (i.e., the \OIII\ and the \NII\ emission comes from areas within the \HII\ region that have different ionization parameter). Hence we derived a relationship between ${\rm OH}_{O3N2}$ and ${\rm OH}_{N2}$, that is similar to those relations derived by \citep{2008ApJ...681.1183K}. This relation was already presented by \citet[][see their Figure 1]{2016ApJ...818L..19M} as a linear fit
\begin{equation}
{\rm OH}_{O3N2} = 1.15 (\pm 0.09) -1.23(\pm 0.77)\times {\rm OH}_{N2}.
\label{MARfit}
\end{equation}

Therefore, we use the above expression to derive OH$_{O3N2}$ from OH$_{N2}$ in the 26 regions lacking of $O3N2$ values.

\subsection{Radial oxygen abundance distributions}

Once the values for OH$_{O3N2}$ in the 86 \HII\ regions within our 28 sample galaxies are obtained, we derive the radial distributions of the oxygen abundances and their corresponding radial metallicity gradients for each galaxy. Next we assign a local value of the oxygen abundance at the position of each SN~Ia.  We proceed as follows:
\begin{enumerate}
\item When several regions within a galaxy have been measured, we derive an abundance radial gradient using a linear fit. If this radial gradient is appropriate (i.e., a good correlation between GCDs and metallicities is measured), we use it to compute the metallicity value for the location of the SN~Ia. 
\item If this is not the case, we adopt as metallicity of the SN~Ia the oxygen abundance derived in the closest \HII\ region. This may be caused by two reasons. First, there are not enough points to derive a gradient (usually only one region has been observed within the galaxy, or when only two very nearby regions are observed). Second, the dispersion of the data is so high that the derived metallicity gradient seems not to be real, or it seems highly inaccurate, e.g., when a inverted metallicity gradient (that is considered to be non-realistic in a normal galaxy) is found. 
\end{enumerate}

We note that we use the deprojected radial distances for this, a critical measure if we want to assign a proper oxygen abundance value for each region in which the supernova is. Deprojecting a galaxy requires two parameters: the position angle (PA), which is the angle between the line of nodes of the projected image and the north, and the ratio between the galaxy's axis (b/a)) (both parameters are shown in Table \ref{tab:paper_table_1}). If the galaxy is seen face-on, b/a will be 0. 
Once obtained the {\it deprojected} GCD, we have the metallicity gradient of each galaxy from all observed \Hii\,regions and, this way, we estimate a local abundance for the SNe~Ia regions. For that, we assume that both all measured \Hii~regions and all the SNe~Ia are located in the galactic plane.

It is important to remark that we are assuming an unique metallicity radial gradient for in each galaxy. In other words, it does not matter the orientation of the slit, because the gradient is the same in every radial direction. There are not many studies of possible azimuthal variations \citep[but see][]{2015A&A...573A.105S}, besides in some few cases when galaxy interactions are observed \citep[e.g.][]{2009A&A...508..615L,2012ApJ...750..122B, 2015MNRAS.450.3381L}. Azimuthal variations in metallicity could be due to the spiral wave effect or the bar role which may move the gas in a non-asymmetric way and create zones with different elemental abundances even being located at the same GCD. This interesting subject is being studied now with data coming from the on-going Integral Field Spectroscopy (IFS) surveys \citep{2014A&A...570A...6S, 2014A&A...563A..49S}, as they allow this type of analysis, but it is out of the scope of our work.

\begin{figure*} 
\centering
\includegraphics[width=0.48\linewidth]{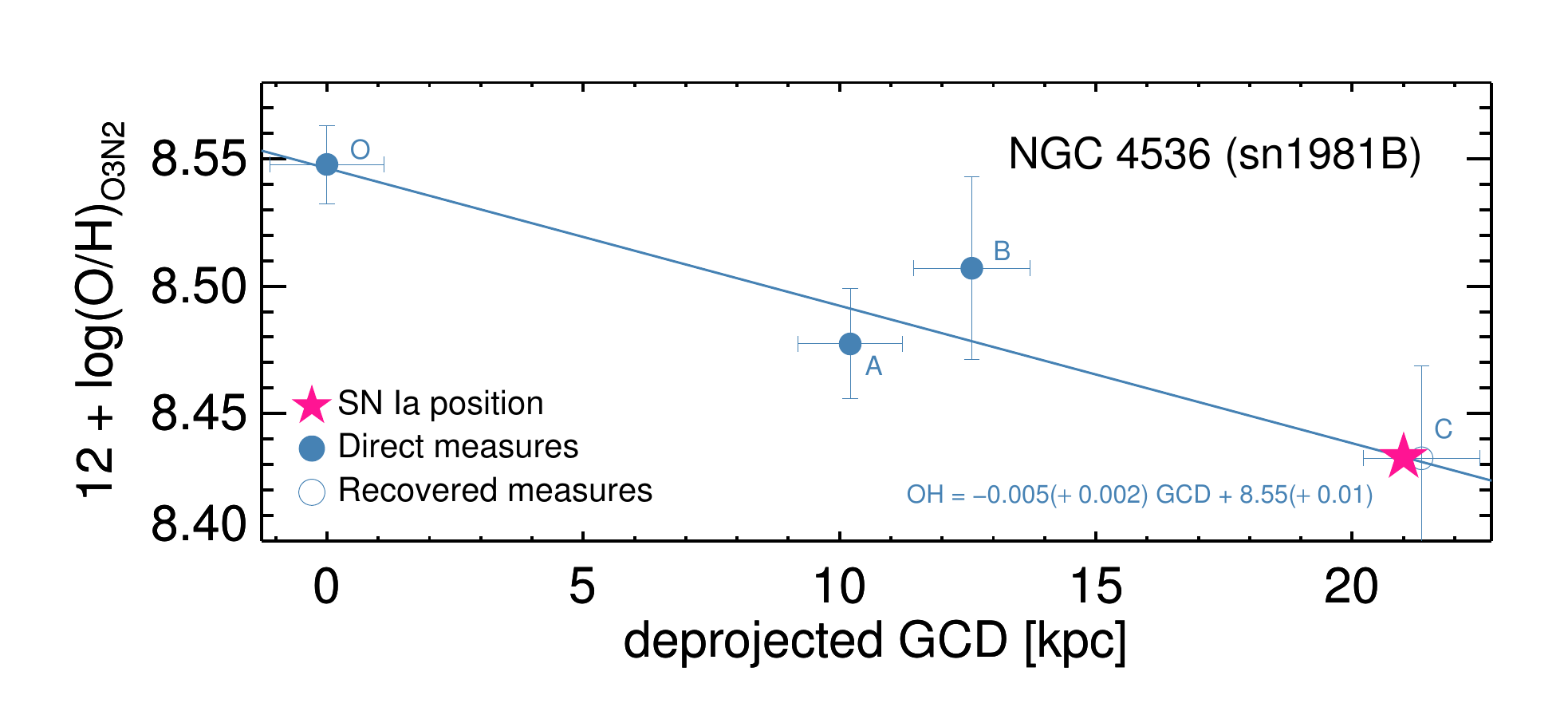}
\includegraphics[width=0.48\linewidth]{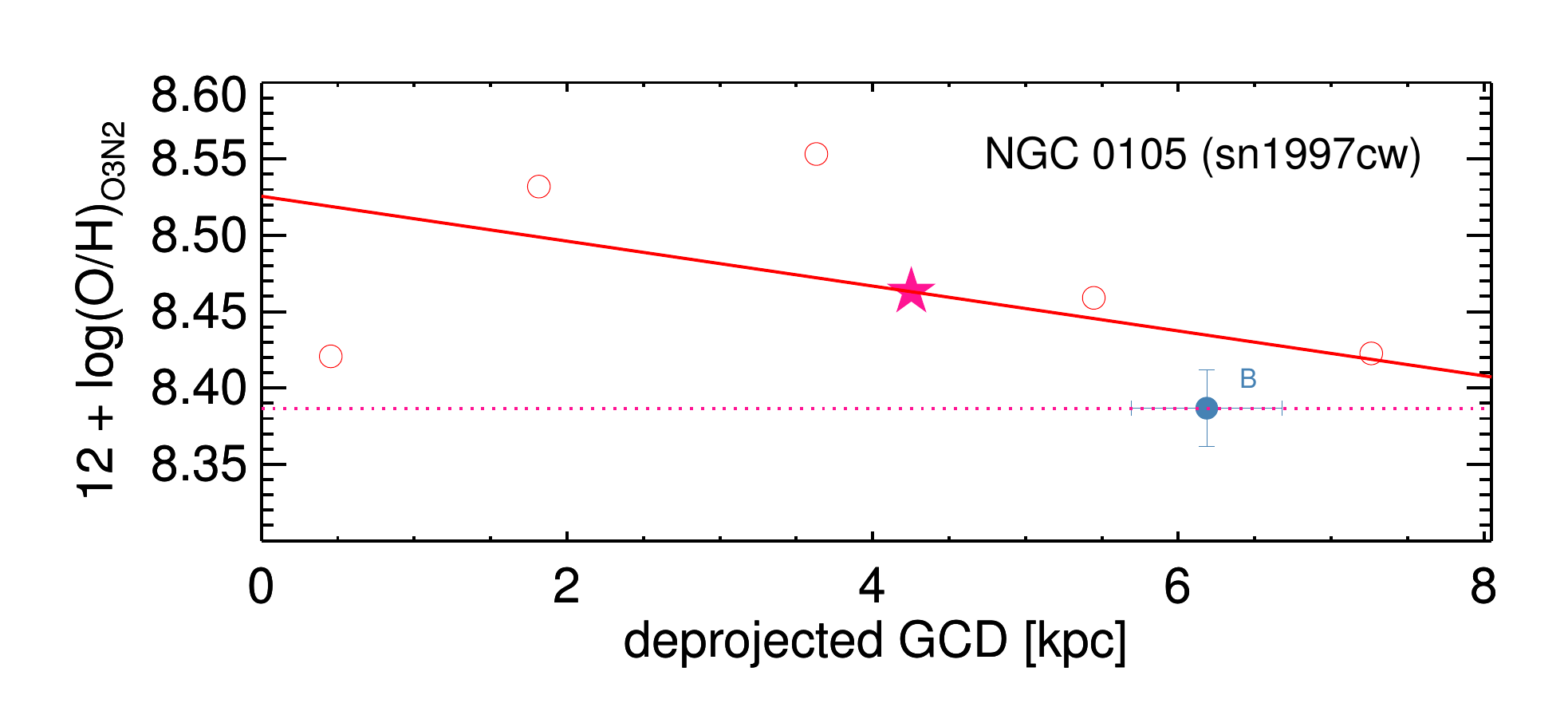}    
\caption{Left: Metallicity gradient for NGC\,4536. Filled points represent metallicities derived using the $O3N2$ parameter directly, while open points are those for which the ${\rm OH}_{O3N2}$  value was recovered from $N2$ parameter following Eq.~\ref{MARfit}. The blue solid line traces the metallicity gradient, while the pink star represents the SN~Ia location with its derived metallicity. Right: For the galaxy NGC\,0105 only one \HII\ region is available (blue circle) We adopt the value (pink star) given by the data from \citet{2016arXiv160307808G} (red circles).}
\label{gradientes_primeros}
\end{figure*}

Left panel of Figure \ref{gradientes_primeros} shows a typical abundance radial distribution, where the radial gradient is well determined and have enough points to be sure of the reliability of its slope. Filled points are the values obtained directly with the $O3N2$ parameter, while open points represent those values of  OH$_{O3N2}$ which have been derived from Equation \ref{MARfit}. The (red) star marks the position of the SN~Ia with the derived value of oxygen abundance. The right panel of Figure  \ref{gradientes_primeros} represents a case for which the radial gradient is undefined (in some cases it may be not reliable enough) and for which the value adopted for the SN~Ia corresponds to that given by its closest \HII\ region.

\begin{figure*} 
\centering
\includegraphics[width=0.48\linewidth]{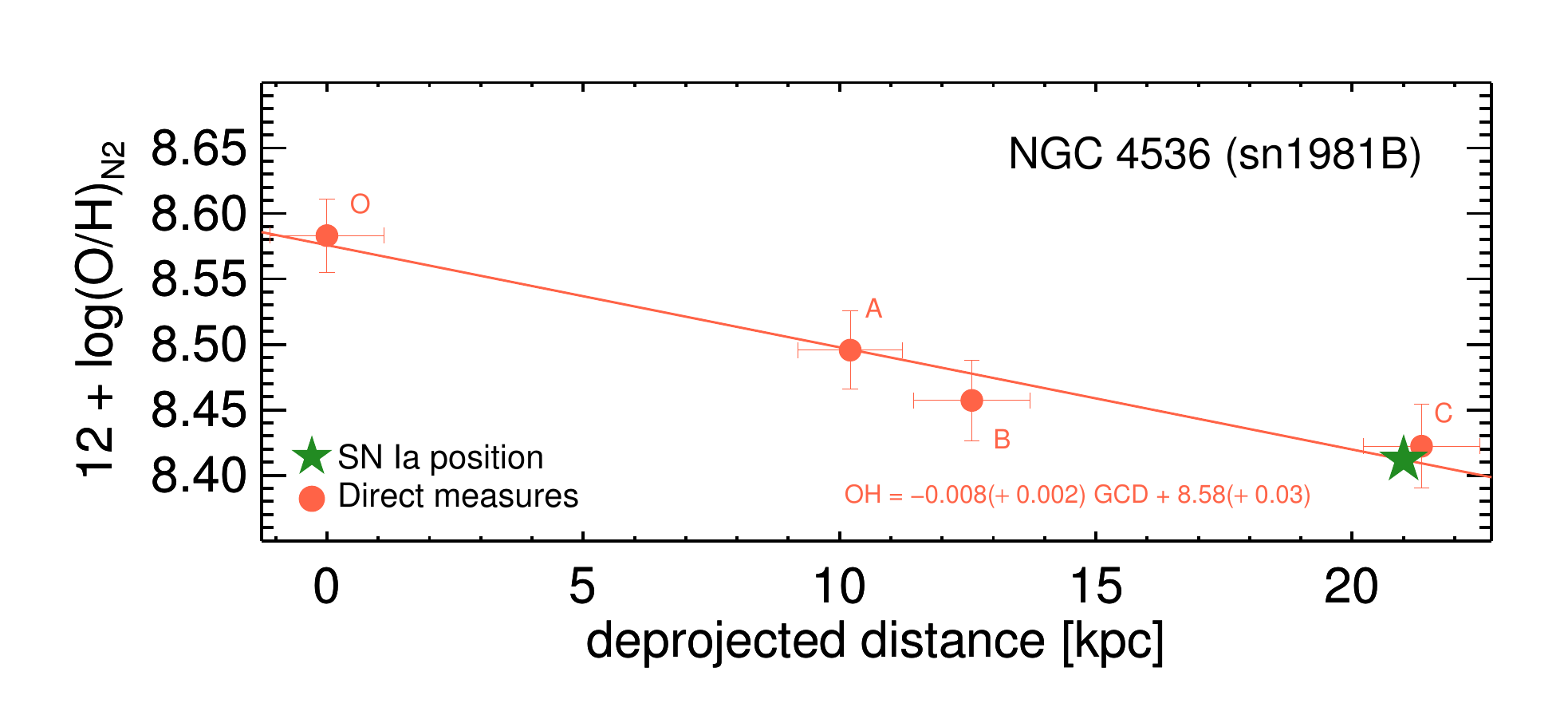}
\includegraphics[width=0.48\linewidth]{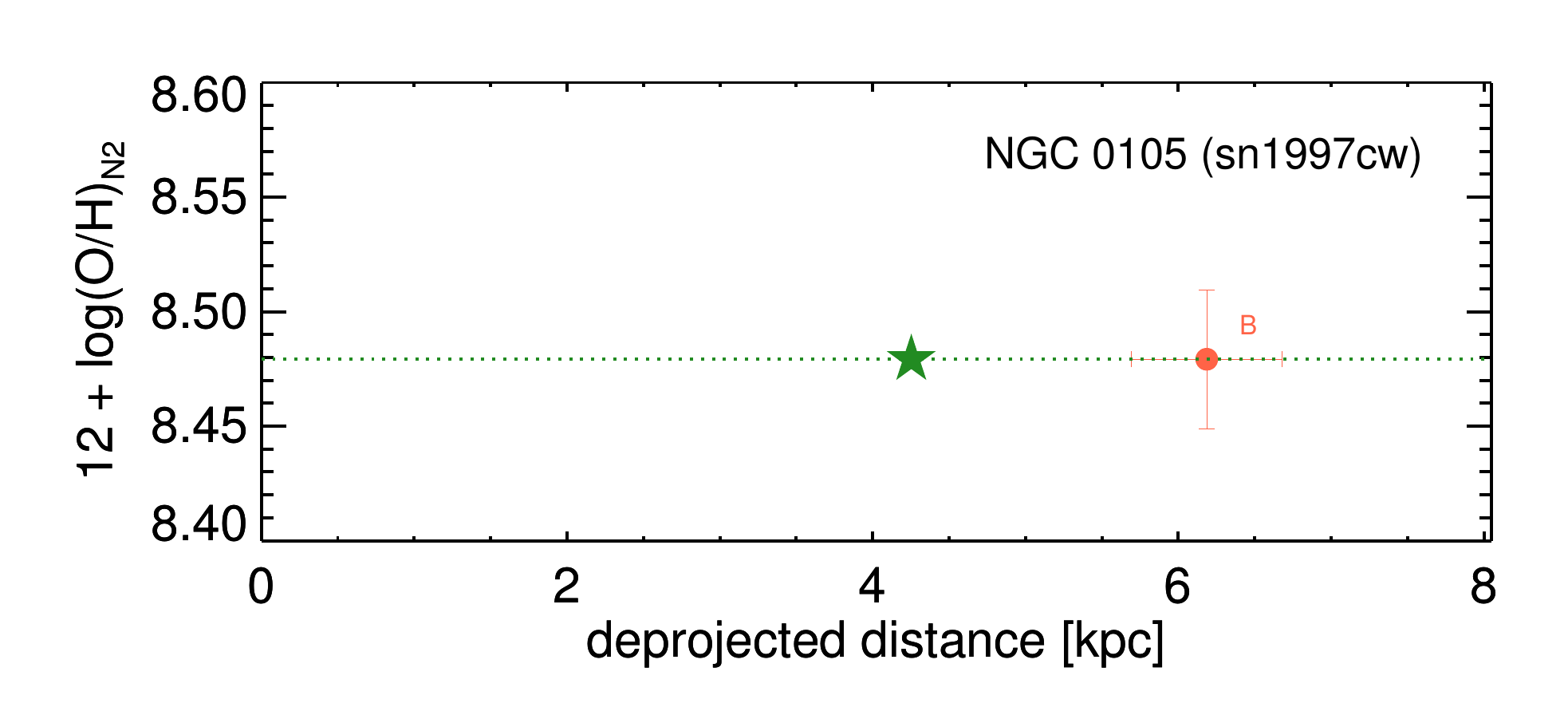}    
\caption{Left panel: Metallicity gradient for NGC\,4536. Red circles represent metallicities derived using the $N2$ parameter. The orange solid line traces the metallicity gradient, while the green star represents the SN~Ia location with its derived metallicity. Right panel: The red circle represents the value of the only \HII\ region observed in NGC\,0105, which is assumed to be the same that all the galaxy (dotted green line). The differences with respect to ${\rm OH}_{O3N2}$ are below 0.10~dex.}
\label{gradienteN2}
\end{figure*}

To check the consistency of these fits, we also derive all oxygen abundances using the $N2$ parameter. Figure~\ref{gradienteN2} shows the abundance radial distributions 
for the same two galaxies shown in Figure \ref{gradientes_primeros}. As we see, the disagreement between both fits is minimal, being identical within the uncertainties. This is further evidence of the robustness of our metallicity values.

Appendix~\ref{App:B} compiles all radial distributions derived for our galaxy sample. We derive accurate metallicity gradients for 21 galaxies, that we use to provide the  metallicity of the SN~Ia. For the remaining 7 galaxies we use the oxygen abundance value of the closest \HII\ region as that of the SN~Ia. 
When the distance between the SN~Ia and the observed \HII\ region is larger than 3~kpc, the metallicity of the SN~Ia will actually be an upper or a lower limit (depending on the location of the SN~Ia with respect to the \HII\ region). We briefly describe the situation in these 7 cases in Appendix~\ref{App:B}.

\begin{table}
\caption{Derived abundances for the environment regions of SNe Ia.}
\begin{tabular}{c@{\hspace{6pt}} c@{\hspace{8pt}} c@{\hspace{8pt}}  l@{\hspace{8pt}} c}
\noalign{\smallskip}
\hline
\hline
{Host Galaxy} & {SN Ia} &{${\rm OH}_{gradient}$} &{${\rm OH}_{closest}$} &{${\rm OH}_{final}$} \\
 \hline 
\noalign{\smallskip}
M 82      		&	2014J  	&	8.59	$\pm$	0.15  	&	8.58	$\pm$	0.01 	&	8.59	$\pm$	0.15		\\
MCG-02-16-02	&	2003kf 	&	8.38	$\pm$	0.14  	&	8.44	$\pm$	0.06 	&	8.38	$\pm$	0.14		\\
NGC 0105  	&	1997cw 	&	8.46	$\pm$	0.08$^{\dagger}$		&	8.38  $\pm$	0.02	&	8.46	$\pm$	0.08		\\
NGC 1275  	&	2005mz 	&	...	 				&	8.54  $\pm$	0.01	&	8.54	$\pm$	0.01		\\
NGC 1309  	&	2002fk 	&	8.41	$\pm$	0.19  	&	8.42	$\pm$	0.04 	&	8.41	$\pm$	0.19		\\
NGC 2935  	&	1996Z  	&	... 	 				&	8.62	$\pm$	0.05	&	8.62	$\pm$	0.05		\\
NGC 3021  	&	1995al 	&	8.57	$\pm$	0.05 		&	8.55	$\pm$	0.01 	&	8.57	$\pm$	0.05		\\
NGC 3147  	&	1997bq 	&	...	 				&	8.66	$\pm$	0.02	&	8.66	$\pm$	0.02		\\
NGC 3169	 &	2003cg 	&	8.75	$\pm$	0.06 		&	8.66	$\pm$	0.02 	&	8.75	$\pm$	0.06		\\
NGC 3368  	&	1998bu 	&	...	 				&	8.51	$\pm$	0.03	&	8.51	$\pm$	0.03		\\
NGC 3370  	&	1994ae 	&	8.30	$\pm$	0.08 		&	8.23	$\pm$	0.02 	&	8.30	$\pm$	0.08		\\
NGC 3672  	&	2007bm 	&	8.63	$\pm$	0.07 		&	8.65	$\pm$	0.05 	&	8.63	$\pm$	0.07		\\
NGC 3982  	&	1998aq 	&	8.54	$\pm$	0.05$^{\dagger}$		&	8.58	$\pm$	0.05 	&	8.54	$\pm$	0.05		\\
NGC 4321  	&	2006X  	&	8.60	$\pm$	0.12  	&	8.59	$\pm$	0.01 	&	8.60	$\pm$	0.12		\\
NGC 4501  	&	1999cl 	&	8.43	$\pm$	0.22  	&	8.43	$\pm$	0.03 	&	8.43	$\pm$	0.22		\\
NGC 4527  	&	1991T  	&	8.50	$\pm$	0.08	 	&	8.54	$\pm$	0.05 	&	8.50	$\pm$	0.08		\\
NGC 4536  	&	1981B  	&	8.43 $\pm$	0.06 		&	8.43	$\pm$	0.06	&	8.43	$\pm$	0.06		\\
NGC 4639  	&	1990N  	&	8.16	$\pm$	0.20  	&	8.40	$\pm$	0.03	&	8.16	$\pm$	0.20		\\
NGC 5005  	&	1996ai 	&	8.60	$\pm$	0.08 		&	8.60	$\pm$	0.01	&	8.60	$\pm$	0.08		\\
NGC 5468  	&	1999cp 	&	8.23	$\pm$	0.08 		&	8.32	$\pm$	0.02	&	8.23	$\pm$	0.08		\\
NGC 5584  	&	2007af 	&	8.34	$\pm$	0.08 		&	8.47	$\pm$	0.02 	&	8.34	$\pm$	0.08		\\
UGC 00272   	&	2005hk 	&	8.25	$\pm$	0.12  	&	8.30	$\pm$	0.04 	&	8.25	$\pm$	0.12		\\
UGC 03218  	&	2006le 	&	8.59	$\pm$	0.09 		&	8.59	$\pm$	0.05 	&	8.59	$\pm$	0.09		\\
UGC 03576  	&	1998ec 	&	8.57	$\pm$	0.16  	&	8.58	$\pm$	0.06 	&	8.57	$\pm$	0.16		\\
UGC 03845  	&	1997do 	&	8.49	$\pm$	0.04 		&	8.54	$\pm$	0.01 	&	8.49	$\pm$	0.04		\\
UGC 04195  	&	2000ce 	&	8.46	$\pm$       0.05$^{\dagger}$ 		&	8.57	$\pm$	0.02	&	8.46	$\pm$	0.05		\\
UGC 09391	 &	2003du 	&	8.27	$\pm$	0.10  	&	8.31	$\pm$	0.07 	&	8.27	$\pm$	0.10		\\
UGCA 017   	&	1998dm 	&	8.32	$\pm$	0.07	 	&	8.29	$\pm$	0.04 	&	8.32	$\pm$	0.07		\\
\hline
\hline
\end{tabular}
\label{tab:abundancias}
$\dagger$: Gradients obtained from \citet{2016arXiv160307808G}
\end{table}

Table ~\ref{tab:abundancias} lists the oxygen abundances for all SN~Ia indicating for each case the given value from the radial gradient (if it does exist), the value of the closest region, and the final adopted value. We note that values from the closest region and from the gradient are in agreement. 
A linear fit to the data, which has a correlation coefficient of $r=0.8947$, provides a slope of $0.915\pm0.054$ (i.e., very near to 1).

It is important to take into account potential biases introduced by the difference between SN~Ia progenitor metallicity (Z$_{\rm Ia}$) at the time of its formation and the metallicity of the host galaxy (Z$_{\rm HOST}$) measured after the explosion. \citet{2011MNRAS.414.1592B} presented a theoretical study concluding that SN~Ia progenitor metallicity can be reasonably estimated by the host galaxy metallicity, and that is better represented by the gas-phase than the stellar host metallicity. In addition, quoting these authors: `for active galaxies, the dispersion of Z$_{\rm Ia}$ is quite small, meaning that Z$_{\rm HOST}$ is a quite good estimator of the supernova metallicity, while passive galaxies present a larger dispersion'. On the other hand, \citet{2016arXiv160307808G} found that the gas-phase metallicity at the locations of a sample of SNe~Ia observed with IFS is on average 0.03 dex higher than the total galaxy metallicity in the MAR13 scale. The differences between local and global stellar metallicities are not significant for SNe~Ia, and this enables the use of the environmental metallicity as a proxy for the SN~Ia progenitor metallicity.

\subsection{Comparing with other results}

\begin{figure*} 
\centering
\includegraphics[width=0.48\linewidth]{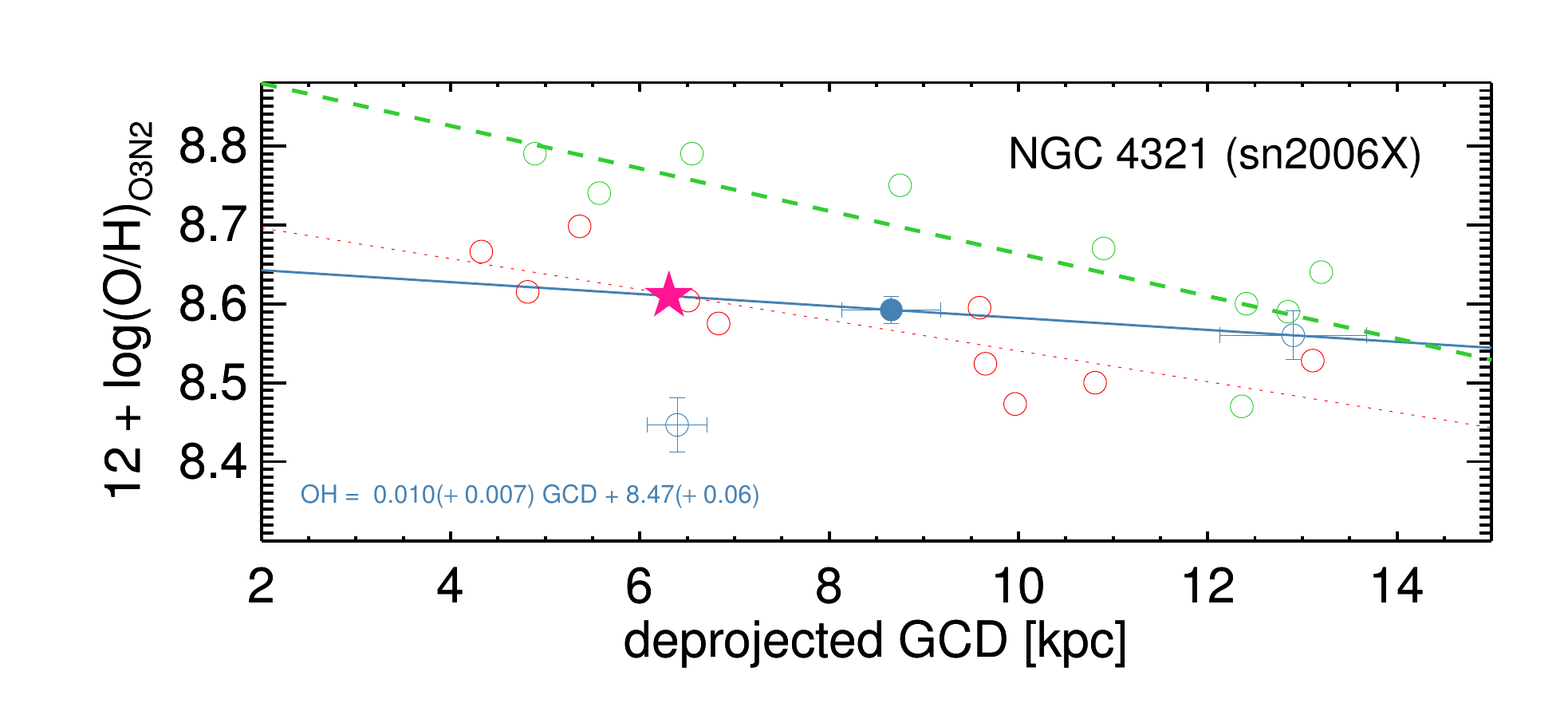}
\includegraphics[width=0.48\linewidth]{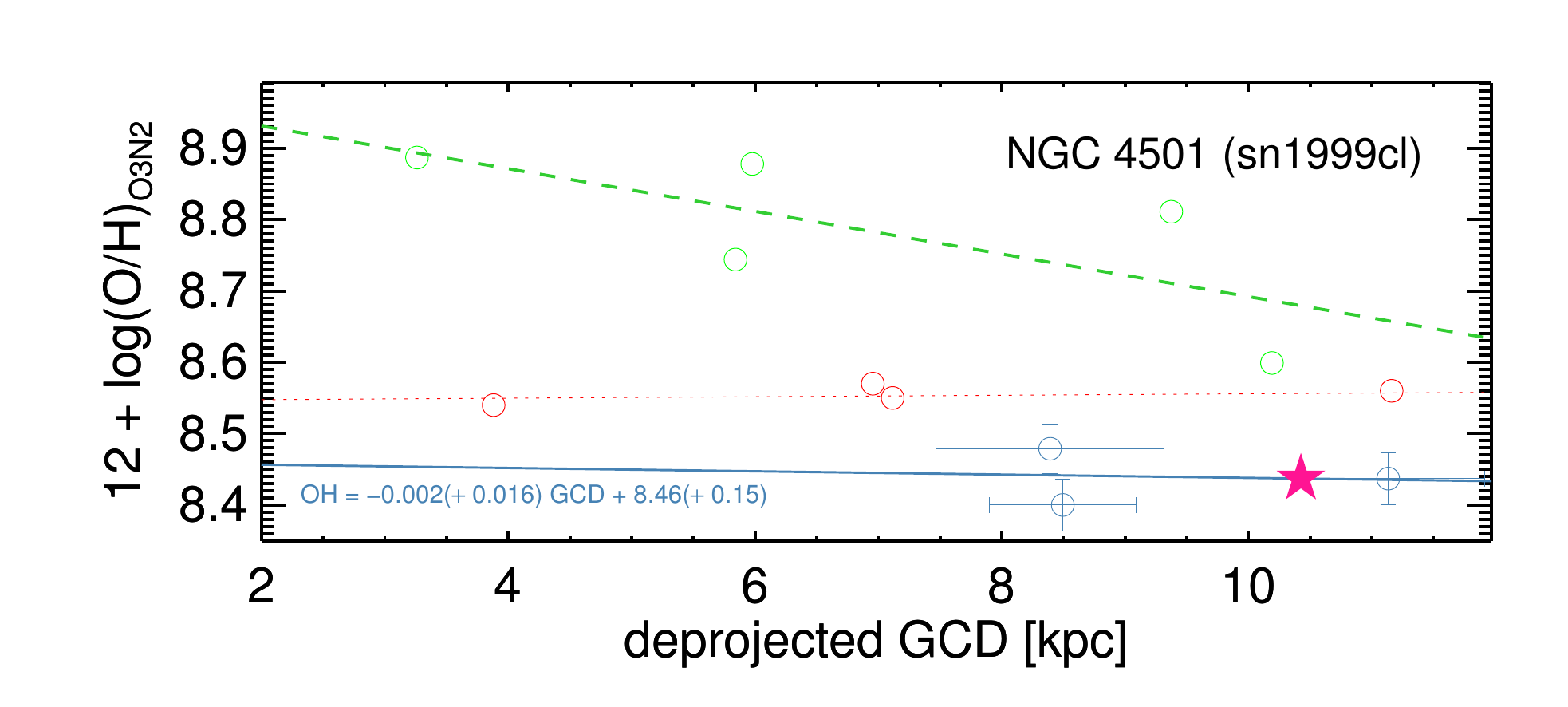}    
\caption{Metallicity gradients for NGC\,4321(left) and NGC\,4501(right). Symbols are the same than in Fig. \ref{gradientes_primeros}, except for the open green circles, which represent values from \citet{2002A&A...383...14P} and the dashed green line, being the gradient that these authors derived from these points. Red points correspond to abundances obtained applying MAR13 calibration to the original data gathered from \citet{1985ApJS...57....1M, 1991ApJ...371...82S, 1996ApJ...462..147S}, and the red dotted line is the linear fit to the data.}
\label{gradientes_pilyugin}
\end{figure*}

The metallicities of two galaxies of our sample,  NGC\,4321 and NGC\,4501, were studied by \citet{2002A&A...383...14P} using the $R_2$ and $R_{23}$ parameters and the empirical calibration provided by  \citet{2001A&A...369..594P}. Figure~\ref{gradientes_pilyugin} compares the gradients for both galaxies derived by these authors  with those found in this work. As we can see, the abundances derived by \citet{2002A&A...383...14P} are systematically higher than those derived here. Such a difference is mainly due to the use of different empirical calibrations. Indeed, the new empirical calibrations derived by MAR13 make it very difficult to get $12 + \log({\rm O / H}) \geq 8.69$, values which are easily reached by the empirical calibration derived by   \citet{2001A&A...369..594P}. As a second cross check, we have gathered data from \citet{1985ApJS...57....1M, 1991ApJ...371...82S, 1996ApJ...462..147S} to get the original emission line fluxes for these two galaxies and we have applied MAR13 calibrations. Figure~\ref{gradientes_pilyugin} also shows in red the gradients obtained with MAR13 calibration. They are in agreement with our results. That stresses the discrepancy between calibrations.

On the other hand, three galaxies for which we have not derived a metallicity gradient (NGC\,0105, NGC\,3982 and UGC\,04195) have been recently observed as part of an extended CALIFA program.  The analysis of these data, presented in \citet{2016arXiv160307808G}, provides a metallicity gradient for each of these galaxies which is made up by approximately $\sim$1,000 spectra. For these three galaxies we were not able to obtain metallicity gradients, so we took the values given in \citet{2016arXiv160307808G}, which help us to improve the quality of the data (see Table \ref{tab:abundancias} and Appendix \ref{App:B} for more details).

\subsection{SN~Ia LC parameters}

We have gathered LC data for these SNe~Ia from different sources (listed in Table \ref{tab:cosmo}). These data have been fitted with SiFTO \citep{2008ApJ...681..482C} to obtain both the LC stretch and the color. SiFTO is a powerful and versatile LC fitter that manipulates spectral templates to simultaneously match the multi-band photometric data of a particular SN with heliocentric redshift $z_{hel}$. A stretch parameterization is used to describe the shape of the LC. The SiFTO color, $C$, is obtained adjusting the spectral energy distribution to observed colors corrected for Milky Way extinction wit values from \citet{SF11}. See \citet{2014ApJ...795..142G} for a detailed procedure of how we obtain $s$ and $C$. The LC stretch is defined in $B$-band.

We classify SNe~Ia regarding their color ($C$) in this way: if $C < 0.2$, we label them Normal SNe, if $C > 0.2$ we consider they are reddened SNe~Ia. Those SNe~Ia that have $C < 0.2$ and a extremely low luminosity (considering the SN~Ia is no reddened) are labeled as subluminous SNe~Ia. For the subsequent analysis, subluminous SNe~Ia are not longer considered in our study, since their behavior cannot be explained as the rest of normal SNe~Ia.

\subsection{Absolute magnitudes}

Using the distance to the galaxies $D$ measured in Sec. \ref{sec:sample_sel} considering data independent from the SNe~Ia method, and the apparent magnitude $m_B$ determined with SiFTO, we calculated the absolute magnitudes $M_{B}$ using,
\begin{equation}
\label{eq:magnitude_distance}
M_{B}=m_{B,ext}+5-5\times\log(D),
\end{equation}
where $D$ is in parsecs.

We note that the $m_B$ determined by SiFTO is already corrected by the MW extinction, and any contribution of the extinction of the host galaxy as well as intrinsic color variations, are retained in the $C$ SiFTO parameter.
Correcting for SN host galaxy extinction is a complicated issue highly debated specially in SN cosmological studies, due to the difficulty to estimate its magnitude. Many assumptions have to be done in order to get proper estimations, including extinction laws, different $R_V$ than the standard 3.1 in the Milky Way, etc.
However, if the SN~Ia is located in the outer parts of the host galaxy, the extinction of its host galaxy seems to be negligible \citep{2012ApJ...755..125G}, meaning that correcting for MW extinction is enough for our purposes. This can be checked by looking at the color of the SNe~Ia.

Table \ref{tab:cosmo} provides a summary of the measurements presented in this subsection. Column~1 lists  the name of the galaxy, while the identification of each SN~Ia is shown in Column~2. The next three columns list the parameters given by SiFTO (SN peak apparent magnitude, color $C$, and stretch $s$). In column~4 we give $M_B$, computed considering only SN~Ia independent distance indicators (see sec.~\ref{sec:sample_sel}). Column~5 compiles the oxygen abundance we derived for each SN~Ia location. Column 7 lists the sources for each SN LC photometry.


\section{Discussion}

In this section we analyze the absolute magnitudes of SNe~Ia in their maximum of the LC, $M_{B}$, and their relation with the oxygen abundances. We also analyze these results in the context of the SNe~Ia parameters as their Colors and their LC shape.

\subsection{Covariance Matrix: $M_{B}-C-{\rm OH}-s$}

In some traditional methods, only two parameters have been considered for reducing the systematics in the SNe~Ia distance modulus, i.e., SN color and LC stretch. They have been used to correct $M_{B}$, which is considered fixed for all SNe~Ia. Here we want to check if the environmental properties of the host galaxies may affect the SNe~Ia luminosities. One such property is the metallicity, using the gas-phase oxygen abundance as its proxy. Our first analysis of this study was already presented in  \citet{2016ApJ...818L..19M}, where we discussed the metallicity dependence of the $M_B$ of the SN~Ia for not-reddened objects, i.e., those with  $C < 0.2$, to avoid possible biases with the SN~Ia color. We found that, for those not-reddened SNe~Ia, a metallicity-dependence does appear when plotting $M_{B}$ as a function of OH$_{O3N2}$, in the sense that metal-rich galaxies host less luminous SNe~Ia than metal-poor galaxies, which host brightest SN~Ia. We now want to check if this metallicity dependence is still found when the whole galaxy sample is considered. Hence, we seek to find the relationship between all these parameters. To do this we  perform a principal component analysis (PCA), calculating the covariance matrix involving these four parameters following this order: the  SNe~Ia absolute magnitude ($M_{B}$), the SNe~Ia color ($C$), the SNe~Ia environment metallicity (${\rm OH}$), and the SNe~Ia LC stretch ($s$).

\begin{table*}
\scriptsize
\caption{SNe Ia parameters. From columns 3 to 6: apparent magnitud corrected for MW extinction; color; stretch; and derived absolute magnitudes, respectively. Column 7 shows environment oxygen abundances. SN Ia classes are shown in column 8: 1 for Normal SNe Ia, 2 for reddened SNe Ia and 3 for subluminous SNe Ia. Sources for SNe ia parameters are listed in column 9. }
\begin{tabular}{c@{\hspace{6pt}} c@{\hspace{8pt}} c@{\hspace{8pt}}  c@{\hspace{8pt}} c@{\hspace{8pt}} c@{\hspace{8pt}}@{\hspace{8pt}} c@{\hspace{8pt}} c@{\hspace{8pt}}l}
\noalign{\smallskip}
\hline
\hline
{Host Galaxy}&{SN Ia}&{$m_{B}$} &{C}&{$s$}&{$M_{B}$}&{OH}&{Class}&{Source} \\
 \hline 
\noalign{\smallskip}
M 82      		&	2014J  	&	11.332 $\pm$     0.056 &      1.167 $\pm$     0.0090 &      1.070 $\pm$    0.004 &      -16.567 $\pm$     0.056 &       8.59 $\pm$     0.07 &        2		&   \citet{2015ApJ...798...39M,2014MNRAS.443.2887F,2014CoSka..44...67T}    \\
MCG-02-16-02	&	2003kf 	&	13.238 $\pm$     0.132 &     -0.027 $\pm$     0.0490 &      1.055 $\pm$    0.013 &      -18.528 $\pm$     0.132 &       8.44 $\pm$     0.13 &        1 		&   \citet{2009ApJ...700..331H}      \\
NGC 0105  	&	1997cw 	&	15.939 $\pm$     0.057 &      0.348 $\pm$     0.0240 &      1.105 $\pm$    0.031 &      -18.099 $\pm$     0.057 &       8.39 $\pm$     0.03 &        2		&    \citet{2006AJ....131..527J}    \\
NGC 1275  	&	2005mz 	&	16.434 $\pm$     0.067 &      0.325 $\pm$     0.0180 &      0.644 $\pm$    0.006 &      -17.507 $\pm$     0.067 &       8.55 $\pm$     0.03 &        3		&    \citet{2009ApJ...700..331H}   \\
NGC 1309  	&	2002fk 	&	13.147 $\pm$     0.018 &     -0.175 $\pm$     0.0200 &      1.013 $\pm$    0.005 &      -19.186 $\pm$     0.018 &       8.38 $\pm$     0.11 &        1 		&    \citet{2009ApJ...700..331H}    \\
NGC 2935  	&	1996Z  	&	14.348 $\pm$     0.095 &      0.315 $\pm$     0.0230 &      0.915 $\pm$    0.085 &      -17.903 $\pm$     0.095 &       8.62 $\pm$     0.05 &        2		&      \citet{1999AJ....117..707R}   \\
NGC 3021  	&	1995al 	&	13.310 $\pm$     0.017 &      0.097 $\pm$     0.0200 &      1.074 $\pm$    0.025 &      -18.789 $\pm$     0.017 &       8.58 $\pm$     0.06 &        1 		&    \citet{1999AJ....117..707R}    \\
NGC 3147  	&	1997bq 	&	14.388 $\pm$     0.024 &      0.080 $\pm$     0.0180 &      0.917 $\pm$    0.010 &      -18.671 $\pm$     0.024 &       8.66 $\pm$     0.02 &        1		&    \citet{2006AJ....131..527J}    \\
NGC 3169	 &	2003cg 	&	15.798 $\pm$     0.014 &      1.154 $\pm$     0.0100 &      0.983 $\pm$    0.004 &      -15.363 $\pm$     0.014 &       8.75 $\pm$     0.06 &        2 		&    \citet{2006MNRAS.369.1880E,2010ApJS..190..418G,2009ApJ...700..331H}     \\
NGC 3368  	&	1998bu 	&	12.106 $\pm$     0.013 &      0.267 $\pm$     0.0090 &      0.973 $\pm$    0.009 &      -18.091 $\pm$     0.013 &       8.51 $\pm$     0.03 &        2		&    \citet{1999AJ....117.1175S}   \\
NGC 3370  	&	1994ae 	&	12.948 $\pm$     0.019 &     -0.079 $\pm$     0.0130 &      1.054 $\pm$    0.008 &      -19.244 $\pm$     0.019 &       8.30 $\pm$     0.08 &        1 		&    \citet{2004MNRAS.349.1344A,2005ApJ...627..579R}   \\
NGC 3672  	&	2007bm 	&	14.411 $\pm$     0.210 &      0.474 $\pm$     0.0130 &      0.938 $\pm$    0.005 &      -17.379 $\pm$     0.210 &       8.64 $\pm$     0.08 &        2 		&    \citet{2009ApJ...700..331H, 2011AJ....142..156S}    \\
NGC 3982  	&	1998aq 	&	12.320 $\pm$     0.009 &     -0.133 $\pm$     0.0080 &      0.986 $\pm$    0.005 &      -19.340 $\pm$     0.009 &       8.59 $\pm$     0.06 &        1 		&    \citet{2005ApJ...627..579R}    \\
NGC 4321  	&	2006X  	&	15.373 $\pm$     0.016 &      1.393 $\pm$     0.0170 &      1.000 $\pm$    0.003 &      -15.572 $\pm$     0.016 &       8.61 $\pm$     0.06 &        2		&    \citet{2010ApJS..190..418G}    \\
NGC 4501  	&	1999cl 	&	14.891 $\pm$     0.017 &      1.129 $\pm$     0.0100 &      0.939 $\pm$    0.008 &      -16.687 $\pm$     0.017 &       8.44 $\pm$     0.12 &        2 		&    \cite{2010ApJS..190..418G,2006AJ....131..527J,2006AJ....131.1639K}    \\
NGC 4527  	&	1991T  	&	11.491 $\pm$     0.023 &      0.050 $\pm$     0.0210 &      1.068 $\pm$    0.011 &      -19.210 $\pm$     0.023 &       8.50 $\pm$     0.08 &        1 		&    \citet{2004MNRAS.349.1344A, 1993AJ....106.1101F,1998AJ....115..234L}   \\
NGC 4536  	&	1981B  	&	                 ...                    &	                    ...                    &                          ...              &                            ...                 &       8.43 $\pm$     0.06 &        1 		&    ...   \\
NGC 4639  	&	1990N  	&	12.675 $\pm$     0.015 &      0.020 $\pm$     0.0110 &      1.064 $\pm$    0.008 &      -19.066 $\pm$     0.015 &       8.17 $\pm$     0.08 &        1		&    \citet{1998AJ....115..234L}     \\
NGC 5005  	&	1996ai 	&	16.892 $\pm$     0.012 &      1.676 $\pm$     0.0140 &      1.097 $\pm$    0.024 &      -14.931 $\pm$     0.012 &       8.60 $\pm$     0.09 &        2		&      \citet{1999AJ....117..707R}   \\
NGC 5468  	&	1999cp 	&	13.933 $\pm$     0.011 &     -0.047 $\pm$     0.0070 &      0.994 $\pm$    0.006 &      -19.157 $\pm$     0.011 &       8.23 $\pm$     0.08 &        1		&    \citet{2010ApJS..190..418G,2000ApJ...539..658K}     \\
NGC 5584  	&	2007af 	&	13.156 $\pm$     0.087 &      0.058 $\pm$     0.0130 &      0.970 $\pm$    0.003 &      -18.976 $\pm$     0.087 &       8.34 $\pm$     0.09 &        1		&    \citet{2009ApJ...700..331H, 2011AJ....142..156S}     \\
UGC 00272   	&	2005hk 	&	15.891 $\pm$     0.011 &      0.293 $\pm$     0.0180 &      0.867 $\pm$    0.003 &      -18.018 $\pm$     0.011 &       8.26 $\pm$     0.10 &        3 		&    \citet{2009ApJ...700..331H,2014ApJ...786..134M,2007PASP..119..360P}    \\
UGC 03218  	&	2006le 	&	14.751 $\pm$     0.168 &     -0.133 $\pm$     0.0440 &      1.074 $\pm$    0.005 &      -19.104 $\pm$     0.168 &       8.60 $\pm$     0.07 &        1 		&    \citet{2009ApJ...700..331H}     \\
UGC 03576  	&	1998ec 	&	16.189 $\pm$     0.053 &      0.176 $\pm$     0.0220 &      0.997 $\pm$    0.033 &      -18.518 $\pm$     0.053 &       8.57 $\pm$     0.08 &        1 		&     \citet{2010ApJS..190..418G,2006AJ....131..527J}    \\
UGC 03845  	&	1997do 	&	14.288 $\pm$     0.037 &      0.019 $\pm$     0.0200 &      0.975 $\pm$    0.020 &      -18.639 $\pm$     0.037 &       8.50 $\pm$     0.04 &        1 		&    \citet{2006AJ....131..527J}     \\
UGC 04195  	&	2000ce 	&	17.038 $\pm$     0.038 &      0.495 $\pm$     0.0170 &      1.038 $\pm$    0.026 &      -17.445 $\pm$     0.038 &       8.57 $\pm$     0.02 &        2		&     \citet{2006AJ....131..527J,2001AJ....122.1616K}    \\
UGC 09391	 &	2003du 	&	13.462 $\pm$     0.007 &     -0.145 $\pm$     0.0140 &      1.023 $\pm$    0.004 &      -19.050 $\pm$     0.007 &       8.28 $\pm$     0.11 &        1 		&    \citet{2005AA...429..667A,2009ApJ...700..331H,2005ApJ...632..450L}    \\
UGCA 017   	&	1998dm 	&	14.663 $\pm$     0.020 &      0.264 $\pm$     0.0090 &      1.044 $\pm$    0.006 &      -17.434 $\pm$     0.020 &       8.33 $\pm$     0.08 &        3 		&     \citet{2010ApJS..190..418G,2006AJ....131..527J}  \\
\hline
\hline
\end{tabular}
\label{tab:cosmo}
\end{table*}

This PCA provides the following Covariance Matrix:

\begin{eqnarray}
\label{eq:covMat}
\Sigma = \left(\begin{array}{cccc}
1.0	&	0.9749	&	0.5464	&  -0.0222\\
		&	1.0	&	0.4780	&  0.0303\\
          	&			&	1.0	&  -0.2251\\
		&			&			&  1.0\\
\end{array}\right)
\end{eqnarray}

The elements on the principal diagonal represent the correlation of each parameter with itself, which is always equal to 1. The crossed terms in the covariance matrix indicate how well two parameters are correlated, and have values ranging from 1 to $-$1. We label the matrix elements in Eq~\ref{eq:covMat} as $a_{n,m}$, being $n$ the row in the matrix and $m$ the column. Hence, element, i.e. $a_{4,3}$ corresponds to the element placed in the fourth row and the third column, which relates $s$ with ${\rm OH}$.

The first row, corresponding to $M_{B}$ gives us that color (in the second column) is the most influential parameter, as $a_{1,2} = 0.9749$, (as $a_{2,1}$, not shown in Eq~\ref{eq:covMat} since the matrix is symmetric). The terms $a_{1,3} = 0.5464$ and $a_{1,4} =-0.0222$ relate $M_{B}$ with ${\rm OH}$ and $s$, respectively. Negative values imply that the parameters are inversely correlated. One can see that none of these terms are negligible. In addition, metallicity correlates with $M_{B}$ almost twice than the LC stretch does. The non-zero values of $a_{1,3}$ let us consider to quantify this effect with metallicity. 

The value $a_{1,4} =-0.0222$ gives information about the low dispersion of $s$ within the sample. In fact, in our sample, $s$ goes from 0.9 to 1.1, which is a narrow range compared to the reported values in the literature (between 0.4 to 1.6). Thus, the low correlation between $M_{B}$ and $s$ must be taken with caution. On the other hand, 70\% of the total SNe~Ia population is within this short range \citep{2006ApJ...648..868S}. Therefore, a no correlation between $s$ and OH within this range and a possible correlation between them for a wider range of $s$ values may be interesting and significant by itself, although we let this study for a next work.

\subsection{$M_B$ vs. color}

\begin{figure} 
\centering
\includegraphics[width=0.99\linewidth]{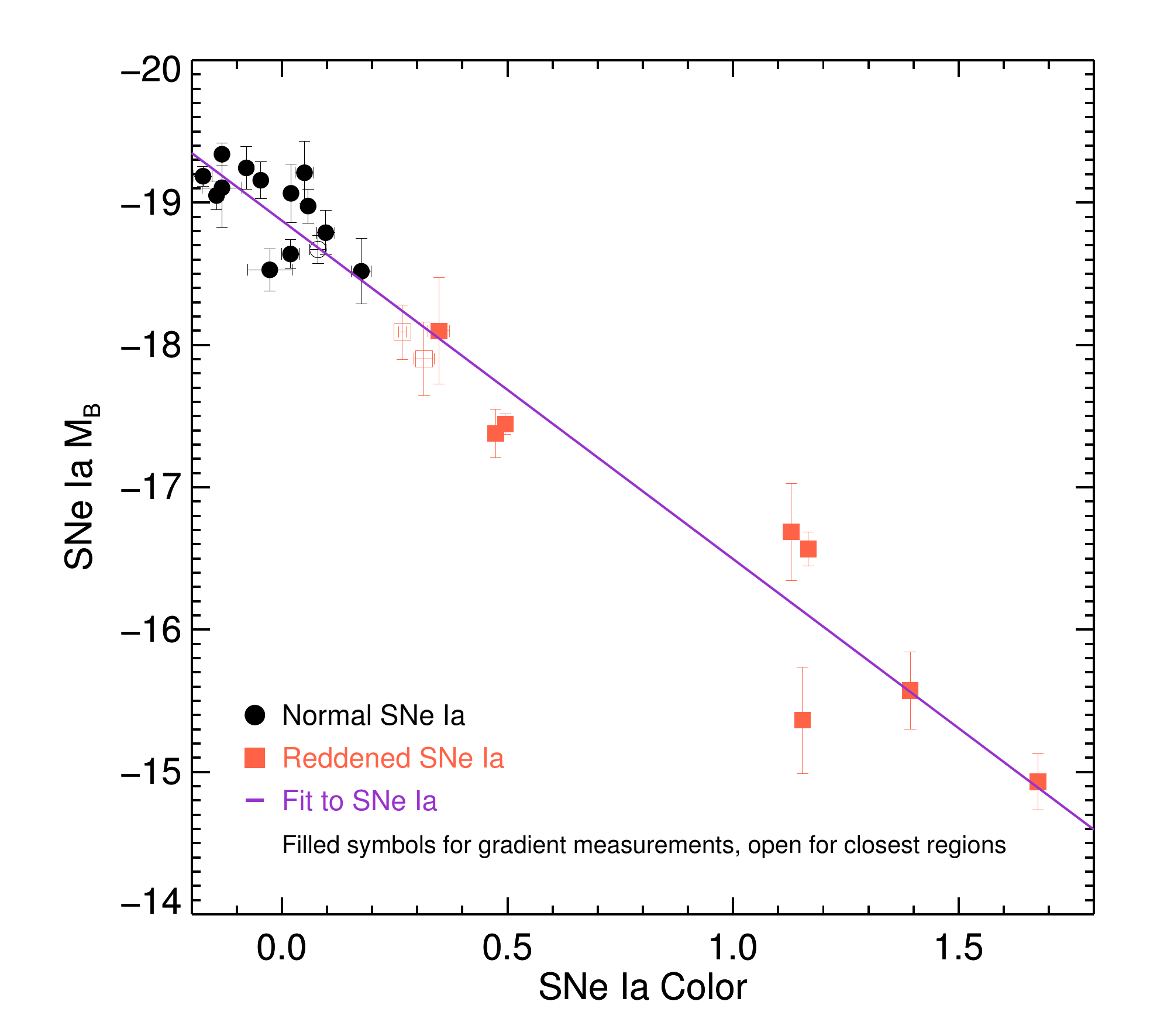}
\caption{SNe~Ia absolute magnitudes, M$_{B}$, as a function of SNe~Ia color $C$. Black circles are "normal" SNe~Ia, red squares are reddened SNe~Ia. The purple line represents a linear fit to all SNe~Ia. 
Open symbols indicate objects for which their metallicities were adopted from a proper gradient estimation, whereas open symbols show objects for which the metallicity was assumed to be the same that the nearest \HII\ region.}
\label{fig:magnitud_color}
\end{figure}

As we discussed in \citet{2016ApJ...818L..19M}, we found that metal-rich galaxies host fainter SNe~Ia than those exploding in metal-poorer galaxies. This metallicity dependence of the SN~Ia luminosity could be attributed to the {\it color} correction, a term already included in the cosmological methods to estimate the distance modulus (and implicitly in the determination of the $M_{B}$ of each SN~Ia). Actually, the color of SNe~Ia shows a dependence on the oxygen abundance (as it is seen in Covariance Matrix). And our data show a good correlation between SNe~Ia magnitudes and their colors, as we show in Figure \ref{fig:magnitud_color}. 

Indeed, this figure shows a linear behavior of $M_B$ with color, which follows this equation:
\begin{equation}
M_{B}=(-18.87\pm0.07)+(2.38\pm0.12)\times C,
\label{eq:magcolor}
\end{equation}
with a correlation coefficient of $r=0.9749$.

Both fitting parameters, $M_{B}=-18.87$ and $\beta=2.38$, are in agreement with those reported by \citet{2014A&A...568A..22B}. We note that the dispersion in $M_{B}$  without the color effect (that should be $M_{B} - \beta~C$, being $\beta$  the slope in Eq~\ref{eq:magcolor}) is reduced when it is compared with that shown by the distribution of $M_{B}$.

\subsection{Relation between metallicity and color}

\begin{figure}
\centering
\includegraphics[width=0.99\linewidth]{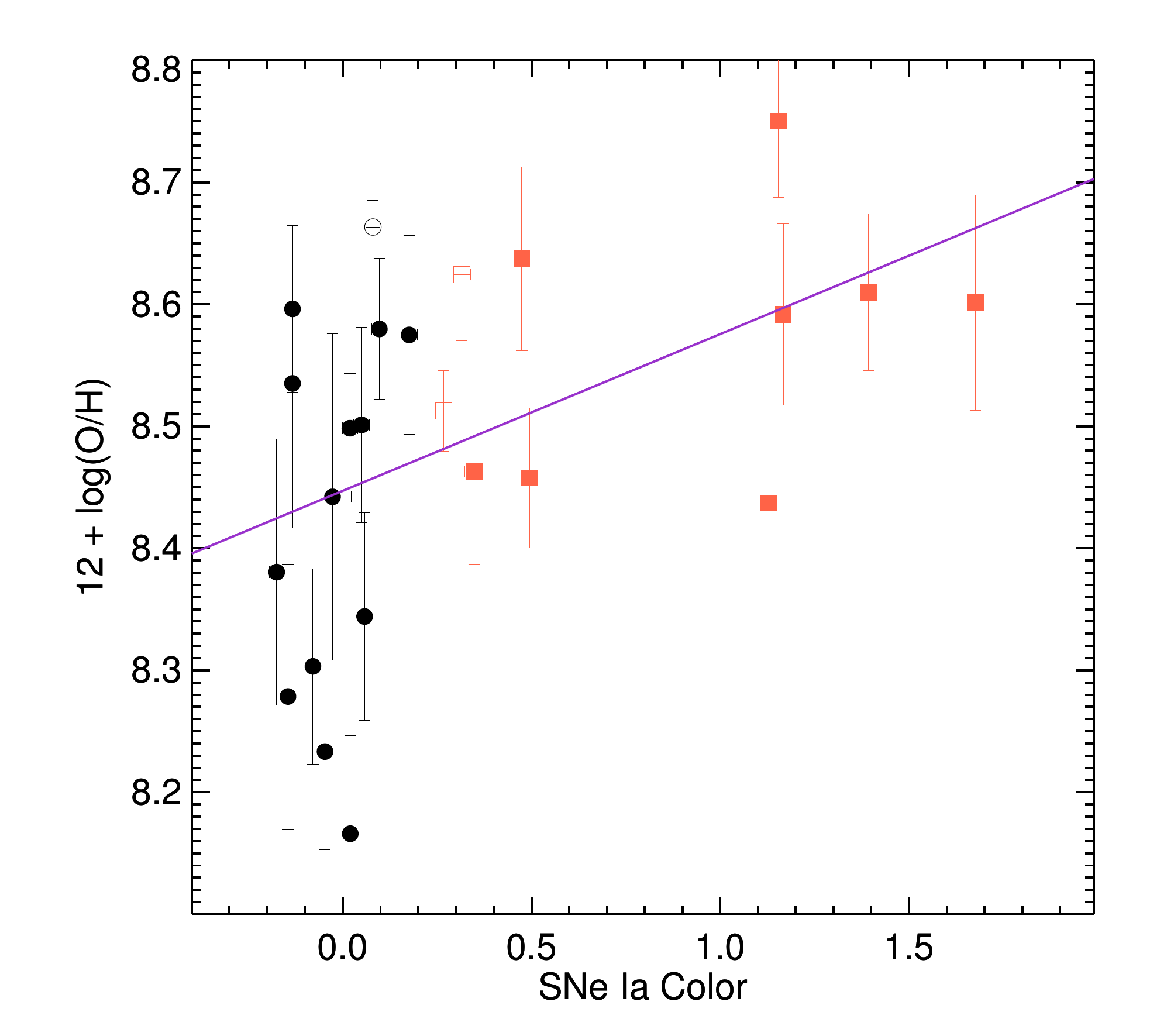}    
\caption{Oxygen abundance as a function of SNe~Ia color. Symbols are the same than in Fig. \ref{fig:magnitud_color}.}
\label{fig:color_oh}
\end{figure}

Before proceeding to the next step, which is eliminating the metallicity dependence, first we have to discuss the dependence between metallicity and color. The term $a_{2,3} = 0.4780$ indicates that there is a correlation between these two parameters, meaning that the parameters are no orthogonal. Figure \ref{fig:color_oh} shows the relation between color and OH, and a linear fit provides:
\begin{equation}
{\rm OH}=(8.45\pm0.03)+(0.13\pm0.05)\times C.
\label{eq:oh_color}
\end{equation}

As we do not want to consider twice the color dependence we create a linear combination between color and OH which is orthogonal to $C$. The new orthogonal parameter is ${\rm {OH}_{pure}} = {\rm OH} - \delta \times C$, where $\delta$ is the slope in Eq. \ref{eq:oh_color}.  The correlation coefficient between $C$ and OH$_{\rm pure}$ is $r = 0.00641$.

\subsection{$M_{B} - \beta~C$ vs. metallicity}

\begin{figure}
\centering
\includegraphics[width=0.99\linewidth]{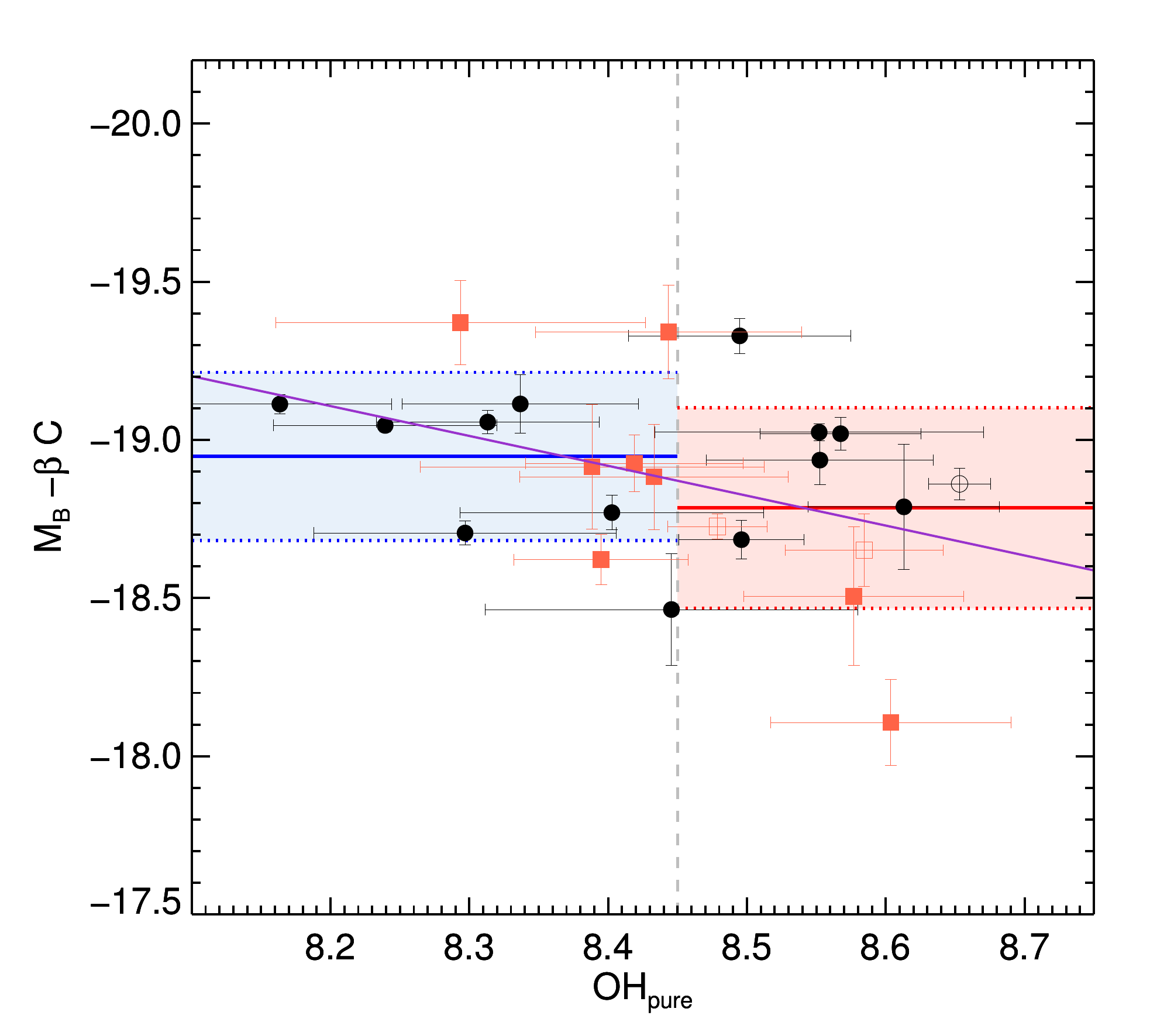}    
\caption{\label{fig:magnitud_color_oh}SNe~Ia absolute magnitude M$_{B}$ without the color effect  ($M_{B} - \beta~C$) as a function of oxygen abundance OH$_{O3N2}$. Symbols are the same than in Fig. \ref{fig:magnitud_color}. The blue and red horizontal solid lines provide the averaged value in the low- and high-metallicity regimes, respectively, with their 1$\sigma$ uncertainty shown with the pale blue and red areas.}
\end{figure}

Since $a_{1,3}$ is non-zero in the Covariance Matrix, we should expect than a metallicity dependence does exists. Figure~\ref{fig:magnitud_color_oh} compares $M_{B} - \beta~C$ ($y$-axis) with the oxygen abundance (${\rm OH}_{pure}$ in $x$-axis), showing that, indeed, $M_{B} - \beta~C$ becomes fainter when the metallicity increases.
This result agrees with the trend we reported in \citet{2016ApJ...818L..19M}, which claims that metal-rich galaxies host fainter SNe~Ia than those observed in metal-poor galaxies.

A linear fit to the data plotted in Figure \ref{fig:magnitud_color_oh} yields 
\begin{equation}
M_{B} - \beta C=(-26.85\pm3.77)+(0.94\pm0.45)\times ({\rm OH_{pure}-8.45}),
\label{eq:magnitude_color_oh}
\end{equation}
being the correlation coefficient of this fit $r$=0.4113.

We divide the abundances into low-metallicity, OH$_{O3N2} < 8.45$ with \mbox{$M_{B}=-18.95\pm0.27$~mag}, and high-metallicity, OH$_{O3N2} > 8.45$ with \mbox{$M_{B}=-18.78\pm0.32$~mag}, regimes (blue and red horizontal lines in Fig.~\ref{fig:magnitud_color_oh}) resulting in a shift of $\sim$0.17 mag. This result agrees with the shift of \mbox{$0.14~{\rm mag}$} in $M_{B}$ for unreddened SNe~Ia found in \citet{2016ApJ...818L..19M}, with high (low) metallicity galaxies hosting less (more) luminous SNe~Ia. This result depends in part on the faintest object in Figure \ref{fig:magnitud_color_oh}, SN2003cg (around $-$18 mag). The two brightest reddened SNe~Ia (around -19.4 in the same figure) have also an effect. These three SNe~Ia are reddened objects, however we have already demonstrated that the non-redenned SNe~Ia show a significant correlation between $M_{B}$ and OH. In addition, one can see at first glance in this figure, that taking only the reddened SNe~Ia, a correlation with OH already exists. Therefore, even with caution, we think that dependence on OH is clear, without any reason to discard one or more points.

\subsection{$M_{B} - \beta~C - \gamma ({\rm OH}_{\rm pure} - 8.45)$ vs. stretch}

\begin{figure}
\centering
\includegraphics[width=0.99\linewidth]{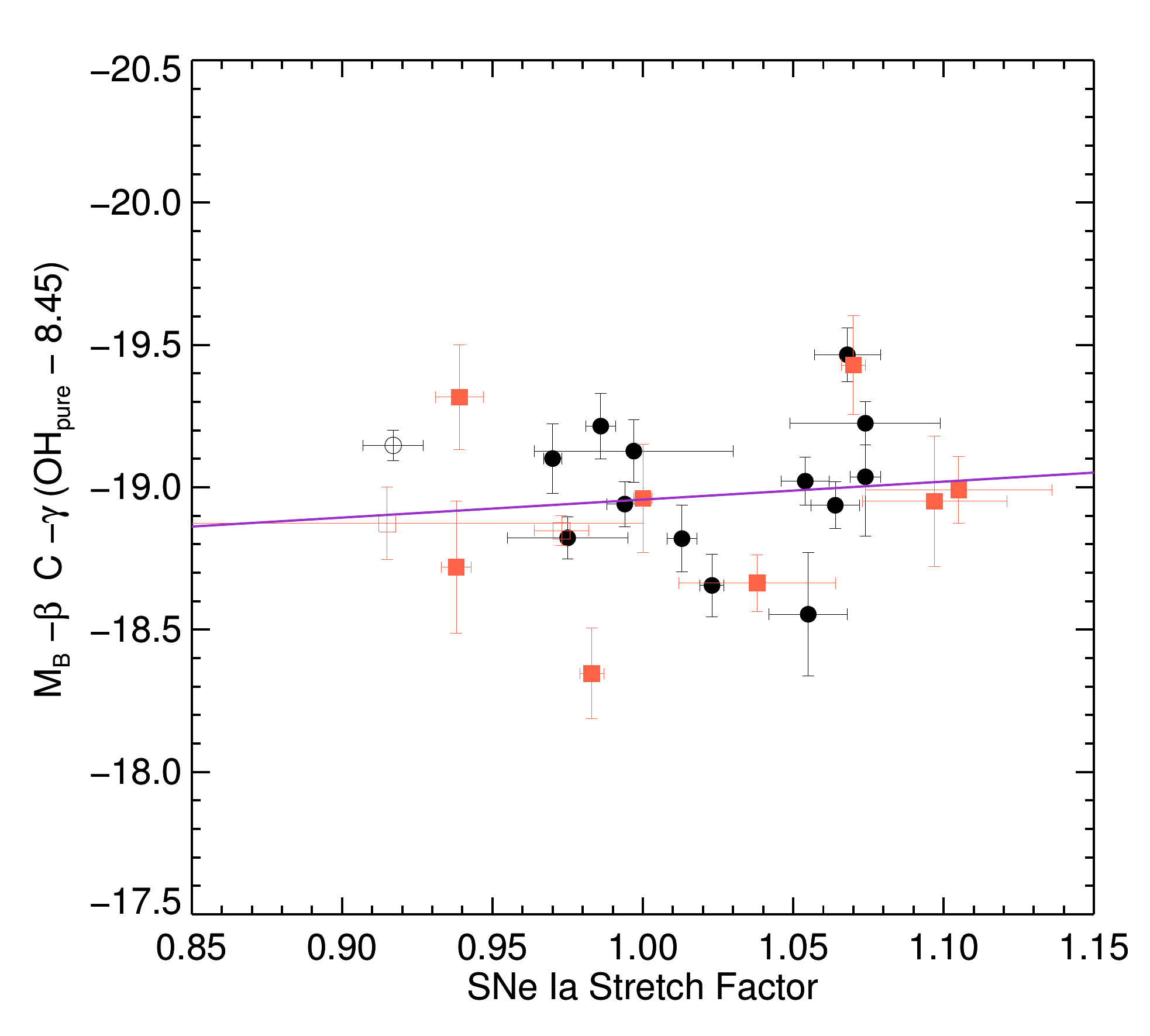}    
\caption{\label{fig:magnitud_color_oh_stretch} SNe~Ia absolute magnitude as a function of LC stretch, as $s-1$. Symbols are the same than in Fig. \ref{fig:magnitud_color}.}
\end{figure}

Once again, we can reduce the dispersion in $M_{B}-\beta\,C$ if we also remove the metallicity effect over magnitudes, that is $M_{B} - \beta C - \gamma ({\rm OH}_{\rm pure} - 8.45)$.
The LC stretch $s$ is also considered in all traditional models (Eq~\ref{eq:salt}) that correct the absolute magnitude. 
Figure~\ref{fig:magnitud_color_oh_stretch} plots the relation between absolute magnitude after the color and metallicity effects are removed, 
as $M_{B} - \beta~C - \gamma ({\rm OH} - 8.45)$, where $\gamma$ is the slope in Eq~\ref{eq:magnitude_color_oh}. 
The factor $8.45$ is obtained minimizing the dispersion in $M_{B}$ and refers to a non-correction when ${\rm OH}_{\rm pure}=8.45$. 
It also coincides with the limit value between the low- and high-metallicity regimes adopted by \citet{2016ApJ...818L..19M}. 
Figure~\ref{fig:magnitud_color_oh_stretch} seems to show that a dependence does exist. A linear fit to the data yields
\begin{equation}\label{eq:magnitud_color_oh_stretch}
M_{B} - \beta~C - \gamma ({\rm OH_{pure}} - 8.45)=(-18.96\pm0.06)-(0.63\pm1.01)\times (s-1),
\end{equation}
where the the term $(s - 1)$ refers to a non-correction when $s = 1$. The correlation coefficient of this linear fit is \mbox{$r$=-0.2984.}
With this, by adopting $M_{B} - \beta C - \gamma ({\rm OH} - 8.45) + \alpha (s - 1)$ (being $\alpha$ the slope found in the Eq~\ref{eq:magnitud_color_oh_stretch}), we have removed for $M_{B}$ all these parameters dependencies

\begin{figure}
\centering
\includegraphics[width=0.99\linewidth]{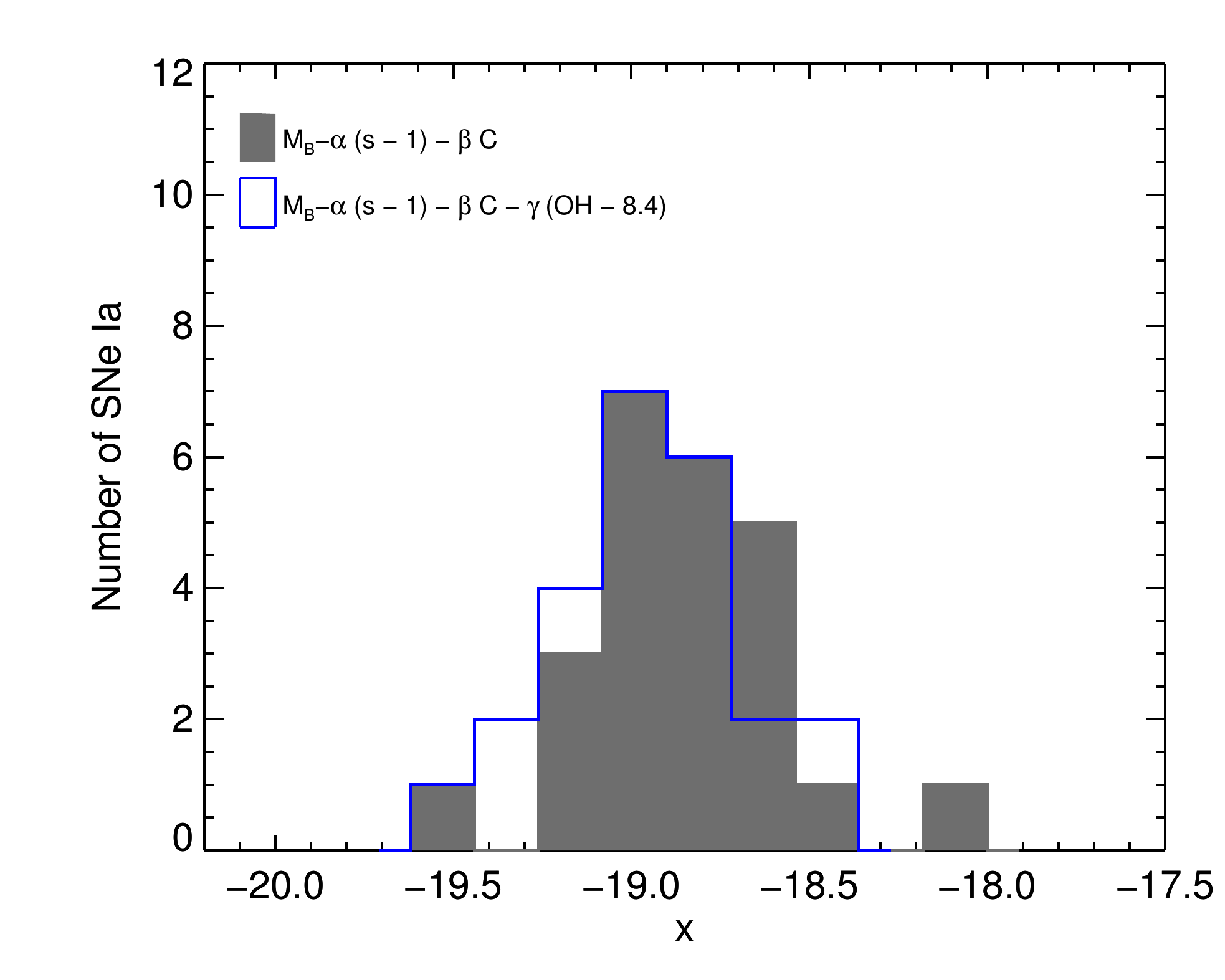}    
\caption{Histogram showing the distribution for $M_{B}$, labeled as $x$ in the horizontal axis, with $C$ and $s$ dependence subtraction (grey) and removing as well metallicity dependence (blue).}
\label{fig:histogram}
\end{figure}

With this last step, we have decreased further the dispersion of the data.  
Figure~\ref{fig:histogram} shows the distribution of magnitudes having been subtracted from traditional parameters (grey filled distribution) compared to the distribution obtained after eliminating the three dependencies (color, LC stretch, and metallicity, blue distribution). 
The standard deviation in our sample in the latter case is now reduced by  a factor of 5.1\% when compared with the standard deviation computed when only the two traditional parameters (color and LC stretch) are considered. Therefore including the metallicity dependence in the analysis eliminates a small, but measurable, luminosity scatter.
This result implies a possible improvement in reducing the Hubble residuals when using SN~Ia techniques.

We have proven how independent measurements of $M_{B}$ depend on these three parameters: $C$, $s$ and ${\rm OH}_{\rm pure}$. A forthcoming paper will include a proper cosmological analysis taking into account all these parameters at the same time.


\section{Conclusions} \label{Section5}

We present an analysis of the gas-phase oxygen abundances for 28 galaxies that has experienced a SN~Ia event. We used intermediate-resolution long-slit spectroscopy obtained at the 4.2m WHT to derive reliable oxygen abundances of \HII\ regions within the host galaxies. SNe~Ia absolute magnitudes have been calculated with parameters given by modern techniques, such SALT2. We seek  for a possible dependence of the SNe~Ia luminosities on the metallicity of the regions where exploded. Below we summarize the main conclusions found in our \mbox{analysis:}
\begin{enumerate}

\item We applied the most recent empirical calibrations by MAR13, which involve the  $N2$ and $O3N2$ indexes and are appropriate for measuring oxygen abundances in well defined \HII~regions, to obtain consistent abundance radial distributions and their corresponding gradient, with which proper oxygen abundances for the locations within their host galaxies  where SNe~Ia exploded, are estimated. This is important, since we have obtained local values of oxygen abundance for each SN~Ia, instead of what in literature has been done before, assuming a global metallicity from a unique spectra (which means a center region abundance estimate) or, even worst, through the use of a mass-metallicity relation. Our technique improves by $\sim$0.2~dex the accuracy of the oxygen abundances determined in SN~Ia host galaxies.

\item In our analysis we consider  both normal and reddened SNe~Ia, and analyzed the color effect over the $M_{B}$ . That is an improvement of the results found in \citet{2016ApJ...818L..19M}, where we only used normal (i.e., not-reddened) SNe~Ia.

\item We demonstrate that the effect of the metallicity cannot be neglected in SN~Ia studies that use their magnitudes, $M_{B}$, or distance moduli, $\mu$, since its dependence on the oxygen abundance does not have a zero value in the covariance matrix. This correction appears as a second order correction after the color correction, since, for our sample, the dependence on LC stretch is low.

\item Our conclusions agree with those presented in \citet{2016ApJ...818L..19M}, where fainter SNe~Ia are found to exist in metal-rich galaxies, whereas brightest SNe~Ia occur in galaxies with a poor content in metals when a subsample of non reddened objects was used. Hubble residuals measured using the common SNe~Ia standardization technique (color and stretch corrections) are found to be correlated to the metal abundance of their host galaxies, in the direction of SNe~Ia in metal rich galaxies are brighter, implying an overcorrection on those objects. Our results show that this overcorrection may be interpreted as a metallicity dependence which has not taken into account in the standardization.

\item We strongly claim, for our local sample, that including the oxygen abundance dependence the dispersion in $M_{B}$ is reduced to around 5\%. Therefore, we agree with \citet{2016MNRAS.457.3470C}, that using the metallicity as a third parameter might help to reduce the systematic dispersion in cosmological studies of SNe~Ia. This possibility will be checked in a future work (Moreno-Raya et al. in prep.) extending the redshift range of the galaxy sample from the Local Universe presented here.

\end{enumerate}

This is a step forward compared to the existing studies so far. In a near future, it will be possible to perform more detailed analysis using IFS data \citep{2014A&A...572A..38G,2016MNRAS.455.4087G,2016arXiv160307808G}.

The metallicity dependence found in this paper may well be important for present and future supernova cosmology surveys, helping to minimize the scatter on the SN Hubble diagram. This could be achieved by considering extra parameters in the light curve fitting or distance modulus calculations.

\section*{Acknowledgements} 

{\footnotesize
The authors acknowledge the anonymous referee for his/her helpful comments.
We thank Miguel C\'ardenas-Montes, Santiago Gonz\'alez-Gait\'an, Inma Dom\'inguez and Brad Gibson for their inestimable comments and help.
Based on observations made with the 4.2m WHT Telescope operated on the islands of La Palma by  the Isaac Newton Group of Telescopes in the Spanish observatory of Roque de Los Muchachos of the Instituto de Astrof\'\i sica de Canarias.
MEMR thanks the hospitality of the staff at both the Australian Astronomical Observatory (Australia) and Departamento de Astronom\'{\i}a of the Universidad de Chile (Chile) during his stay in 2013 \& 2014, respectively.
ARL-S thanks the support of the "Study of Emission-Line Galaxies with Integral-Field Spectroscopy" (SELGIFS) programme, funded by the EU (FP7-PEOPLE-2013-IRSES-612701).
This work has been partially supported by MINECO-FEDER grant AYA2010-21887-C04-02 and AYA2013--47742-C4-4-P. 
Support for LG is provided by the Ministry of Economy, Development, and Tourism's Millennium Science Initiative through grant IC120009 awarded to The Millennium Institute of Astrophysics (MAS), and CONICYT through FONDECYT grant 3140566.
ACR acknowledges financial support provided by the PAPDRJ CAPE/FAPERJ Fellowship.
This research has made use of the NASA/IPAC Extragalactic Database (NED) which is operated by the Jet Propulsion Laboratory, California Institute of Technology, under contract with the National Aeronautics and Space Administration. 
} 

\bibliographystyle{aa}
\bibliography{bibliography}

\newpage
\appendix

\section{Line intensities}\label{App:A}

This Appendix provides a table with line intensities for the sample spectra. All intensities are given over $I$(\Hb) except those labeled with $^{a}$, in which spectra there were not \Hb emission lines measurables. In these cases, the provided intensities are over $I$(\Ha). 

\begin{deluxetable} {c c c c c c c c c c c}
\tabletypesize{\scriptsize} 
\tablecolumns{11}
\tablenum{A.1}
\tablewidth{0pt}
\tablecaption{Extinction coefficients and de-reddened intensities for the sample. All intensities are over $I$(\Hb). For those regions where $I$(\Hb) was not available, intensities are over $I$(\Ha).}
\tablehead{
\colhead{Galaxy} &
\colhead{Region} &
\colhead{$c$(\Hb)}  &
\colhead{$I$(\Oii)} &
\colhead{$I$(\Hg)} &
\colhead{$I$(\Hb)} &
\colhead{$I$(\Oiii)} &
\colhead{$I$(\Ha)} &
\colhead{$I$(\Nii)} &
\colhead{$I$(\Sii)} &
\colhead{$I$(\Sii)}\\
&  &  & ${\rm \lambda}$3727 & ${\rm \lambda}$4340 & ${\rm \lambda}$4861 & ${\rm \lambda}$5007 & ${\rm \lambda}$6563 & ${\rm \lambda}$6584 & ${\rm \lambda}$6717 & ${\rm \lambda}$6731 
}
\startdata                   
       M\,82						&       A   &		1.93	$\pm$	0.01	&	   	 ... 	   	&	0.67	$\pm$	0.01	&	1.00	$\pm$	0.01	&	0.31	$\pm$	0.01	&	2.86	$\pm$	0.02	&	1.52	$\pm$	0.01	&	0.29	$\pm$	0.03	&	0.33	$\pm$	0.03	\\
       M\,82						&       B   &		2.22	$\pm$	0.01	&	   	 ... 	   	&	0.67	$\pm$	0.03	&	1.00	$\pm$	0.01	&	0.27	$\pm$	0.01	&	2.86	$\pm$	0.05	&	1.53	$\pm$	0.03	&	0.28	$\pm$	0.01	&	0.31	$\pm$	0.01	\\
 MCG-02-16-02				&       B   &		0.60	$\pm$	0.16	&	   	 ... 	   	&	   	 ... 	   	&	1.00	$\pm$	0.16	&	2.05	$\pm$	0.50	&	2.86	$\pm$	0.67	&	0.28	$\pm$	0.07	&	0.31	$\pm$	0.08	&	0.22	$\pm$	0.06	\\
 MCG-02-16-02				&       C   &		0.58	$\pm$	0.12	  &	   	 ... 	   	&	   	 ... 	   	&	   1.00	$\pm$	0.14 	   	&	   	 2.51	$\pm$	0.50 	   	&	2.86	$\pm$	0.03	&	0.21	$\pm$	0.02	&	0.18	$\pm$	0.03	&	0.09	$\pm$	0.06	\\
   NGC\,0105					&       A   &		0.52	$\pm$	0.14	&	   	 ... 	   	&	   	 ... 	   	&	1.00	$\pm$	0.14	&	9.03	$\pm$	1.77	&	2.86	$\pm$	0.56	&	3.98	$\pm$	0.78	&	0.87	$\pm$	0.18	&	0.85	$\pm$	0.18	\\
   NGC\,0105					&       B   &		0.34	$\pm$	0.02	&	   	 ... 	   	&	0.28	$\pm$	0.03	&	1.00	$\pm$	0.02	&	1.27	$\pm$	0.04	&	2.86	$\pm$	0.09	&	0.77	$\pm$	0.02	&	0.47	$\pm$	0.02	&	0.31	$\pm$	0.01	\\
   NGC\,0105					&       C   &		0.94	$\pm$	0.06	&	   	 ... 	   	&	   	 ... 	   	&	1.00	$\pm$	0.06	&	4.25	$\pm$	0.39	&	2.86	$\pm$	0.26	&	0.15	$\pm$	0.02	&	0.29	$\pm$	0.03	&	0.22	$\pm$	0.02	\\
   NGC\,1275					&       C   &		0.02	$\pm$	0.04	&	   	 ... 	   	&	0.34	$\pm$	0.04	&	1.00	$\pm$	0.04	&	0.65	$\pm$	0.04	&	2.86	$\pm$	0.15	&	2.19	$\pm$	0.11	&	0.51	$\pm$	0.03	&	0.32	$\pm$	0.02	\\
   NGC\,1309					&       B   &		0.17	$\pm$	0.01	&	2.38	$\pm$	0.04	&	0.43	$\pm$	0.01	&	1.00	$\pm$	0.01	&	0.73	$\pm$	0.01	&	2.86	$\pm$	0.03	&	0.66	$\pm$	0.01	&	0.51	$\pm$	0.01	&	0.36	$\pm$	0.03	\\
   NGC\,1309					&   B$+$C	&	0.35	$\pm$	0.01	&	2.23	$\pm$	0.04	&	0.44	$\pm$	0.01	&	1.00	$\pm$	0.01	&	0.78	$\pm$	0.01	&	2.86	$\pm$	0.02	&	0.63	$\pm$	0.01	&	0.48	$\pm$	0.03	&	0.35	$\pm$	0.03	\\
   NGC\,1309					&       C   &		0.34	$\pm$	0.01	&	1.56	$\pm$	0.04	&	0.44	$\pm$	0.01	&	1.00	$\pm$	0.01	&	0.89	$\pm$	0.01	&	2.86	$\pm$	0.03	&	0.56	$\pm$	0.01	&	0.42	$\pm$	0.03	&	0.28	$\pm$	0.03	\\
   NGC\,1309					&       D   &		0.23	$\pm$	0.04	&	2.28	$\pm$	0.25	&	0.50	$\pm$	0.07	&	1.00	$\pm$	0.04	&	0.57	$\pm$	0.05	&	2.86	$\pm$	0.18	&	0.85	$\pm$	0.05	&	0.63	$\pm$	0.04	&	0.48	$\pm$	0.03	\\
   NGC\,2935					&       A   &		1.15	$\pm$	0.10	&	   	 ... 	   	&	   	 ... 	   	&	1.00	$\pm$	0.10	&	2.43	$\pm$	0.38	&	2.86	$\pm$	0.42	&	2.76	$\pm$	0.40	&	0.76	$\pm$	0.11	&	0.75	$\pm$	0.11	\\
   NGC\,2935\tablenotemark{a}		&       C   &		        	 ... 	   		&	   	 ... 	   	&	   	 ... 	   	&	   	 ... 	   	&	   	 ... 	   	&	1.00	$\pm$	0.02	&	0.37	$\pm$	0.06	&	   	 ... 	   	&	   	 ... 	   	\\
   NGC\,2935\tablenotemark{a}		&       D   &		        	 ... 	   		&	   	 ... 	   	&	   	 ... 	   	&	   	 ... 	   	&	   	 ... 	   	&	1.00	$\pm$	0.08	&	0.45	$\pm$	0.11	&	0.29	$\pm$	0.24	&	   	 ... 	   	\\
   NGC\,2935\tablenotemark{a}		&       E   &		        	 ... 	   		&	   	 ... 	   	&	   	 ... 	   	&	   	 ... 	   	&	   	 ... 	   	&	1.00	$\pm$	0.04	&	0.46	$\pm$	0.08	&	   	 ... 	   	&	   	 ... 	   	\\
   NGC\,3021					&       B   &		0.53	$\pm$	0.01	&	0.96	$\pm$	0.08	&	0.50	$\pm$	0.02	&	1.00	$\pm$	0.01	&	0.16	$\pm$	0.02	&	2.86	$\pm$	0.06	&	1.07	$\pm$	0.02	&	0.40	$\pm$	0.01	&	0.30	$\pm$	0.01	\\
   NGC\,3021					&       C   &		0.54	$\pm$	0.02	&	1.34	$\pm$	0.05	&	0.56	$\pm$	0.03	&	1.00	$\pm$	0.02	&	0.26	$\pm$	0.02	&	2.86	$\pm$	0.07	&	1.07	$\pm$	0.03	&	0.46	$\pm$	0.01	&	0.33	$\pm$	0.01	\\
   NGC\,3021					&       D   &		0.00  $\pm$	0.01	&	0.62	$\pm$	0.05	&	0.35	$\pm$	0.02	&	1.00	$\pm$	0.01	&	0.29	$\pm$	0.02	&	2.86	$\pm$	0.06	&	1.01	$\pm$	0.02	&	0.50	$\pm$	0.01	&	0.35	$\pm$	0.01	\\
   NGC\,3021					&       E   &		0.68	$\pm$	0.03	&	1.80	$\pm$	0.28	&	0.35	$\pm$	0.06	&	1.00	$\pm$	0.03	&	0.19	$\pm$	0.03	&	2.86	$\pm$	0.13	&	1.02	$\pm$	0.05	&	0.45	$\pm$	0.03	&	0.33	$\pm$	0.02	\\
   NGC\,3147					&       B   &		0.25	$\pm$	0.03	&	   	 ... 	   	&	0.45	$\pm$	0.03	&	1.00	$\pm$	0.03	&	0.06	$\pm$	0.02	&	2.86	$\pm$	0.10	&	0.76	$\pm$	0.03	&	0.27	$\pm$	0.01	&	0.19	$\pm$	0.01	\\
   NGC\,3147					&       C   &		0.55	$\pm$	0.06	&	   	 ... 	   	&	   	 ... 	   	&	1.00	$\pm$	0.06	&	0.16	$\pm$	0.02	&	2.86	$\pm$	0.24	&	0.74	$\pm$	0.07	&	0.18	$\pm$	0.04	&	0.23	$\pm$	0.05	\\
   NGC\,3147					&       D   &		1.03	$\pm$	0.16	&	   	 ... 	   	&	   	 ... 	   	&	1.00	$\pm$	0.16	&	   	 ... 	   	&	2.86	$\pm$	0.65	&	1.03	$\pm$	0.24	&	0.29	$\pm$	0.08	&	0.23	$\pm$	0.06	\\
   NGC\,3147					&       E   &		0.65	$\pm$	0.05	&	   	 ... 	   	&	   	 ... 	   	&	1.00	$\pm$	0.05	&	   	 ... 	   	&	2.86	$\pm$	0.21	&	0.59	$\pm$	0.05	&	0.17	$\pm$	0.03	&	0.13	$\pm$	0.03	\\
   NGC\,3169					&       E   &		0.00  $\pm$	0.02	&	2.22	$\pm$	0.06	&	0.41	$\pm$	0.02	&	1.00	$\pm$	0.02	&	1.31	$\pm$	0.04	&	2.86	$\pm$	0.07	&	0.85	$\pm$	0.02	&	0.44	$\pm$	0.01	&	0.31	$\pm$	0.01	\\
   NGC\,3169					&       G   &		0.34	$\pm$	0.08	&	   	 ... 	   	&	   	 ... 	   	&	1.00	$\pm$	0.08	&	0.09	$\pm$	0.03	&	2.86	$\pm$	0.33	&	1.13	$\pm$	0.13	&	0.48	$\pm$	0.06	&	0.36	$\pm$	0.06	\\
   NGC\,3368					&       B   &		0.32	$\pm$	0.01	&	   	 ... 	   	&	0.45	$\pm$	0.02	&	1.00	$\pm$	0.01	&	0.14	$\pm$	0.01	&	2.86	$\pm$	0.06	&	1.00	$\pm$	0.02	&	0.46	$\pm$	0.01	&	0.32	$\pm$	0.01	\\
   NGC\,3368					&       C   &		0.52	$\pm$	0.12	&	   	 ... 	   	&	   	 ... 	   	&	1.00	$\pm$	0.12	&	   	 ... 	   	&	2.86	$\pm$	0.48	&	0.80	$\pm$	0.14	&	0.50	$\pm$	0.08	&	0.32	$\pm$	0.06	\\
   NGC\,3370					&       B   &		0.17	$\pm$	0.03	&	2.03	$\pm$	0.02	&	0.49	$\pm$	0.03	&	1.00	$\pm$	0.03	&	2.59	$\pm$	0.01	&	2.86	$\pm$	0.01	&	0.29	$\pm$	0.03	&	0.22	$\pm$	0.03	&	0.16	$\pm$	0.03	\\
   NGC\,3370					&       C   &		0.46	$\pm$	0.01	&	1.97	$\pm$	0.08	&	0.42	$\pm$	0.01	&	1.00	$\pm$	0.01	&	1.67	$\pm$	0.03	&	2.86	$\pm$	0.04	&	0.81	$\pm$	0.01	&	0.51	$\pm$	0.01	&	0.39	$\pm$	0.01	\\
   NGC\,3370					&       D   &		0.21	$\pm$	0.03	&	   	 ... 	   	&	0.43	$\pm$	0.03	&	1.00	$\pm$	0.03	&	0.18	$\pm$	0.05	&	2.86	$\pm$	0.11	&	0.95	$\pm$	0.04	&	0.47	$\pm$	0.02	&	0.35	$\pm$	0.02	\\
   NGC\,3370					&       E   &		0.07	$\pm$	0.02	&	1.67	$\pm$	0.09	&	0.39	$\pm$	0.02	&	1.00	$\pm$	0.02	&	0.87	$\pm$	0.03	&	2.86	$\pm$	0.09	&	0.62	$\pm$	0.02	&	0.55	$\pm$	0.02	&	0.36	$\pm$	0.01	\\
   NGC\,3370					&       F   &		0.41	$\pm$	0.01	&	0.96	$\pm$	0.11	&	0.49	$\pm$	0.02	&	1.00	$\pm$	0.01	&	0.11	$\pm$	0.02	&	2.86	$\pm$	0.05	&	0.78	$\pm$	0.02	&	0.39	$\pm$	0.01	&	0.28	$\pm$	0.01	\\
   NGC\,3672					&       A   &		0.84	$\pm$	0.03	&	   	 ... 	   	&	0.72	$\pm$	0.07	&	1.00	$\pm$	0.03	&	0.16	$\pm$	0.03	&	2.86	$\pm$	0.11	&	0.90	$\pm$	0.04	&	0.35	$\pm$	0.01	&	0.27	$\pm$	0.01	\\
   NGC\,3672					&       O   &		1.04	$\pm$	0.14	&	   	 ... 	   	&	   	 ... 	   	&	1.00	$\pm$	0.14	&	   	 ... 	   	&	2.86	$\pm$	0.56	&	1.41	$\pm$	0.29	&	0.46	$\pm$	0.11	&	0.32	$\pm$	0.08	\\
   NGC\,3672\tablenotemark{a}		&       B   &		        	 ... 	   		&	   	 ... 	   	&	   	 ... 	   	&	   	 ... 	   	&	   	 ... 	   	&	1.00	$\pm$	0.01	&	0.37	$\pm$	0.02	&	0.17	$\pm$	0.04	&	0.12	$\pm$	0.06	\\
   NGC\,3982					&       A   &		0.95	$\pm$	0.05	&	   	 ... 	   	&	   	 ... 	   	&	1.00	$\pm$	0.05	&	17.38	$\pm$	1.30	&	2.86	$\pm$	0.21	&	3.33	$\pm$	0.25	&	0.78	$\pm$	0.06	&	1.10	$\pm$	0.08	\\
   NGC\,3982					&       B   &  	0.41	$\pm$	0.02	&	   	 ... 	   	&	0.34	$\pm$	0.03	&	1.00	$\pm$	0.02	&	0.12	$\pm$	0.02	&	2.86	$\pm$	0.07	&	0.95	$\pm$	0.02	&	0.20	$\pm$	0.01	&	0.15	$\pm$	0.01	\\
   NGC\,3982					&       C   &		0.34	$\pm$	0.02	&	   	 ... 	   	&	0.37	$\pm$	0.02	&	1.00	$\pm$	0.02	&	0.09	$\pm$	0.01	&	2.86	$\pm$	0.08	&	1.03	$\pm$	0.03	&	0.41	$\pm$	0.01	&	0.29	$\pm$	0.01	\\
   NGC\,3982					&       D   &		0.45	$\pm$	0.01	&	   	 ... 	   	&	0.48	$\pm$	0.02	&	1.00	$\pm$	0.01	&	0.23	$\pm$	0.04	&	2.86	$\pm$	0.05	&	0.84	$\pm$	0.02	&	0.39	$\pm$	0.01	&	0.29	$\pm$	0.01	\\
   NGC\,3982\tablenotemark{a}		&       E   &		        	 ... 	   		&	   	 ... 	   	&	   	 ... 	   	&	   	 ... 	   	&	   	 ... 	   	&	1.00	$\pm$	0.03	&	0.38	$\pm$	0.01	&	0.13	$\pm$	0.01	&	0.09	$\pm$	0.02	\\
   NGC\,4321					&       B   &		0.43	$\pm$	0.01	&	0.51	$\pm$	0.07	&	0.42	$\pm$	0.02	&	1.00	$\pm$	0.01	&	0.14	$\pm$	0.02	&	2.86	$\pm$	0.04	&	0.75	$\pm$	0.01	&	0.35	$\pm$	0.01	&	0.26	$\pm$	0.01	\\
   NGC\,4321					&       C   &		0.36	$\pm$	0.03	&	   	 ... 	   	&	0.47	$\pm$	0.05	&	1.00	$\pm$	0.03	&	   	 ... 	   	&	2.86	$\pm$	0.13	&	0.61	$\pm$	0.03	&	0.20	$\pm$	0.02	&	0.15	$\pm$	0.02	\\
   NGC\,4321					&       E   &		1.05	$\pm$	0.04	&	   	 ... 	   	&	0.64	$\pm$	0.07	&	1.00	$\pm$	0.04	&	   	 ... 	   	&	2.86	$\pm$	0.15	&	0.97	$\pm$	0.05	&	0.31	$\pm$	0.02	&	0.23	$\pm$	0.02	\\
   NGC\,4501					&       B   &		0.97	$\pm$	0.12	&	   	 ... 	   	&	   	 ... 	   	&	1.00	$\pm$	0.12	&	   	 ... 	   	&	2.86	$\pm$	0.47	&	0.50	$\pm$	0.09	&	0.18	$\pm$	0.04	&	0.16	$\pm$	0.03	\\
   NGC\,4501					&       C   &		0.94	$\pm$	0.10	&	   	 ... 	   	&	   	 ... 	   	&	1.00	$\pm$	0.10	&	   	 ... 	   	&	2.86	$\pm$	0.42	&	0.69	$\pm$	0.10	&	0.21	$\pm$	0.04	&	0.18	$\pm$	0.04	\\
   NGC\,4501\tablenotemark{a}		&       A   &		        	 ... 	   		&	   	 ... 	   	&	   	 ... 	   	&	   	 ... 	   	&	   	 ... 	   	&	1.00	$\pm$	0.01	&	0.21	$\pm$	0.05	&	0.06	$\pm$	0.16	&	0.07	$\pm$	0.20	\\
   NGC\,4527					&       A   &		1.13	$\pm$	0.07	&	   	 ... 	   	&	   	 ... 	   	&	1.00	$\pm$	0.07	&	0.41	$\pm$	0.12	&	2.86	$\pm$	0.28	&	1.01	$\pm$	0.10	&	0.45	$\pm$	0.05	&	0.32	$\pm$	0.03	\\
   NGC\,4527					&       B   &		0.72	$\pm$	0.06	&	   	 ... 	   	&	   	 ... 	   	&	1.00	$\pm$	0.06	&	0.26	$\pm$	0.08	&	2.86	$\pm$	0.24	&	1.04	$\pm$	0.09	&	0.45	$\pm$	0.05	&	0.38	$\pm$	0.04	\\
   NGC\,4527					&       D   &		1.08	$\pm$	0.06	&	   	 ... 	   	&	   	 ... 	   	&	1.00	$\pm$	0.06	&	   	 ... 	   	&	2.86	$\pm$	0.22	&	0.96	$\pm$	0.08	&	0.32	$\pm$	0.03	&	0.25	$\pm$	0.02	\\
   NGC\,4527					&       E   &		0.89	$\pm$	0.15	&	   	 ... 	   	&	   	 ... 	   	&	1.00	$\pm$	0.15	&	   	 ... 	   	&	2.86	$\pm$	0.62	&	0.85	$\pm$	0.19	&	0.39	$\pm$	0.09	&	0.24	$\pm$	0.05	\\
   NGC\,4527\tablenotemark{a}		&       F   &		        	 ... 	   		&	   	 ... 	   	&	   	 ... 	   	&	   	 ... 	   	&	   	 ... 	   	&	1.00	$\pm$	0.02	&	0.32	$\pm$	0.06	&	   	 ... 	   	&	   	 ... 	   	\\
   NGC\,4527\tablenotemark{a}		&       O   &		        	 ... 	   		&	   	 ... 	   	&	   	 ... 	   	&	   	 ... 	   	&	   	 ... 	   	&	1.00	$\pm$	0.03	&	0.53	$\pm$	0.01	&	0.17	$\pm$	0.03	&	0.13	$\pm$	0.03	\\
   NGC\,4536					&       A   &		0.60	$\pm$	0.04	&	   	 ... 	   	&	0.49	$\pm$	0.07	&	1.00	$\pm$	0.04	&	0.51	$\pm$	0.05	&	2.86	$\pm$	0.18	&	0.83	$\pm$	0.05	&	0.51	$\pm$	0.03	&	0.36	$\pm$	0.02	\\
   NGC\,4536					&       B   &		0.33	$\pm$	0.07	&	   	 ... 	   	&	   	 ... 	   	&	1.00	$\pm$	0.07	&	0.31	$\pm$	0.20	&	2.86	$\pm$	0.29	&	0.69	$\pm$	0.08	&	0.54	$\pm$	0.07	&	0.36	$\pm$	0.05	\\
   NGC\,4536					&       O   &		1.33	$\pm$	0.02	&	   	 ... 	   	&	0.56	$\pm$	0.03	&	1.00	$\pm$	0.02	&	0.35	$\pm$	0.02	&	2.86	$\pm$	0.06	&	1.27	$\pm$	0.03	&	0.55	$\pm$	0.01	&	0.45	$\pm$	0.01	\\
   NGC\,4536\tablenotemark{a}		&       C   &		        	 ... 	   		&	   	 ... 	   	&	   	 ... 	   	&	   	 ... 	   	&	   	 ... 	   	&	1.00	$\pm$	0.02	&	0.20	$\pm$	0.05	&	0.22	$\pm$	0.07	&	0.17	$\pm$	0.11	\\
   NGC\,4639					&       A   &		0.23	$\pm$	0.02	&	   	 ... 	   	&	0.45	$\pm$	0.02	&	1.00	$\pm$	0.02	&	0.20	$\pm$	0.02	&	2.86	$\pm$	0.06	&	0.94	$\pm$	0.02	&	0.41	$\pm$	0.04	&	0.30	$\pm$	0.01	\\
   NGC\,4639					&       B   &		0.21	$\pm$	0.06	&	   	 ... 	   	&	   	 ... 	   	&	1.00	$\pm$	0.05	&	0.27	$\pm$	0.08	&	2.86	$\pm$	0.23	&	1.04	$\pm$	0.09	&	0.30	$\pm$	0.05	&	0.24	$\pm$	0.04	\\
   NGC\,4639					&       C   &		0.45	$\pm$	0.07	&	   	 ... 	   	&	   	 ... 	   	&	1.00	$\pm$	0.07	&	1.18	$\pm$	0.16	&	2.86	$\pm$	0.28	&	0.90	$\pm$	0.10	&	   	 ... 	   	&	   	 ... 	   	\\
   NGC\,5005					&       A   &		0.46	$\pm$	0.01	&	0.51	$\pm$	0.07	&	0.44	$\pm$	0.02	&	1.00	$\pm$	0.01	&	0.10	$\pm$	0.03	&	2.86	$\pm$	0.05	&	0.75	$\pm$	0.02	&	0.32	$\pm$	0.01	&	0.23	$\pm$	0.01	\\
   NGC\,5005					&       B   &		0.78	$\pm$	0.02	&	0.99	$\pm$	0.10	&	0.44	$\pm$	0.03	&	1.00	$\pm$	0.02	&	0.15	$\pm$	0.02	&	2.86	$\pm$	0.06	&	1.00	$\pm$	0.02	&	0.35	$\pm$	0.01	&	0.26	$\pm$	0.01	\\
   NGC\,5468					&       A   &		0.04	$\pm$	0.01	&	   	 ... 	   	&	0.37	$\pm$	0.02	&	1.00	$\pm$	0.01	&	0.14	$\pm$	0.02	&	2.86	$\pm$	0.05	&	1.06	$\pm$	0.02	&	0.47	$\pm$	0.01	&	0.38	$\pm$	0.01	\\
   NGC\,5468					&       B   &		 0.00 $\pm$	0.01	&	1.93	$\pm$	0.03	&	0.43	$\pm$	0.01	&	1.00	$\pm$	0.01	&	2.25	$\pm$	0.03	&	2.86	$\pm$	0.03	&	0.31	$\pm$	0.01	&	0.37	$\pm$	0.01	&	0.26	$\pm$	0.01	\\
   NGC\,5468					&       C   &		0.07	$\pm$	0.01	&	1.84	$\pm$	0.05	&	0.45	$\pm$	0.01	&	1.00	$\pm$	0.01	&	1.41	$\pm$	0.02	&	2.86	$\pm$	0.03	&	0.45	$\pm$	0.01	&	0.42	$\pm$	0.01	&	0.29	$\pm$	0.01	\\
   NGC\,5584					&       A   &		0.07	$\pm$	0.04	&	   	 ... 	   	&	   	 ... 	   	&	1.00	$\pm$	0.04	&	0.44	$\pm$	0.04	&	2.86	$\pm$	0.16	&	0.69	$\pm$	0.04	&	0.51	$\pm$	0.03	&	0.29	$\pm$	0.02	\\
   NGC\,5584					&       B   &		0.24	$\pm$	0.09	&	   	 ... 	   	&	   	 ... 	   	&	1.00	$\pm$	0.09	&	0.46	$\pm$	0.10	&	2.86	$\pm$	0.37	&	0.68	$\pm$	0.09	&	   	 ... 	   	&	   	 ... 	   	\\
   NGC\,5584					&       C   &		0.31	$\pm$	0.12	&	   	 ... 	   	&	   	 ... 	   	&	1.00	$\pm$	0.12	&	0.90	$\pm$	0.17	&	2.86	$\pm$	0.48	&	0.57	$\pm$	0.10	&	0.47	$\pm$	0.09	&	0.35	$\pm$	0.07	\\
   NGC\,5584					&       O   &		0.15	$\pm$	0.08	&	   	 ... 	   	&	   	 ... 	   	&	1.00	$\pm$	0.08	&	0.19	$\pm$	0.06	&	2.86	$\pm$	0.33	&	0.92	$\pm$	0.11	&	0.56	$\pm$	0.07	&	0.41	$\pm$	0.06	\\
    UGC\,00272					&       A   &		0.18	$\pm$	0.04	&	   	 ... 	   	&	0.53	$\pm$	0.08	&	1.00	$\pm$	0.04	&	0.84	$\pm$	0.08	&	2.86	$\pm$	0.18	&	0.88	$\pm$	0.06	&	0.80	$\pm$	0.05	&	0.62	$\pm$	0.04	\\
    UGC\,00272					&       C   &		0.04	$\pm$	0.05	&	   	 ... 	   	&	   	 ... 	   	&	1.00	$\pm$	0.05	&	0.69	$\pm$	0.07	&	2.86	$\pm$	0.21	&	0.59	$\pm$	0.06	&	0.71	$\pm$	0.06	&	0.48	$\pm$	0.04	\\
    UGC\,00272					&       D   &		1.30	$\pm$	0.13	&	   	 ... 	   	&	0.83	$\pm$	0.23	&	1.00	$\pm$	0.12	&	1.50	$\pm$	0.28	&	2.86	$\pm$	0.51	&	0.50	$\pm$	0.11	&	0.64	$\pm$	0.12	&	0.46	$\pm$	0.09	\\
    UGC\,00272					&       F   &		 0.00	$\pm$	0.08	&	   	 ... 	   	&	   	 ... 	   	&	1.00	$\pm$	0.07	&	1.50	$\pm$	0.18	&	2.86	$\pm$	0.31	&	0.35	$\pm$	0.06	&	0.60	$\pm$	0.09	&	0.49	$\pm$	0.08	\\
   UGC\,03218\tablenotemark{a}	&       A   &		        	 ... 	   		&	   	 ... 	   	&	   	 ... 	   	&	   	 ... 	   	&	   	 ... 	   	&	1.00	$\pm$	0.03	&	0.40	$\pm$	0.09	&	   	 ... 	   	&	   	 ... 	   	\\
   UGC\,03218\tablenotemark{a}	&       B   &		        	 ... 	 	  	&	   	 ... 	   	&	   	 ... 	   	&	   	 ... 	   	&	   	 ... 	   	&	1.00	$\pm$	0.03	&	0.39	$\pm$	0.06	&	0.30	$\pm$	0.12	&	0.20	$\pm$	0.18	\\
   UGC\,03218\tablenotemark{a}	&       O   &		        	 ... 	 	  	&	   	 ... 	   	&	   	 ... 	   	&	   	 ... 	   	&	   	 ... 	   	&	1.00	$\pm$	0.01	&	0.37	$\pm$	0.03	&	0.14	$\pm$	0.05	&	0.09	$\pm$	0.06	\\
   UGC\,03576\tablenotemark{a}	&       A   &		        	 ... 		   	&	   	 ... 	   	&	   	 ... 	   	&	   	 ... 	   	&	   	 ... 	   	&	1.00	$\pm$	0.02	&	0.39	$\pm$	0.05	&	0.20	$\pm$	0.09	&	0.15	$\pm$	0.10	\\
   UGC\,03576\tablenotemark{a}	&       C   &		        	 ... 		   	&	   	 ... 	   	&	   	 ... 	   	&	   	 ... 	   	&	   	 ... 	   	&	1.00	$\pm$	0.03	&	0.47	$\pm$	0.03	&	   	 ... 	   	&	   	 ... 	   	\\
   UGC\,03576\tablenotemark{a}	&       D   &		        	 ... 	   		&	   	 ... 	   	&	   	 ... 	   	&	   	 ... 	   	&	   	 ... 	   	&	1.00	$\pm$	0.04	&	0.39	$\pm$	0.09	&	   	 ... 	   	&	   	 ... 	   	\\
   UGC\,03845					&       A   &		0.39	$\pm$	0.04	&	   	 ... 	   	&	   	 ... 	   	&	1.00	$\pm$	0.04	&	0.30	$\pm$	0.05	&	2.86	$\pm$	0.18	&	0.99	$\pm$	0.06	&	0.59	$\pm$	0.04	&	0.45	$\pm$	0.03	\\
   UGC\,03845					&       B   &		0.33   $\pm$   0.01    &       3.23    $\pm$   0.10    &       0.03    $\pm$   0.01	&       1.00    $\pm$   0.03    &       0.11    $\pm$   0.02    &       2.86    $\pm$   0.09    &       0.53    $\pm$   0.01    &       0.49    $\pm$   0.01    &       0.36    $\pm$   0.01    \\
   UGC\,04195					&       A   &		0.42	$\pm$	0.05	&	   	 ... 	   	&	0.40	$\pm$	0.07	&	1.00	$\pm$	0.05	&	0.34	$\pm$	0.06	&	2.86	$\pm$	0.22	&	1.48	$\pm$	0.11	&	0.40	$\pm$	0.03	&	0.38	$\pm$	0.03	\\
   UGC\,09391					&       A   &		0.41	$\pm$	0.03	&	0.72	$\pm$	0.12	&	0.31	$\pm$	0.04	&	1.00	$\pm$	0.03	&	2.55	$\pm$	0.10	&	2.86	$\pm$	0.11	&	0.24	$\pm$	0.01	&	0.41	$\pm$	0.03	&	0.28	$\pm$	0.02	\\
   UGC\,09391					&       B   &		0.63	$\pm$	0.09	&	2.23	$\pm$	0.49	&	   	 ... 	   	&	1.00	$\pm$	0.09	&	0.93	$\pm$	0.14	&	2.86	$\pm$	0.36	&	0.32	$\pm$	0.05	&	0.50	$\pm$	0.07	&	0.36	$\pm$	0.05	\\
   UGC\,09391					&       C   &		0.37	$\pm$	0.08	&	1.69	$\pm$	0.36	&	   	 ... 	   	&	1.00	$\pm$	0.08	&	1.00	$\pm$	0.13	&	2.86	$\pm$	0.32	&	0.57	$\pm$	0.07	&	0.61	$\pm$	0.07	&	0.44	$\pm$	0.06	\\
   UGC\,09391					&       D   &		0.72	$\pm$	0.21	&	9.46	$\pm$	2.96	&	   	 ... 	   	&	1.00	$\pm$	0.21	&	1.52	$\pm$	0.47	&	2.86	$\pm$	0.84	&	0.42	$\pm$	0.13	&	0.75	$\pm$	0.22	&	0.53	$\pm$	0.16	\\
    UGCA\,017					&       A   &		0.60	$\pm$	0.07	&	   	 ... 	   	&	   	 ... 	   	&	1.00	$\pm$	0.07	&	2.38	$\pm$	0.24	&	2.86	$\pm$	0.27	&	0.29	$\pm$	0.04	&	0.30	$\pm$	0.04	&	0.21	$\pm$	0.03	\\
    UGCA\,017					&       B   &		0.45	$\pm$	0.04	&	1.11	$\pm$	0.21	&	0.30	$\pm$	0.04	&	1.00	$\pm$	0.04	&	1.37	$\pm$	0.09	&	2.86	$\pm$	0.17	&	0.39	$\pm$	0.02	&	0.50	$\pm$	0.03	&	0.36	$\pm$	0.02	\\
    UGCA\,017					&       C   &		0.49	$\pm$	0.04	&	2.22	$\pm$	0.49	&	0.39	$\pm$	0.06	&	1.00	$\pm$	0.04	&	1.03	$\pm$	0.07	&	2.86	$\pm$	0.16	&	0.67	$\pm$	0.04	&	0.72	$\pm$	0.04	&	0.53	$\pm$	0.03	\\
    UGCA\,017					&       D   &		0.40	$\pm$	0.02	&	1.53	$\pm$	0.12	&	0.36	$\pm$	0.03	&	1.00	$\pm$	0.02	&	0.93	$\pm$	0.03	&	2.86	$\pm$	0.08	&	0.66	$\pm$	0.02	&	0.54	$\pm$	0.02	&	0.38	$\pm$	0.01	\\
    UGCA\,017					&       E   &		0.70	$\pm$	0.07	&	5.89	$\pm$	0.65	&	   	 ... 	   	&	1.00	$\pm$	0.07	&	2.04	$\pm$	0.21	&	2.86	$\pm$	0.29	&	0.47	$\pm$	0.05	&	0.60	$\pm$	0.07	&	0.43	$\pm$	0.05	\\\enddata
\tablenotetext{a}{Due to the lack of \Hb, intensities are over $I$(\Ha).}
\end{deluxetable}

\newpage
\section{Galaxy gradients}\label{App:B}

This Appendix compiles both the images and the metallicity measurements of all the 28 galaxies included in this paper. Figures are made up of two panels. The left panels show the $R-$Band image for the galaxy with the slit position (the slit position is not to scale). The position of the SN~Ia is marked with a yellow circle. Observed regions are represented within the slit, each one has been identified with a capital letter. The regions are drawn in different colors depending on the nature of the emission: blue for star-forming regions; yellow for composite; red for AGNs; and purple for regions without emission lines. The right  panels show the oxygen abundance in function of the GCD. The \HII\ regions for which oxygen abundances are derived are plotted using blue circles. A pink star marks the position and metallicity for each SN~Ia. Metallicity gradients (continuous blue line) are shown when they have been derived. Symbols are the same than in Figure \ref{gradientes_primeros}.

\begin{figure*} 
\centering
\includegraphics[width=0.33\textwidth]{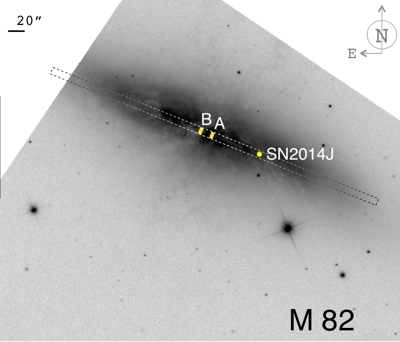}
\includegraphics[width=0.66\textwidth]{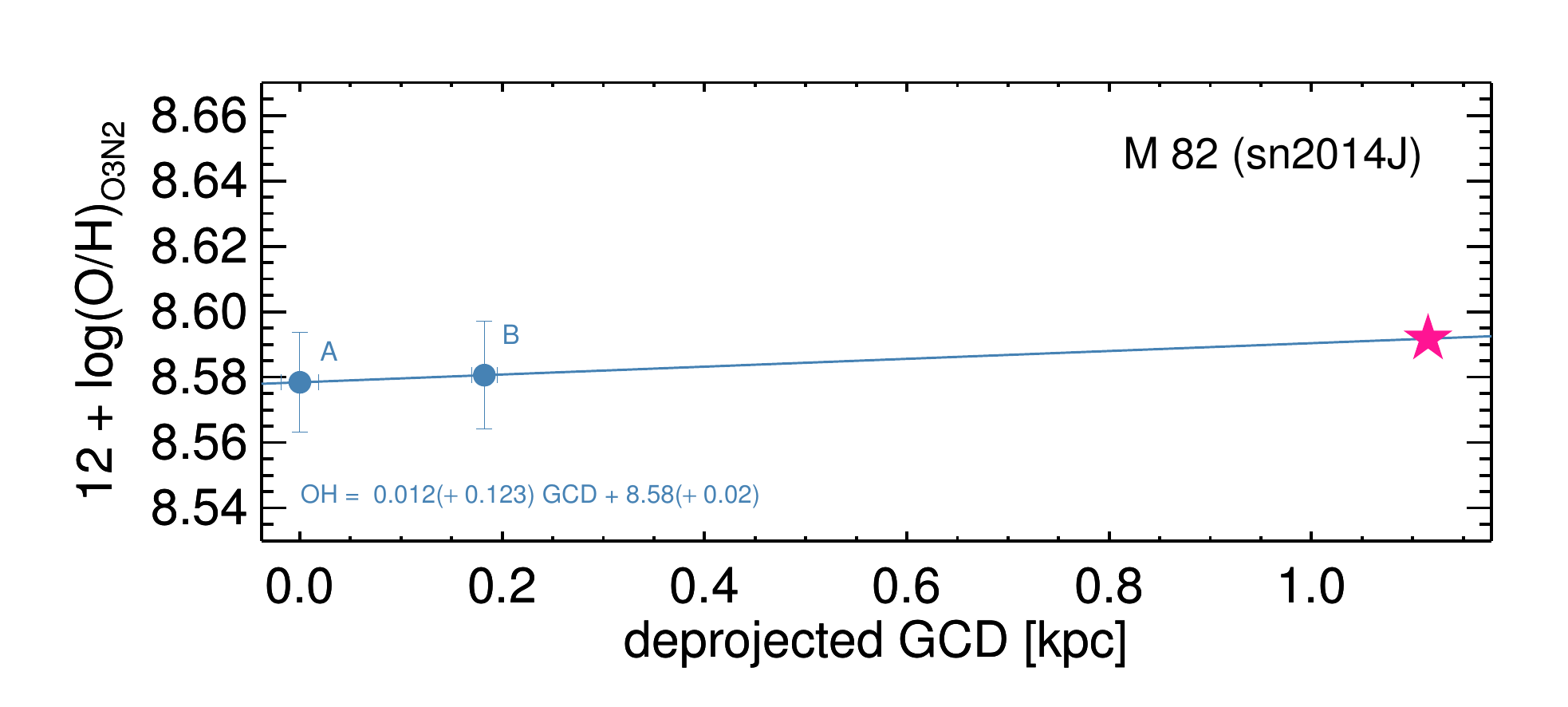}
\includegraphics[width=0.33\textwidth]{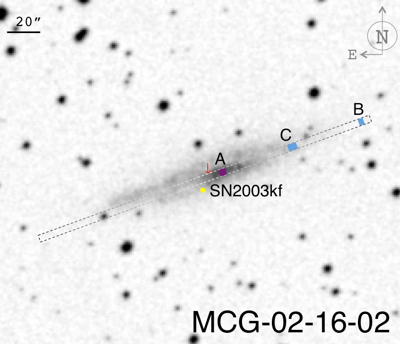}
\includegraphics[width=0.66\textwidth]{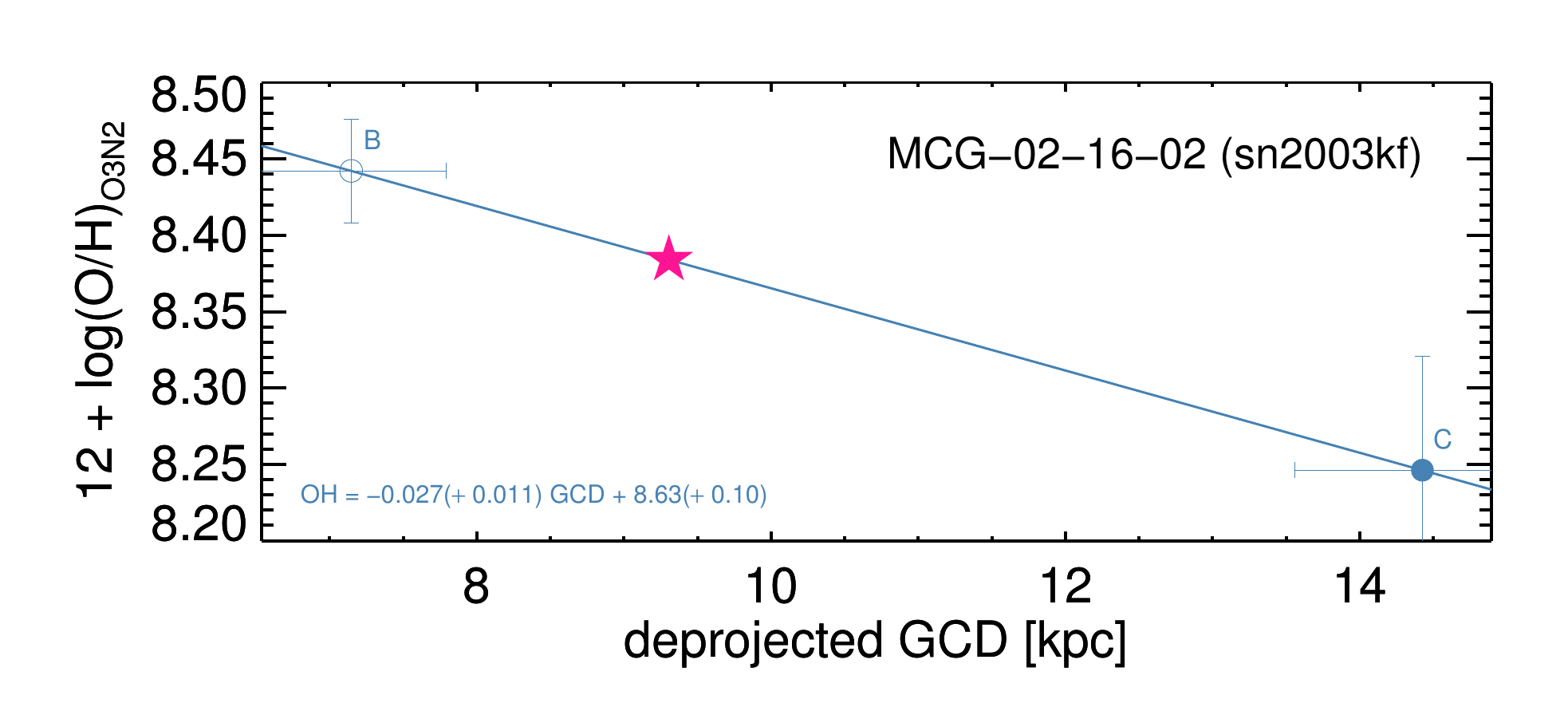}
\includegraphics[width=0.33\textwidth]{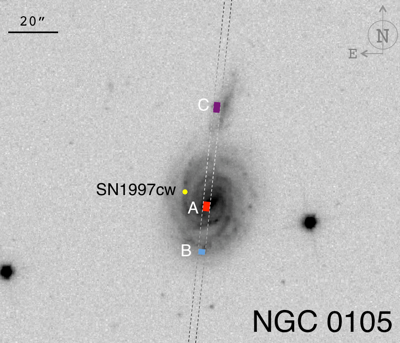}
\includegraphics[width=0.66\textwidth]{gradNGC0105.pdf}
\includegraphics[width=0.33\textwidth]{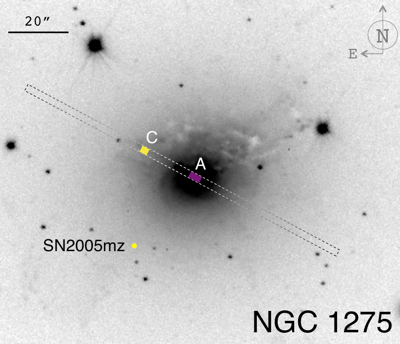}
\includegraphics[width=0.66\textwidth]{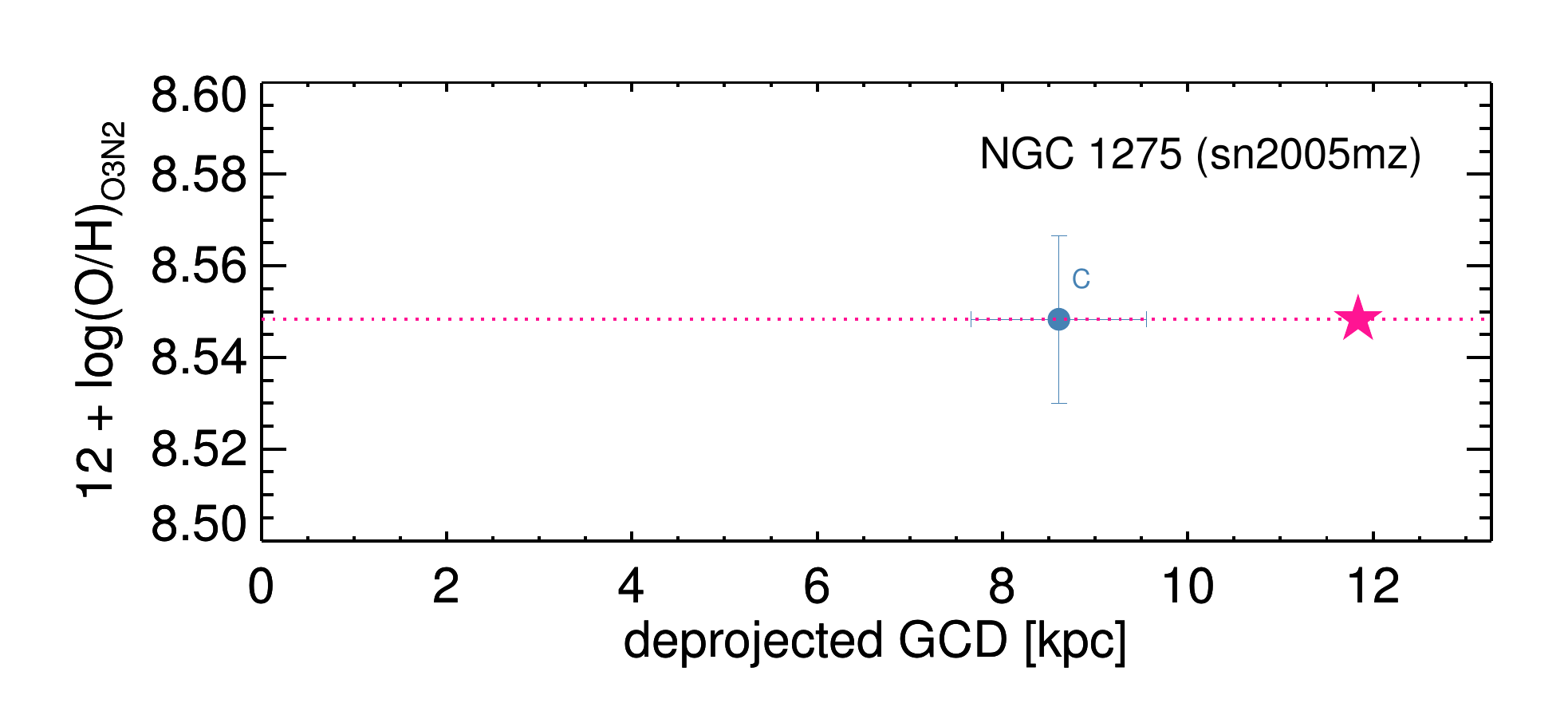}
\caption{Set of galaxies with extracted regions and derived gradients (or closest regions).}
\label{fig:gradientes_todos1}
\end{figure*}

\begin{figure*} 
\centering
\includegraphics[width=0.33\textwidth]{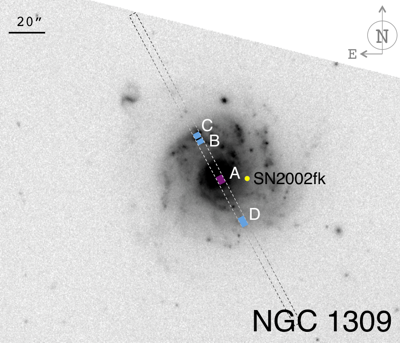}
\includegraphics[width=0.66\textwidth]{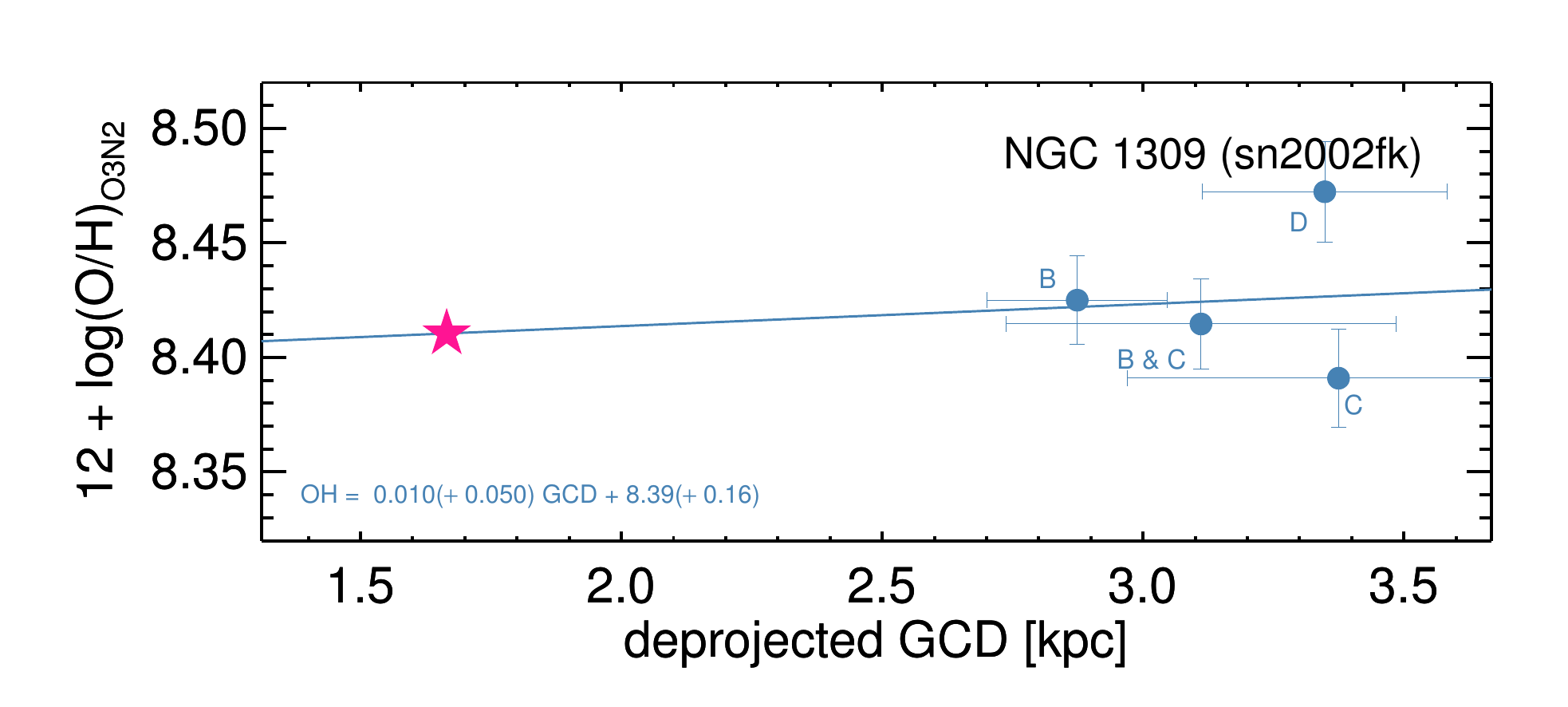}
\includegraphics[width=0.33\textwidth]{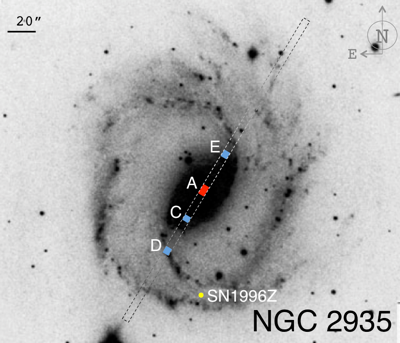}
\includegraphics[width=0.66\textwidth]{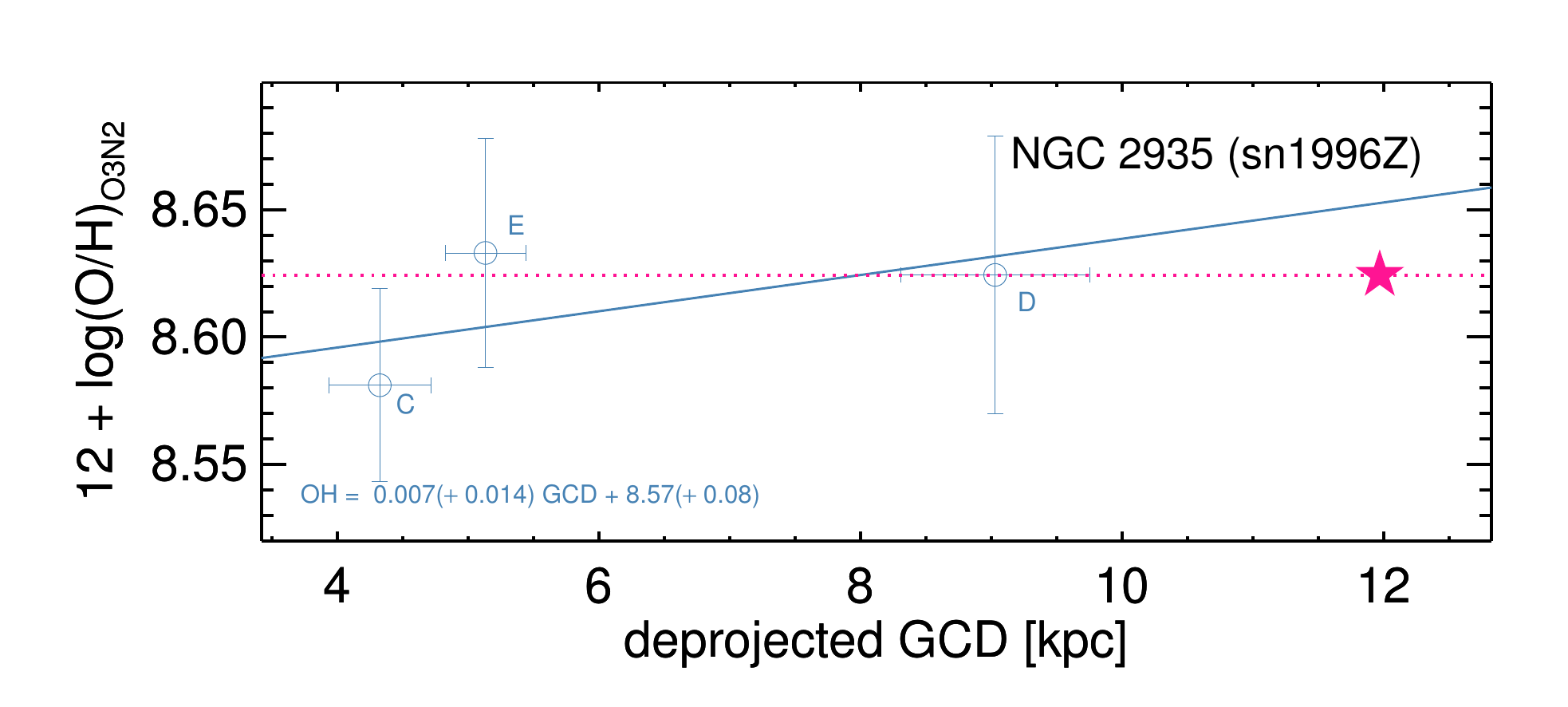}
\includegraphics[width=0.33\textwidth]{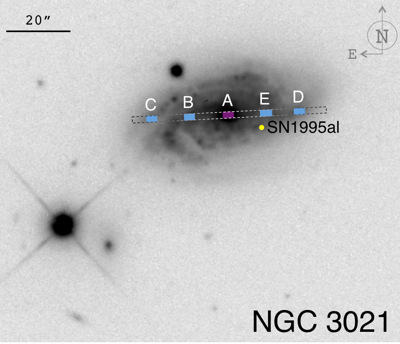}
\includegraphics[width=0.66\textwidth]{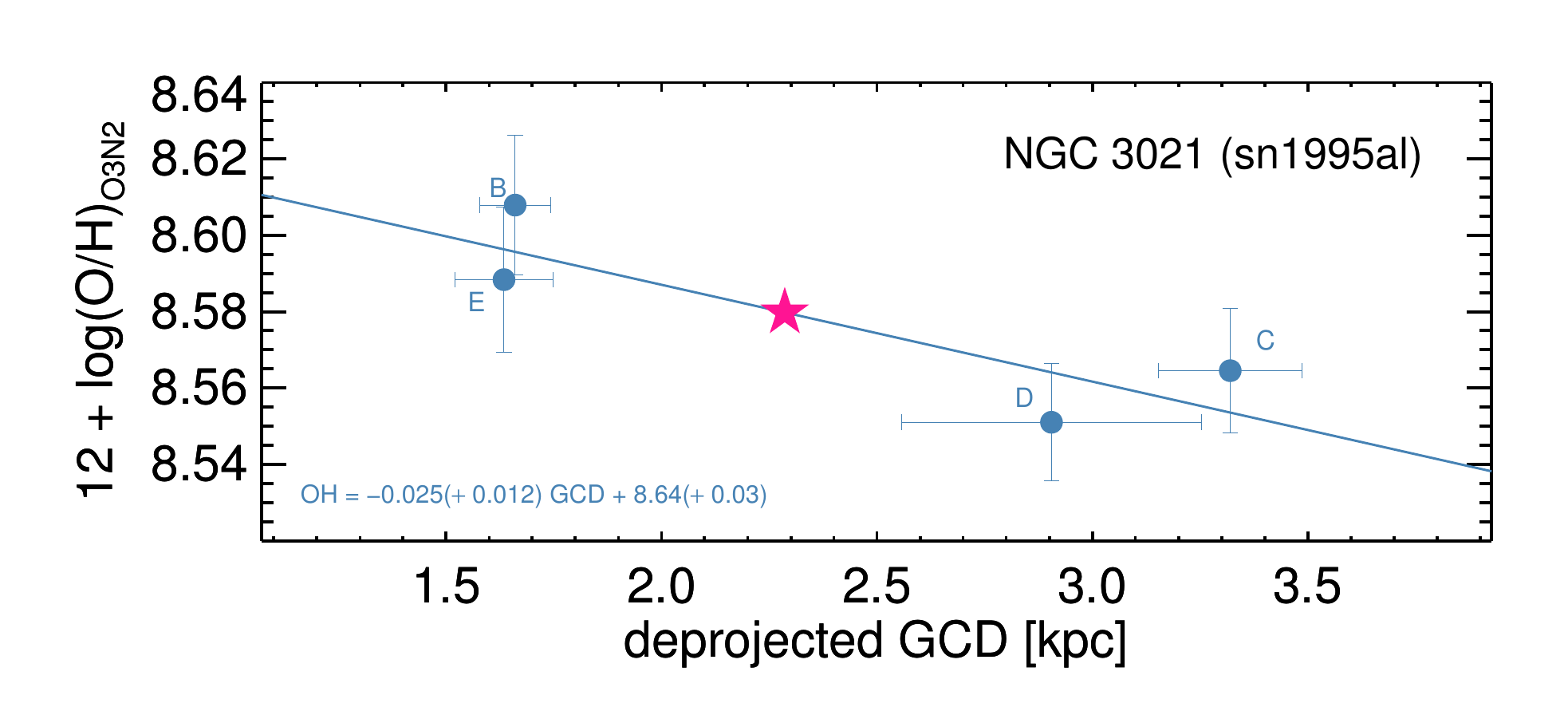}
\includegraphics[width=0.33\textwidth]{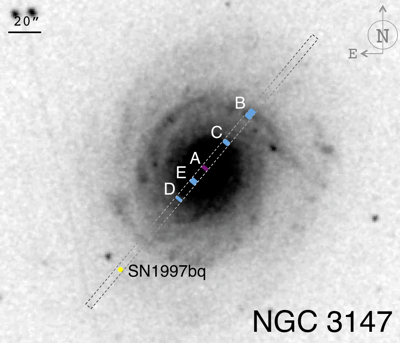}
\includegraphics[width=0.66\textwidth]{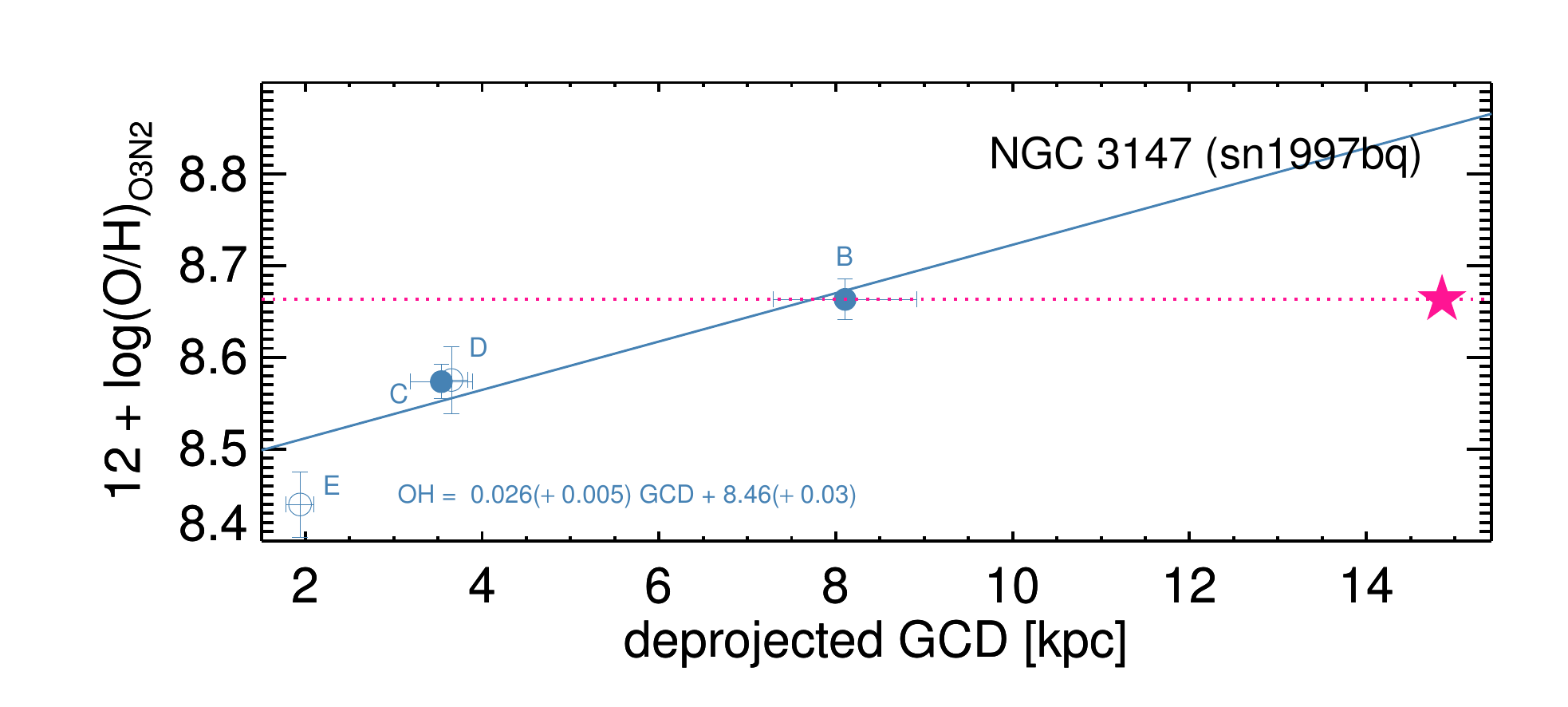}
\caption{Set of galaxies with extracted regions and derived gradients (or closest regions).}
\label{fig:gradientes_todos2}
\end{figure*}

\begin{figure*} 
\centering
\includegraphics[width=0.33\textwidth]{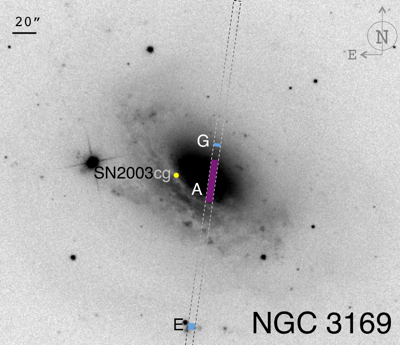}
\includegraphics[width=0.66\textwidth]{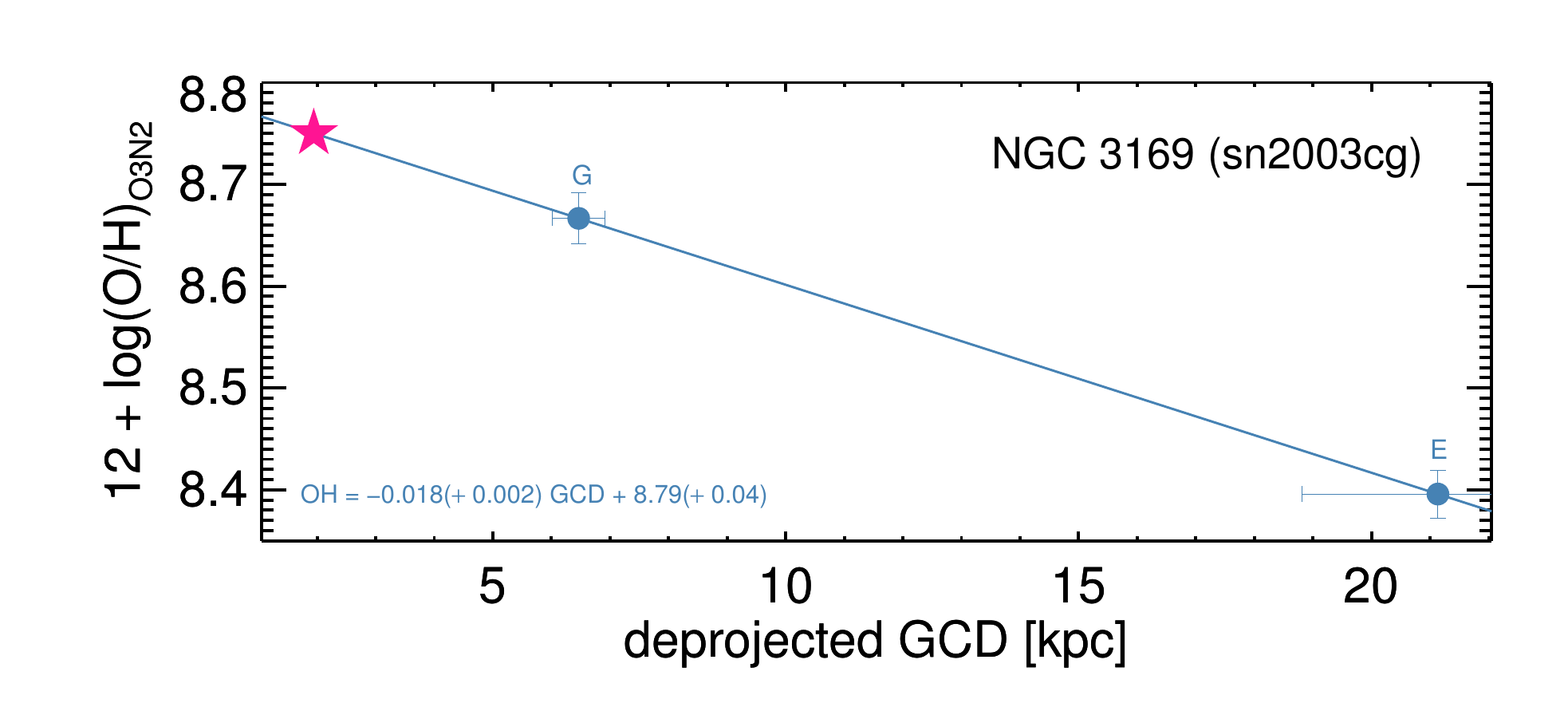}
\includegraphics[width=0.33\textwidth]{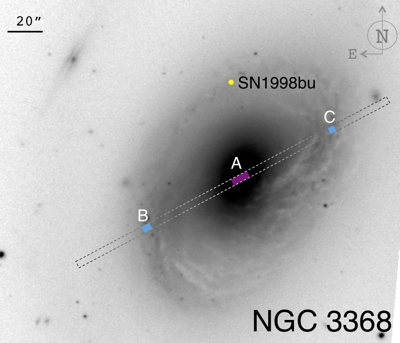}
\includegraphics[width=0.66\textwidth]{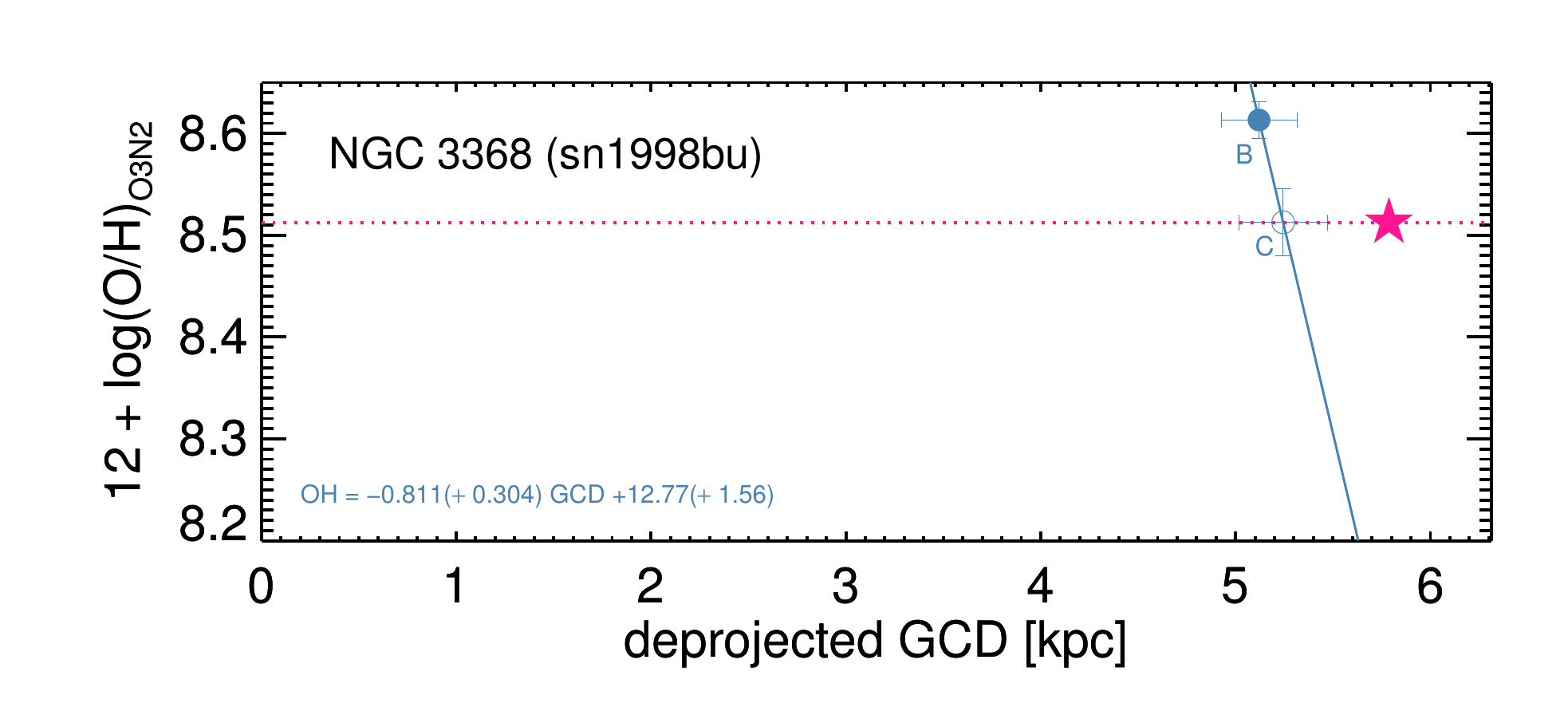}
\includegraphics[width=0.33\textwidth]{ren_NGC3370.png}
\includegraphics[width=0.66\textwidth]{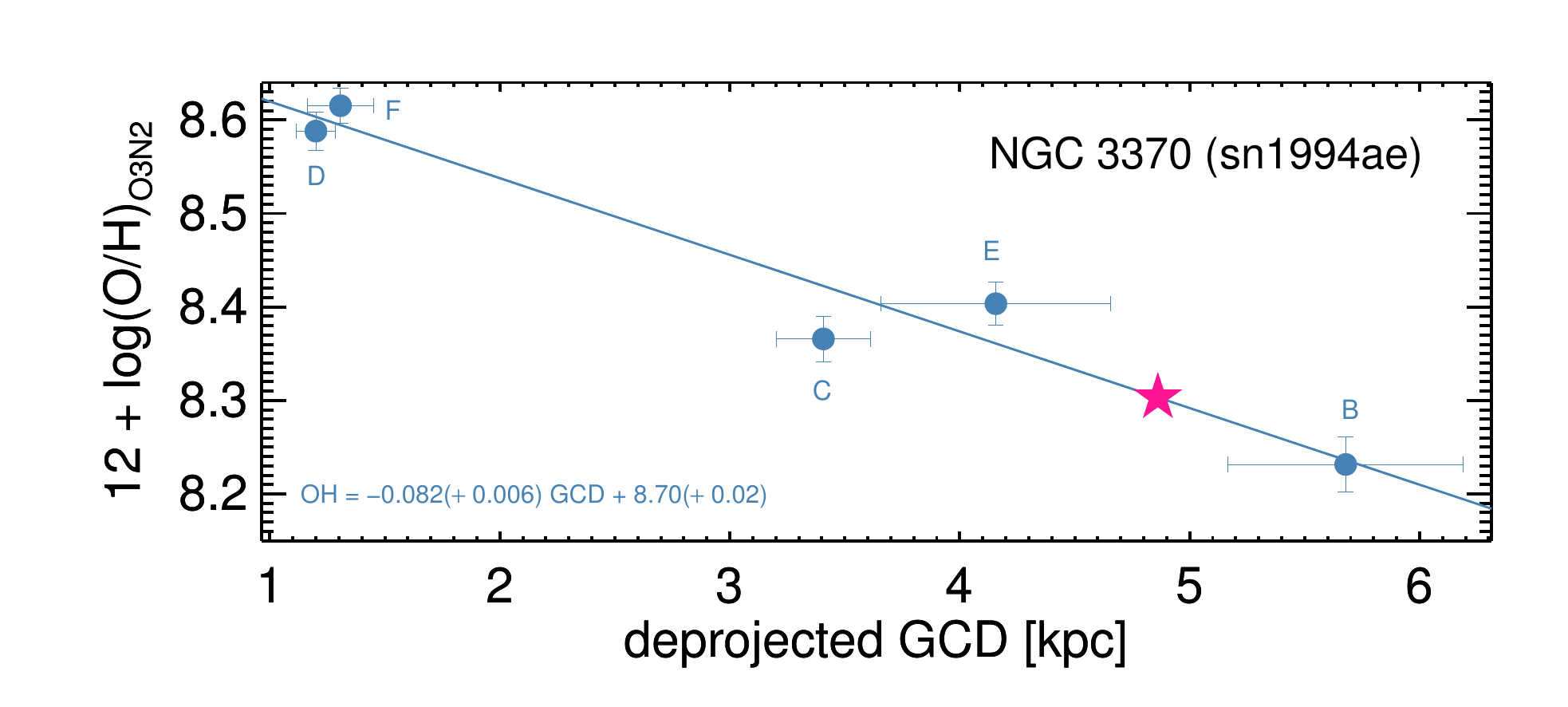}
\includegraphics[width=0.33\textwidth]{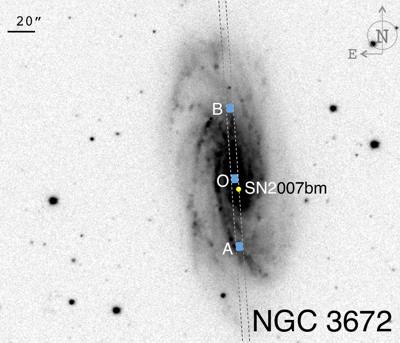}
\includegraphics[width=0.66\textwidth]{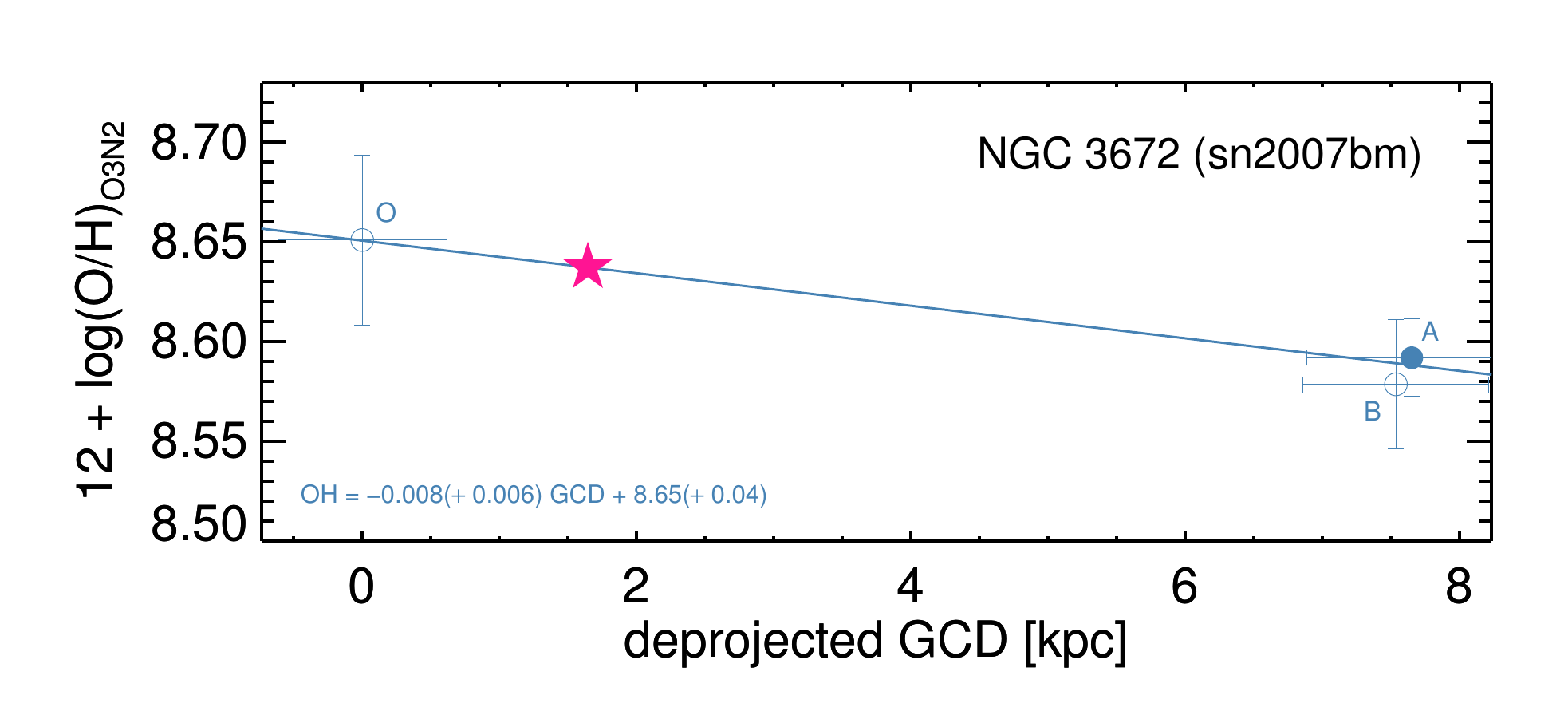}
\caption{Set of galaxies with extracted regions and derived gradients (or closest regions).}
\label{fig:gradientes_todos3}
\end{figure*}

\begin{figure*} 
\centering
\includegraphics[width=0.33\textwidth]{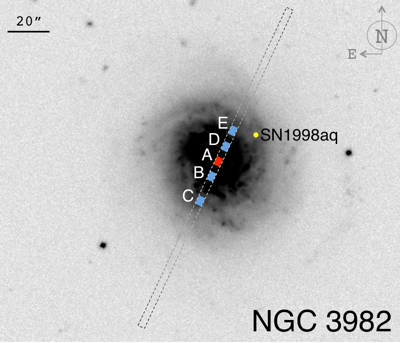}
\includegraphics[width=0.66\textwidth]{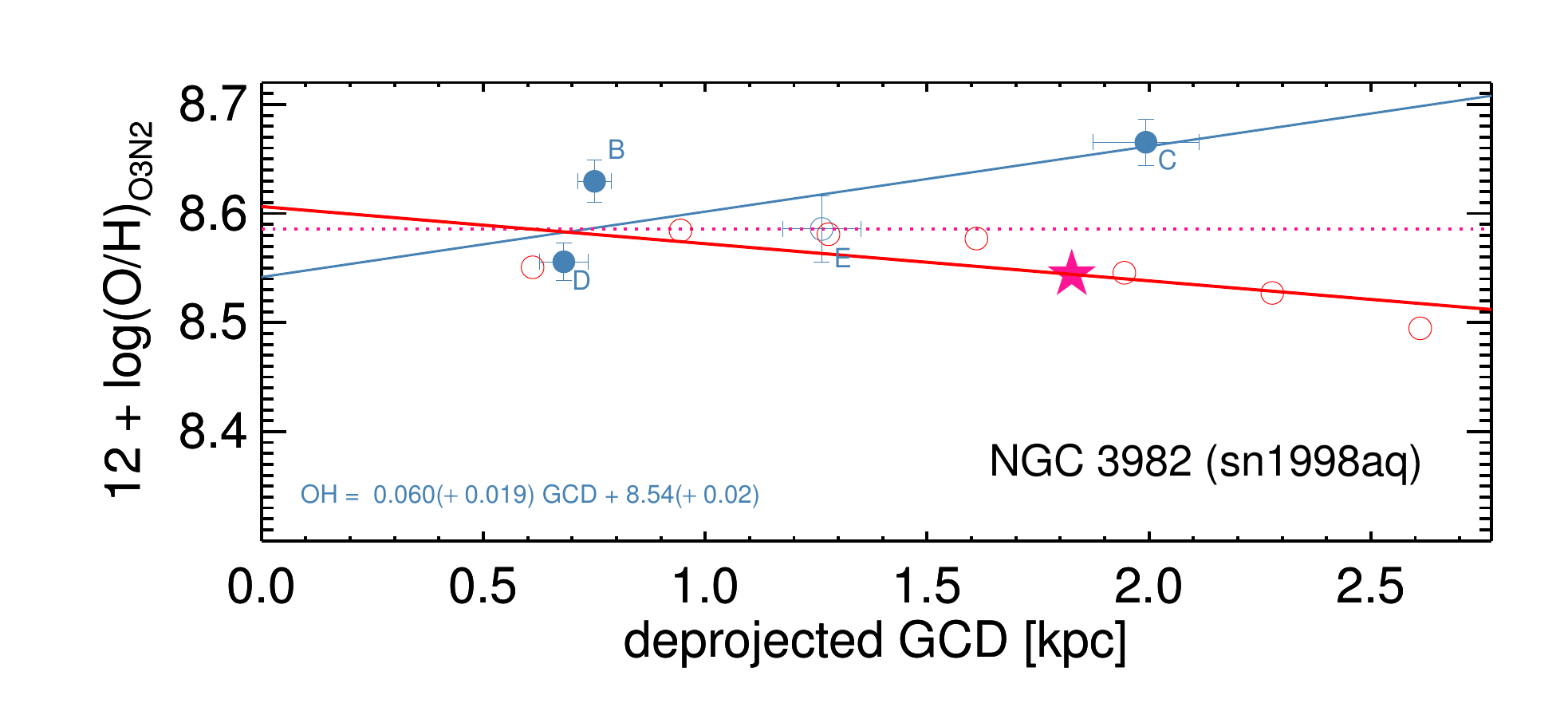}
\includegraphics[width=0.33\textwidth]{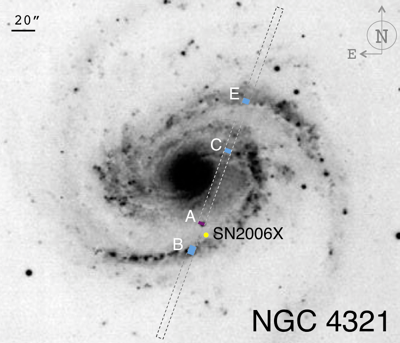}
\includegraphics[width=0.66\textwidth]{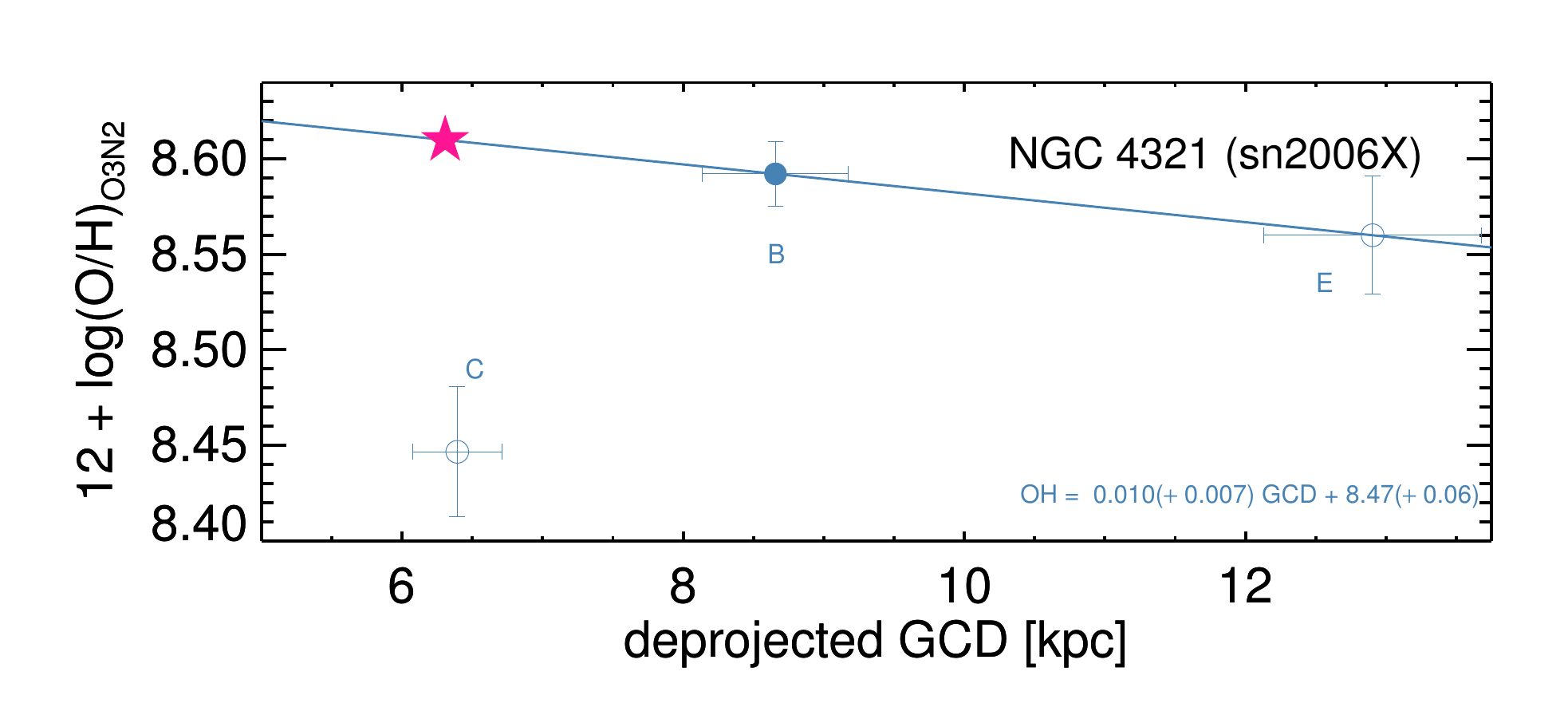}
\includegraphics[width=0.33\textwidth]{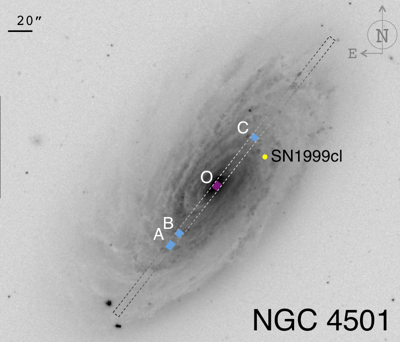}
\includegraphics[width=0.66\textwidth]{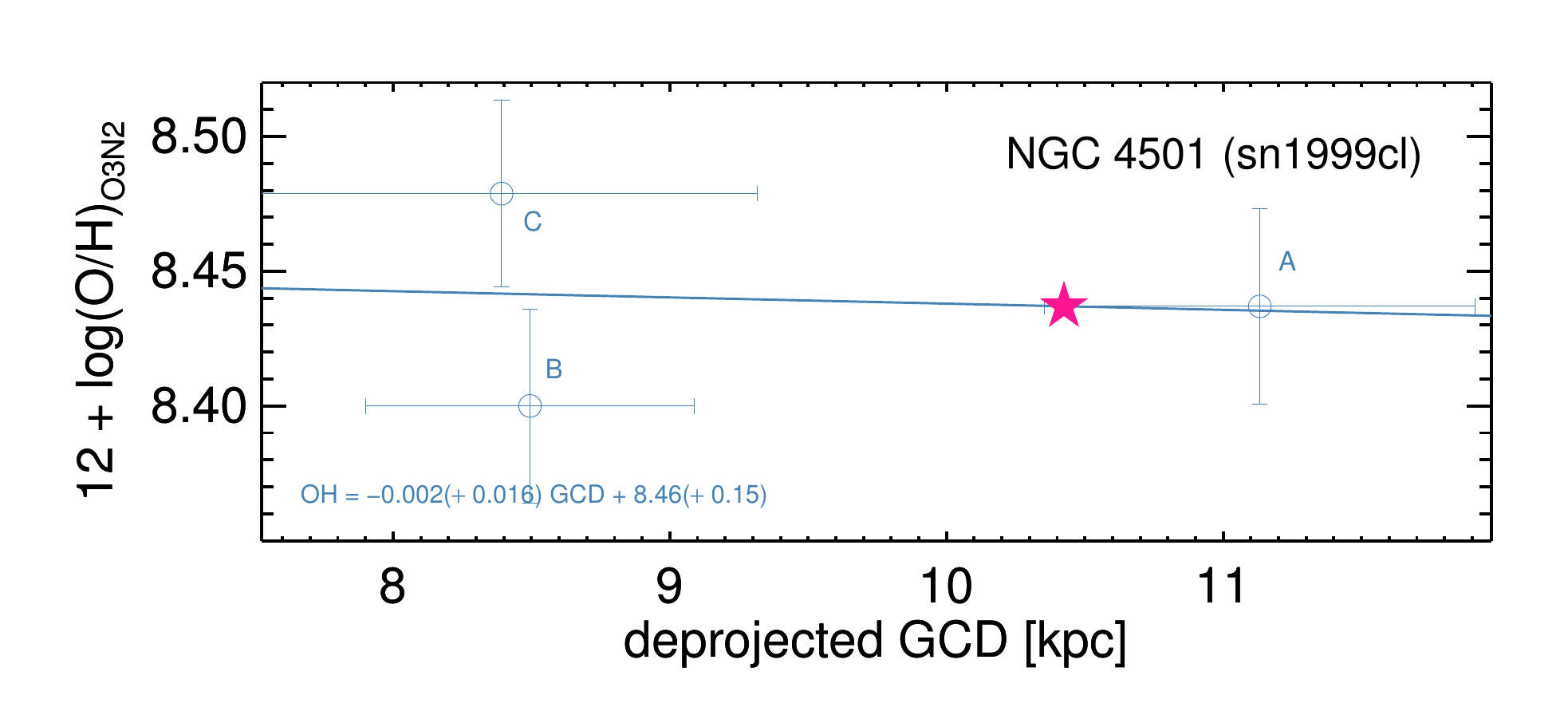}
\includegraphics[width=0.33\textwidth]{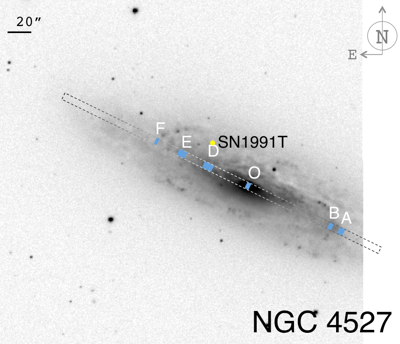}
\includegraphics[width=0.66\textwidth]{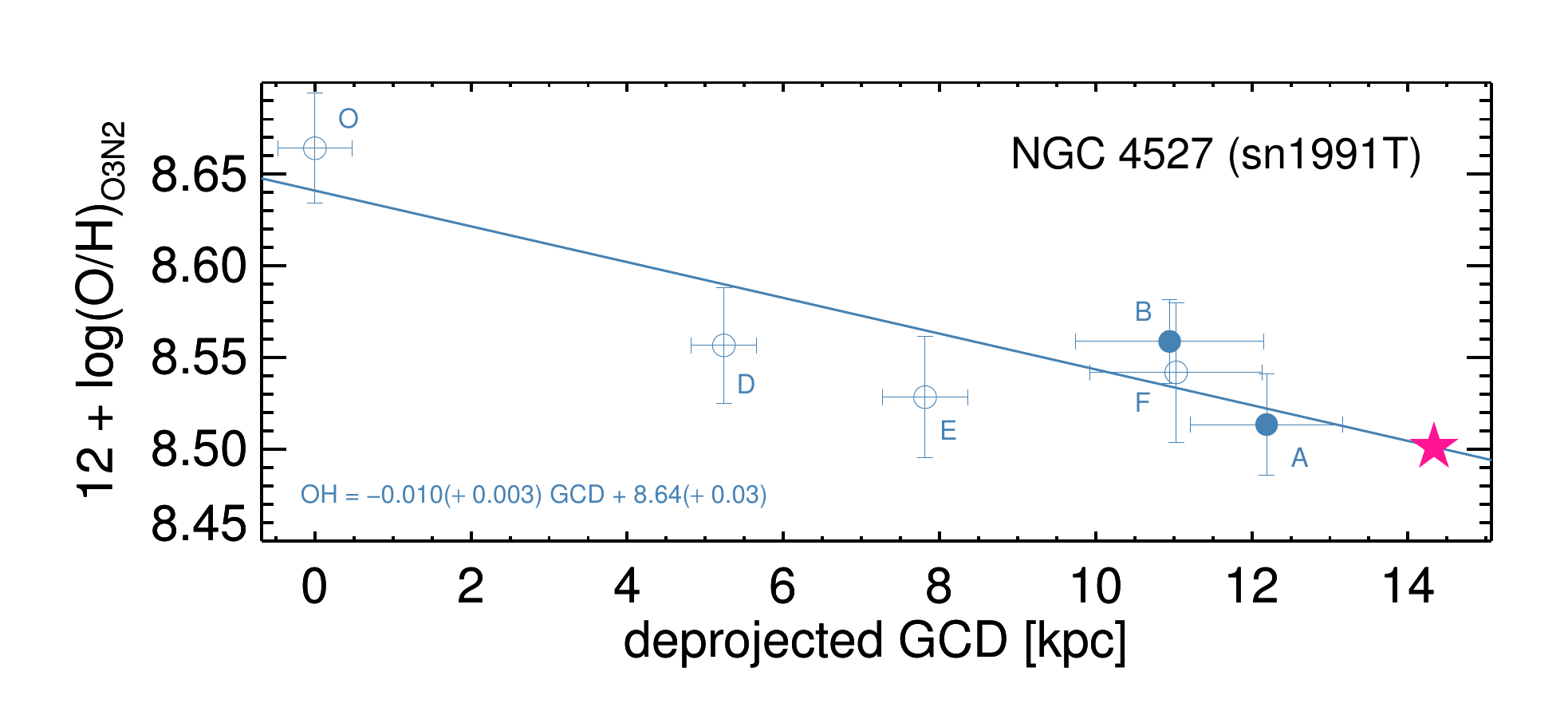}
\caption{Set of galaxies with extracted regions and derived gradients (or closest regions).}
\label{fig:gradientes_todos4}
\end{figure*}

Here we describe as well the situation in the 7 conflicting cases where a metallicity gradient could not be obtained:
\begin{description}

\item{\bf NGC\,0105}: Only one region is measured at a distance of 2\,kpc from the position of the SN~Ia. Since there is a proper metallicity gradient given by \citet{2016arXiv160307808G}, we adopt that value for the oxygen abundance in the environment of SN1997cw (see third panel in Figure \ref{fig:gradientes_todos1}). \\

\item{\bf NGC\,1275}: Only one region is measured at a distance of \mbox{$\sim$3.5\,kpc} from the position of the SN~Ia (see bottom panel in Figure \ref{fig:gradientes_todos1}). \\

\item{\bf UGC\,04195}: Only one region is measured at a distance of $\sim$ 8\,kpc from the SN~Ia position. In this case there is a proper metallicity gradient avalilable in \citet{2016arXiv160307808G}, so we adopt that value for the oxygen abundance in the environment of SN2000ce (see Figure \ref{fig:gradientes_todos7}, second panel).\\ 

\item{\bf NGC\,2935}: Three \HII\ regions are measured in this object, but the gradient is not reliable. All points here have been recovered from $N2$ parameter.  The closest \HII~region to the SN~Ia is at a distance of $\sim$ 3\,kpc (see second panel on Figure \ref{fig:gradientes_todos2}).\\ 

\item{\bf NGC\,3147}: Data for four \HII\ regions are available, but the derived metallicity gradient is very steep and inverted. We do not consider this metallicity gradient to be representative, so we adopt the metallicity of the SN~Ia as that derived in the closest \HII\ region, which is located at a distance of 7\,kpc. Therefore the assigned value actually is an upper limit to the real metallicity of the SN~Ia, as the later one is located at a larger galactocentric radius than the measured \HII\ region (Figure \ref{fig:gradientes_todos2}, bottom panel).\\

\item {\bf NGC\,3368}: Data for two \HII\ regions providing very simular metallicities are available, but they are located at both sides of the galaxy centre and at almost the same GCD, \mbox{$\sim$1\,kpc}, and hence the gradient is not representative. Furthermore, the SN~Ia lies very close to one of these \HII\ regions, so it is easily acceptable that both share the same metallicity  (Figure \ref{fig:gradientes_todos3}, third panel).\\ 

\item{\bf NGC\,3982}: Data of four \HII\ regions provide a very steep inverted metallicity gradient, which do not consider to be real. In this case a proper metallicity gradient is obtained in \citet{2016arXiv160307808G}, so that we can adopt that value for the oxygen abundance in the environment of SN1998aq (first panel in Figure \ref{fig:gradientes_todos4}). 

\end{description}

\begin{figure*} 
\centering
\includegraphics[width=0.33\textwidth]{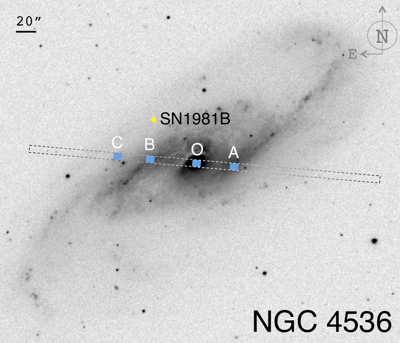}
\includegraphics[width=0.66\textwidth]{gradNGC4536.pdf}
\includegraphics[width=0.33\textwidth]{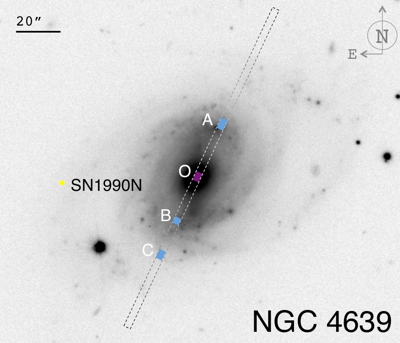}
\includegraphics[width=0.66\textwidth]{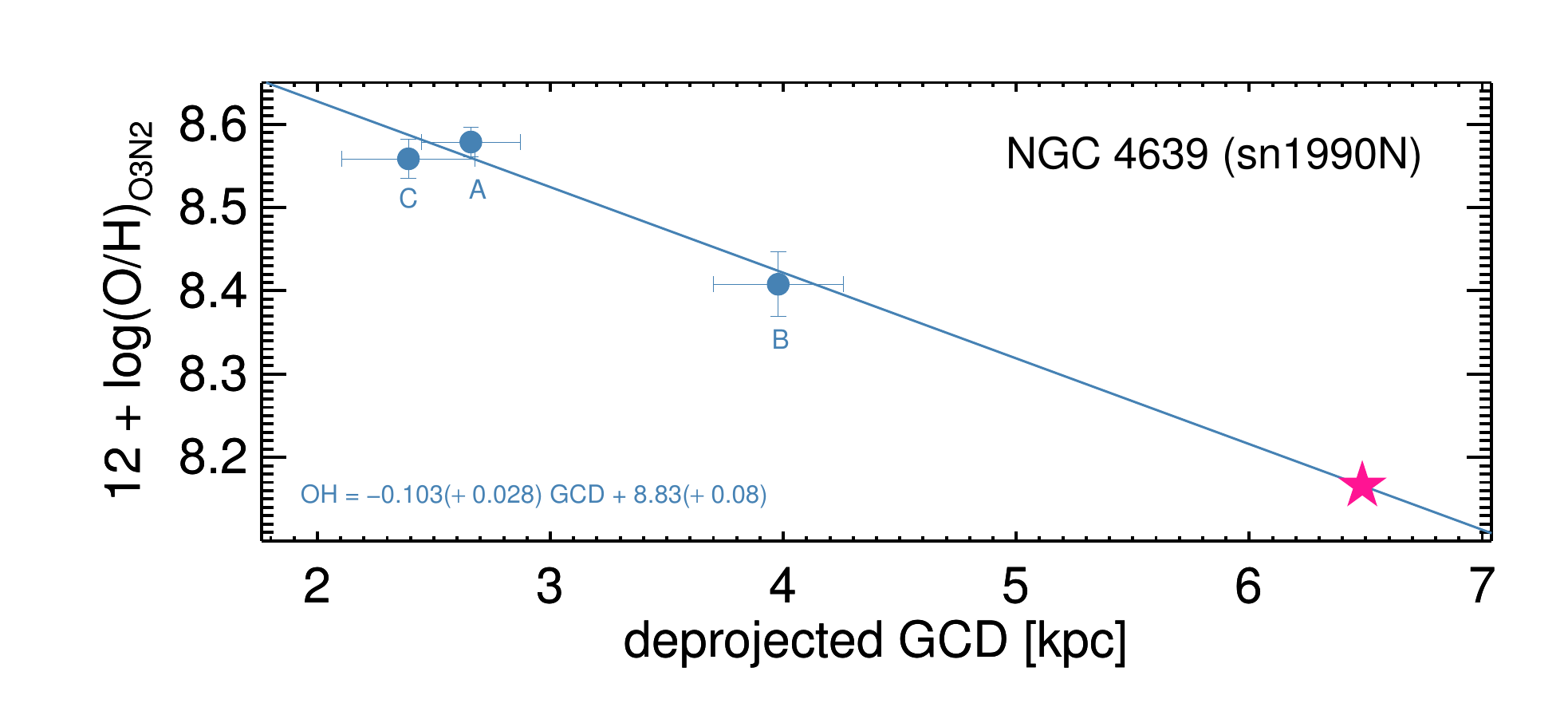}
\includegraphics[width=0.33\textwidth]{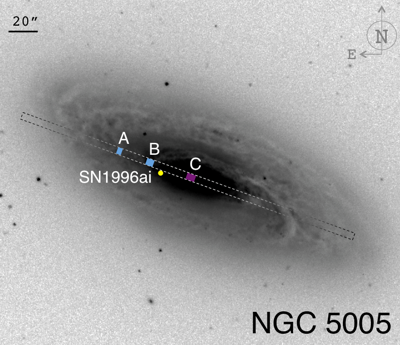}
\includegraphics[width=0.66\textwidth]{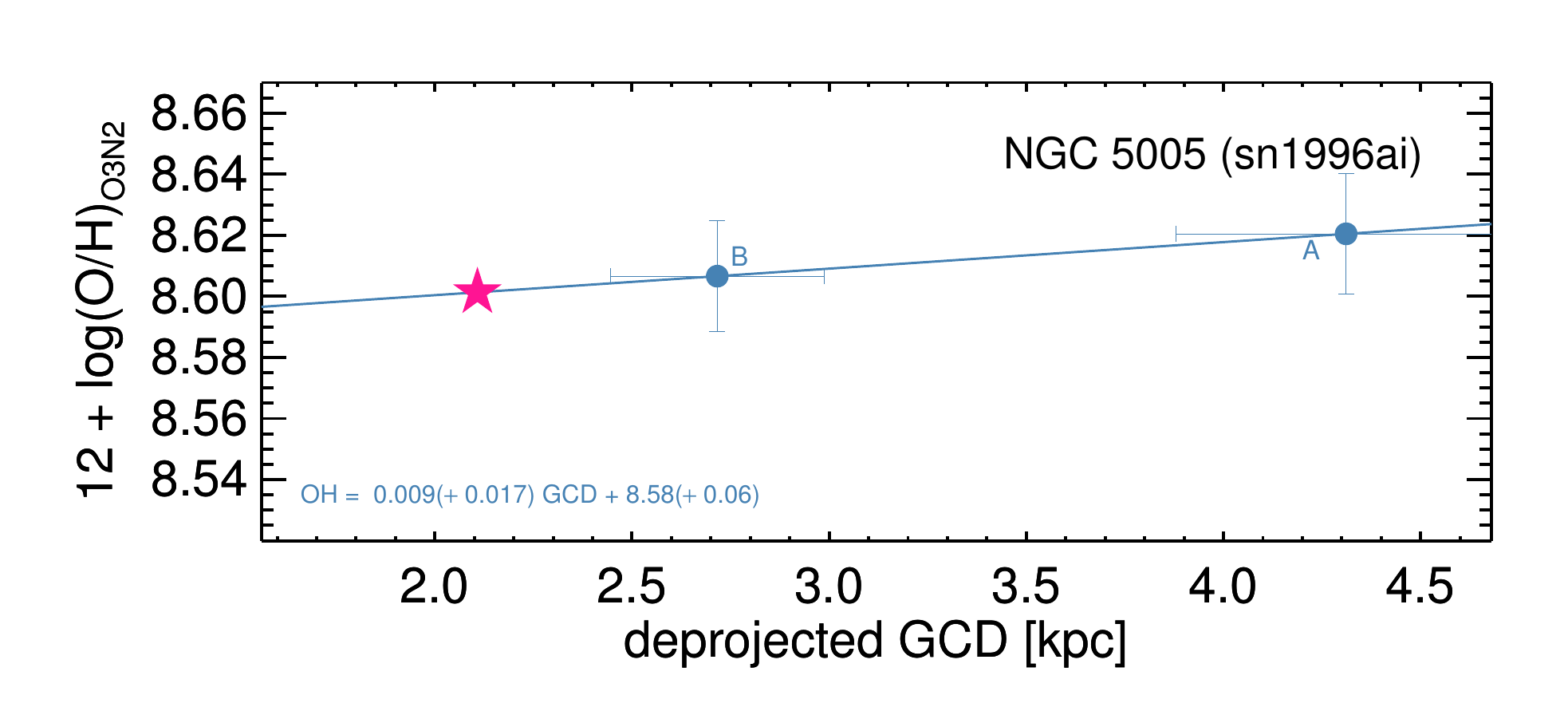}
\includegraphics[width=0.33\textwidth]{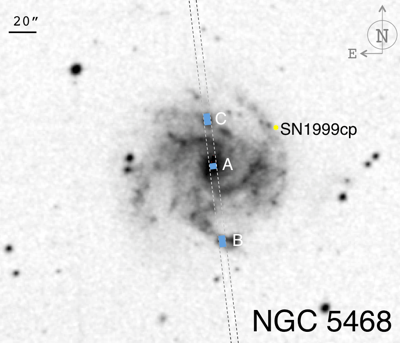}
\includegraphics[width=0.66\textwidth]{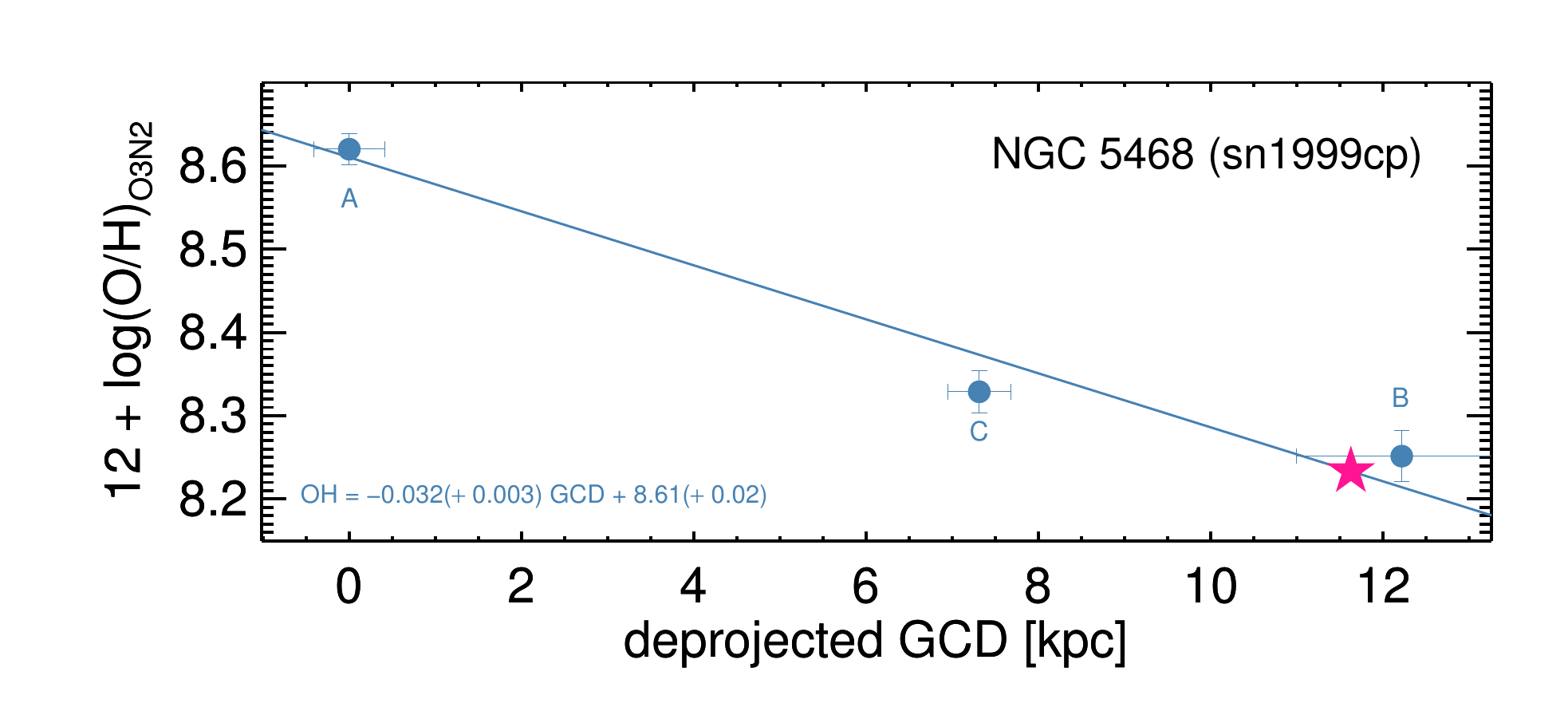}
\caption{Set of galaxies with extracted regions and derived gradients (or closest regions).}
\label{fig:gradientes_todos5}
\end{figure*}

\begin{figure*} 
\centering
\includegraphics[width=0.33\textwidth]{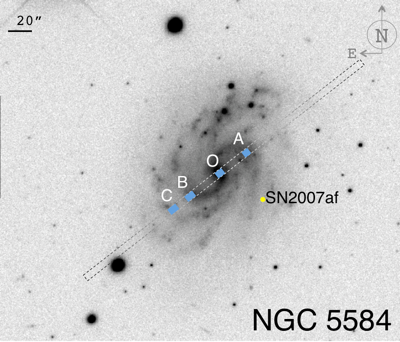}
\includegraphics[width=0.66\textwidth]{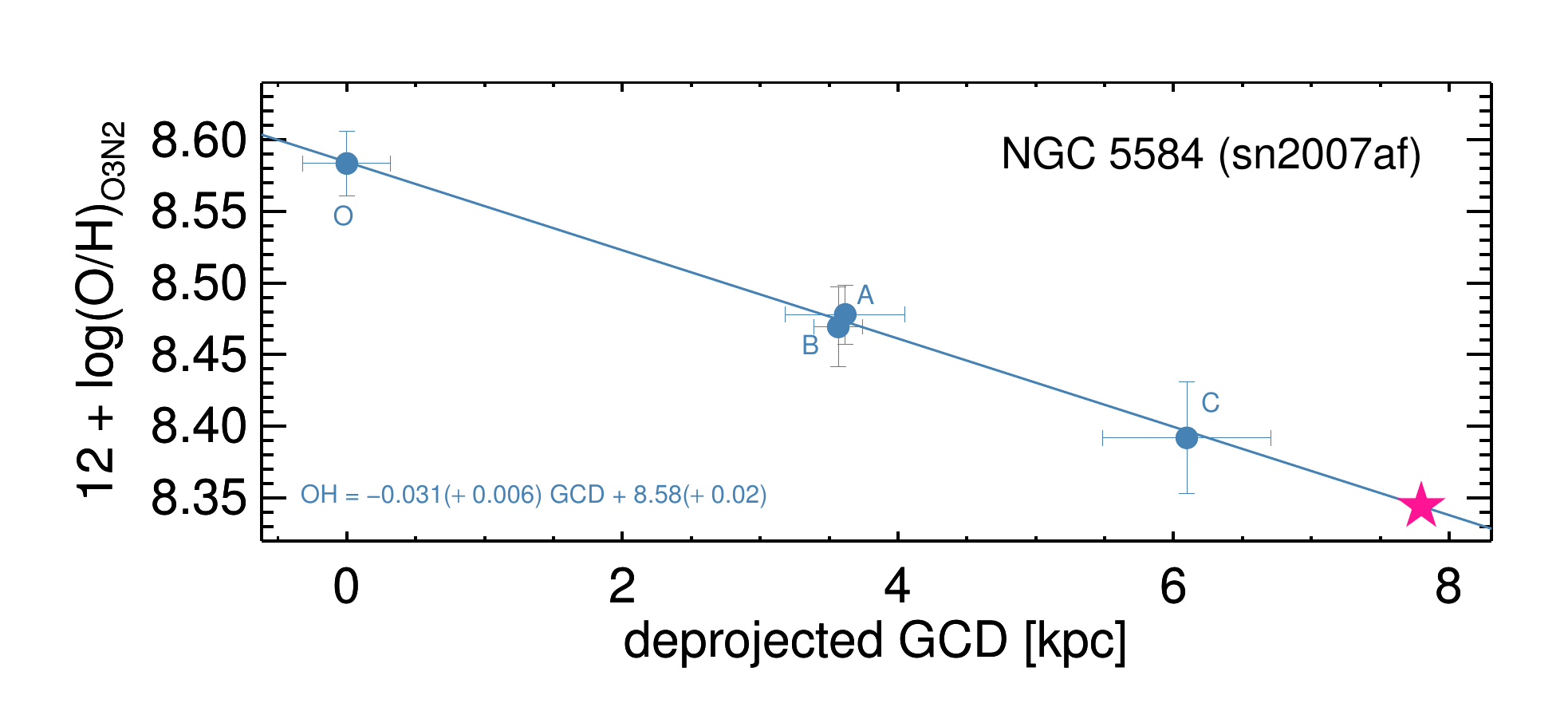}
\includegraphics[width=0.33\textwidth]{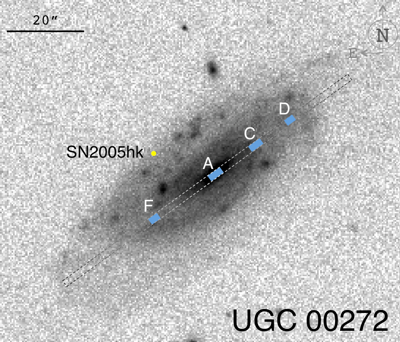}
\includegraphics[width=0.66\textwidth]{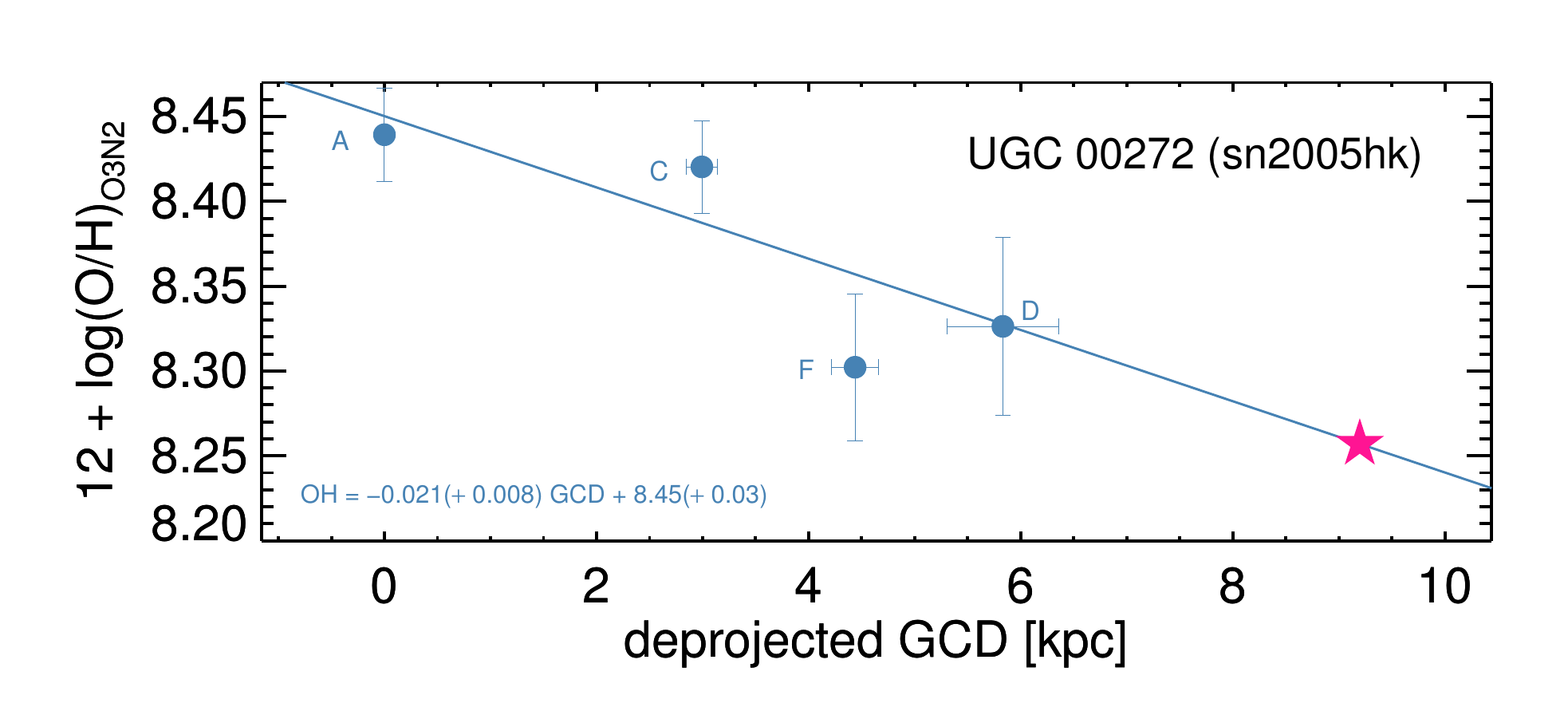}
\includegraphics[width=0.33\textwidth]{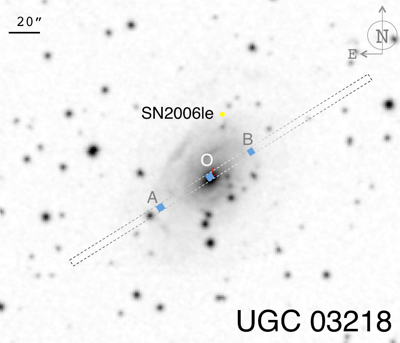}
\includegraphics[width=0.66\textwidth]{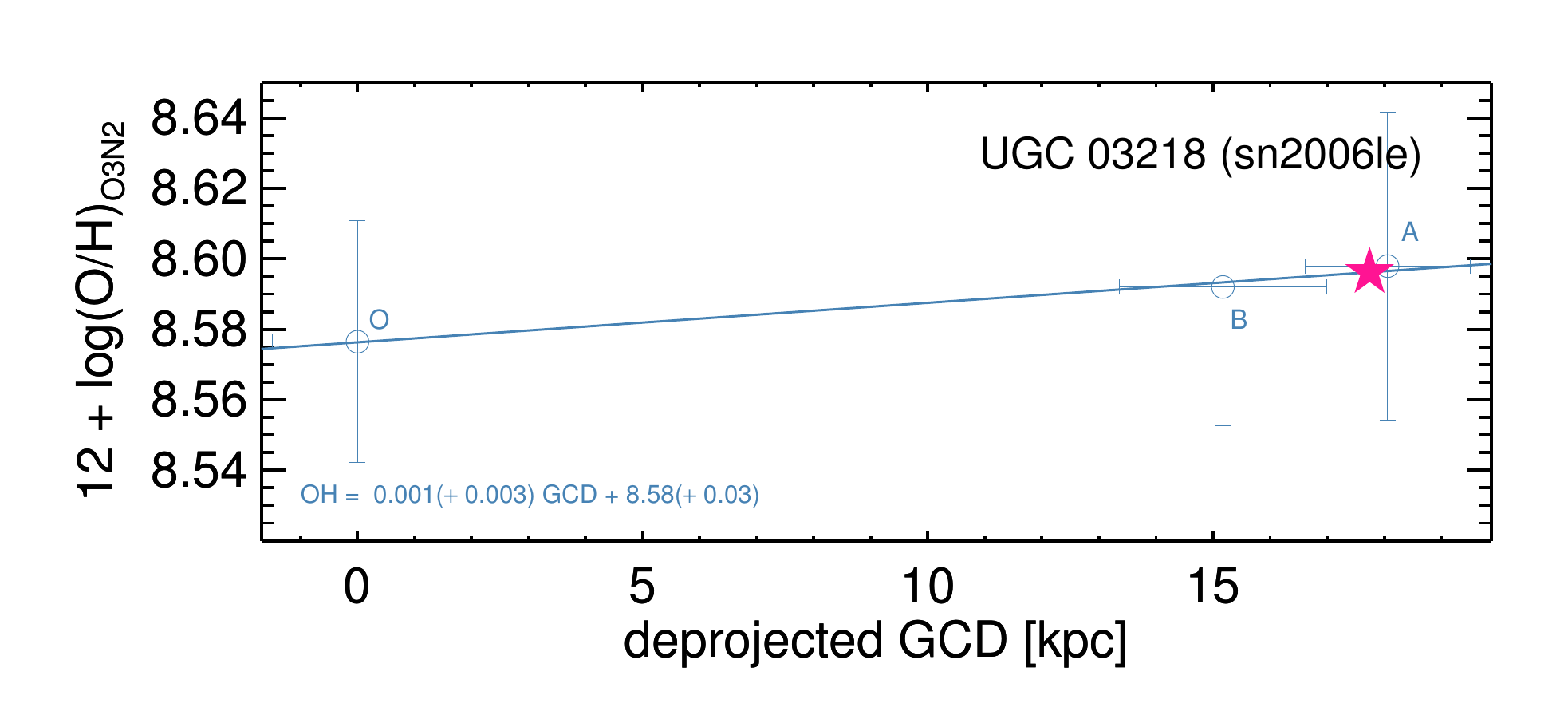}
\includegraphics[width=0.33\textwidth]{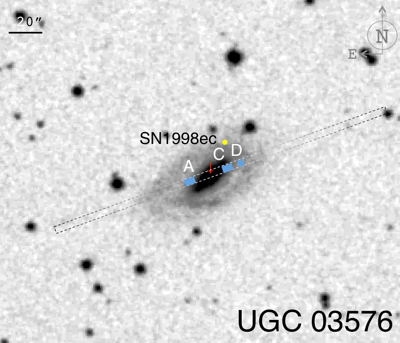}
\includegraphics[width=0.66\textwidth]{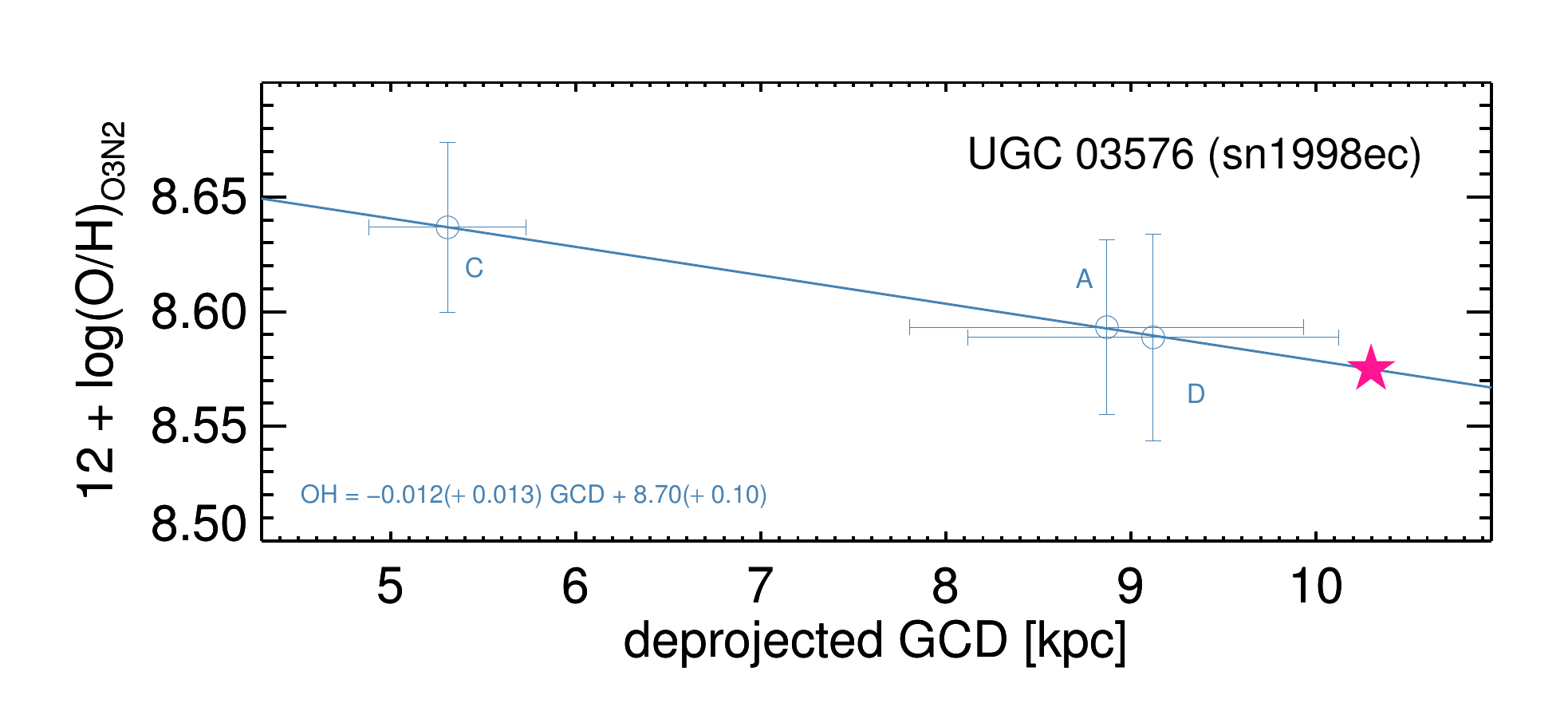}
\caption{Set of galaxies with extracted regions and derived gradients (or closest regions).}
\label{fig:gradientes_todos6}
\end{figure*}

\begin{figure*} 
\centering
\includegraphics[width=0.33\textwidth]{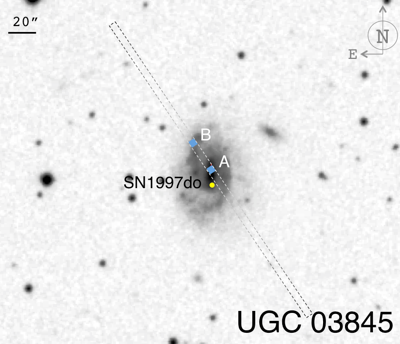}
\includegraphics[width=0.66\textwidth]{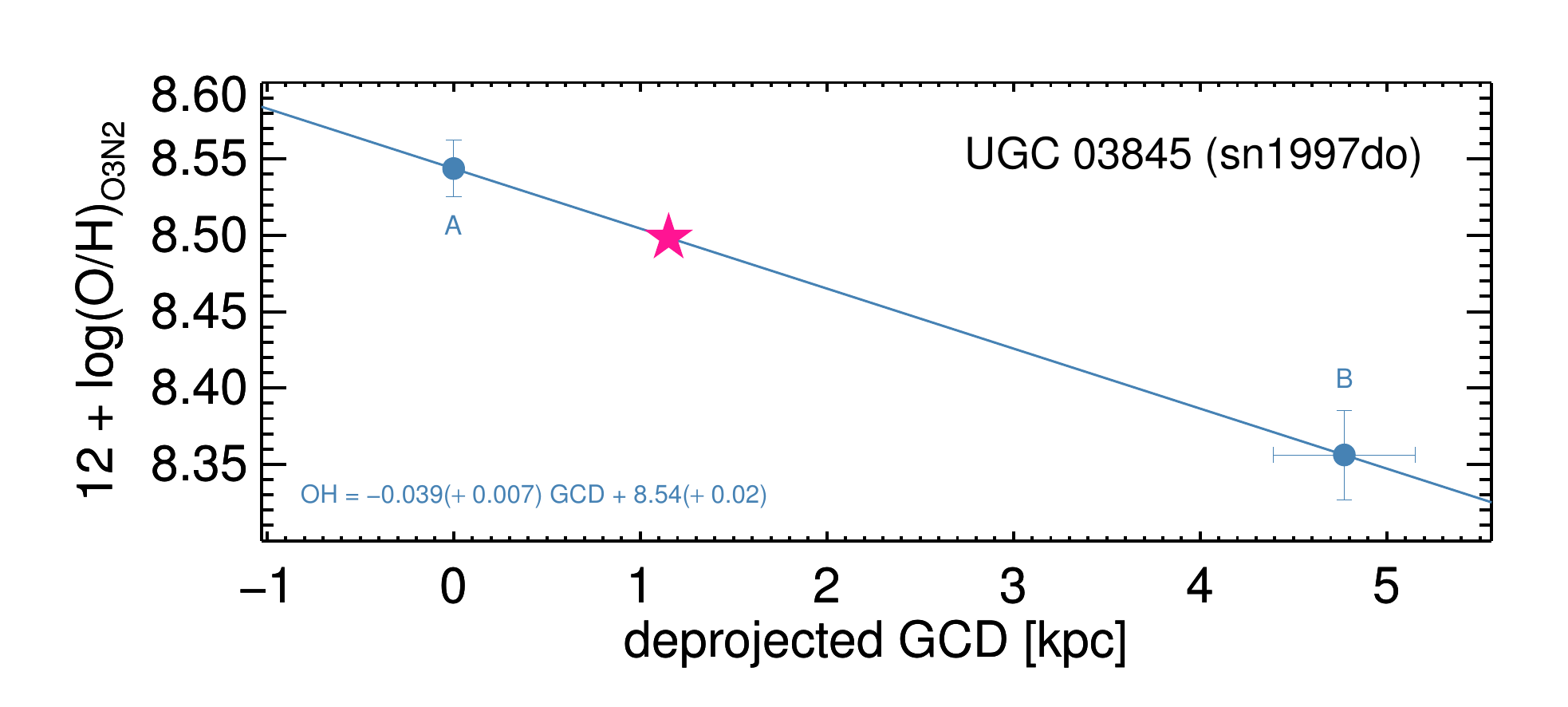}
\includegraphics[width=0.33\textwidth]{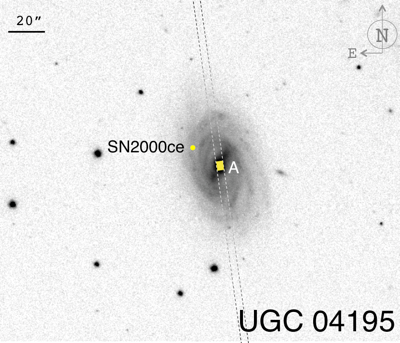}
\includegraphics[width=0.66\textwidth]{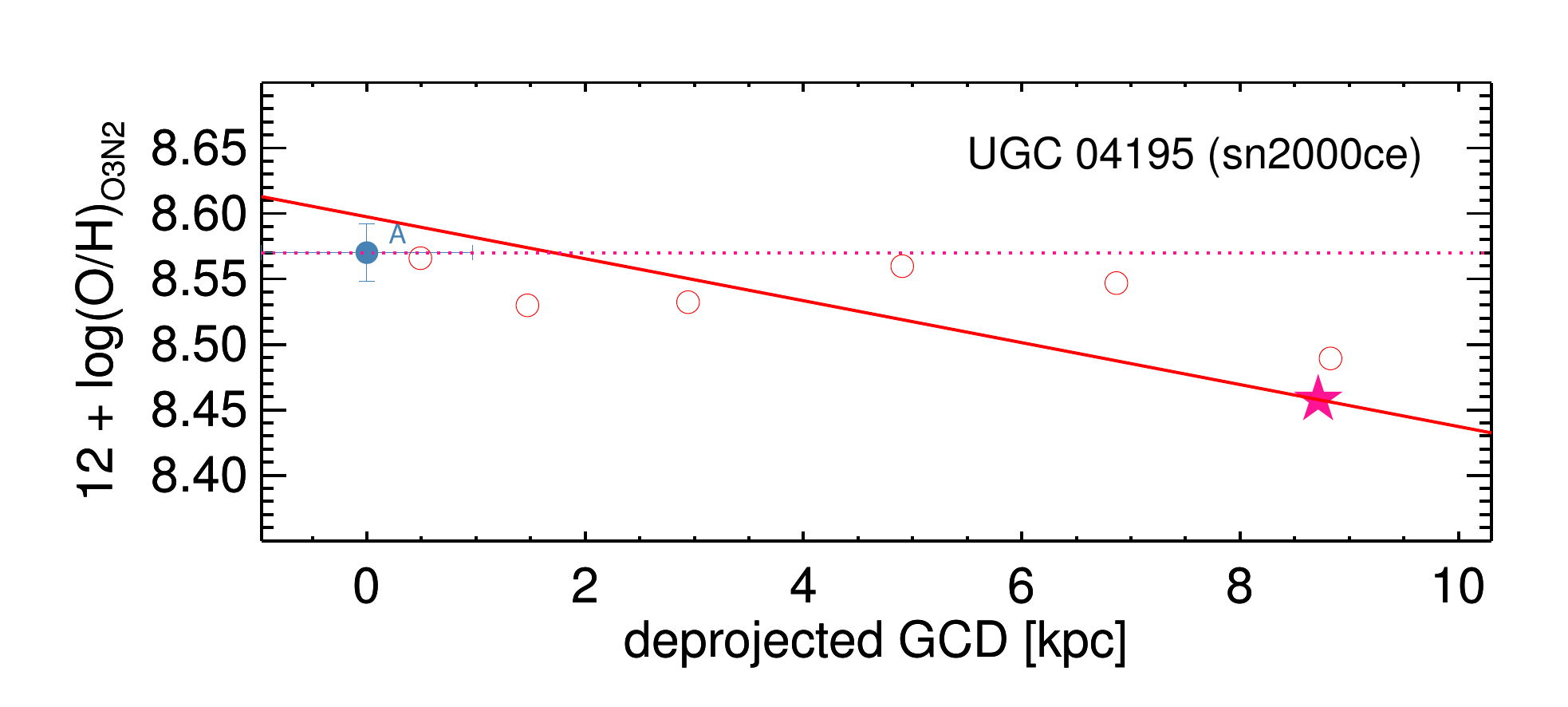}
\includegraphics[width=0.33\textwidth]{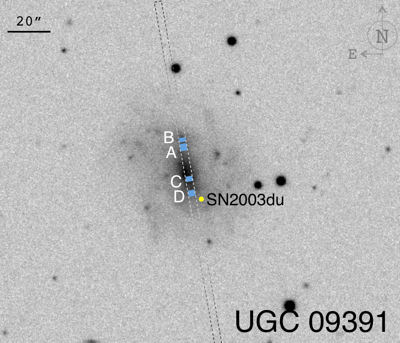}
\includegraphics[width=0.66\textwidth]{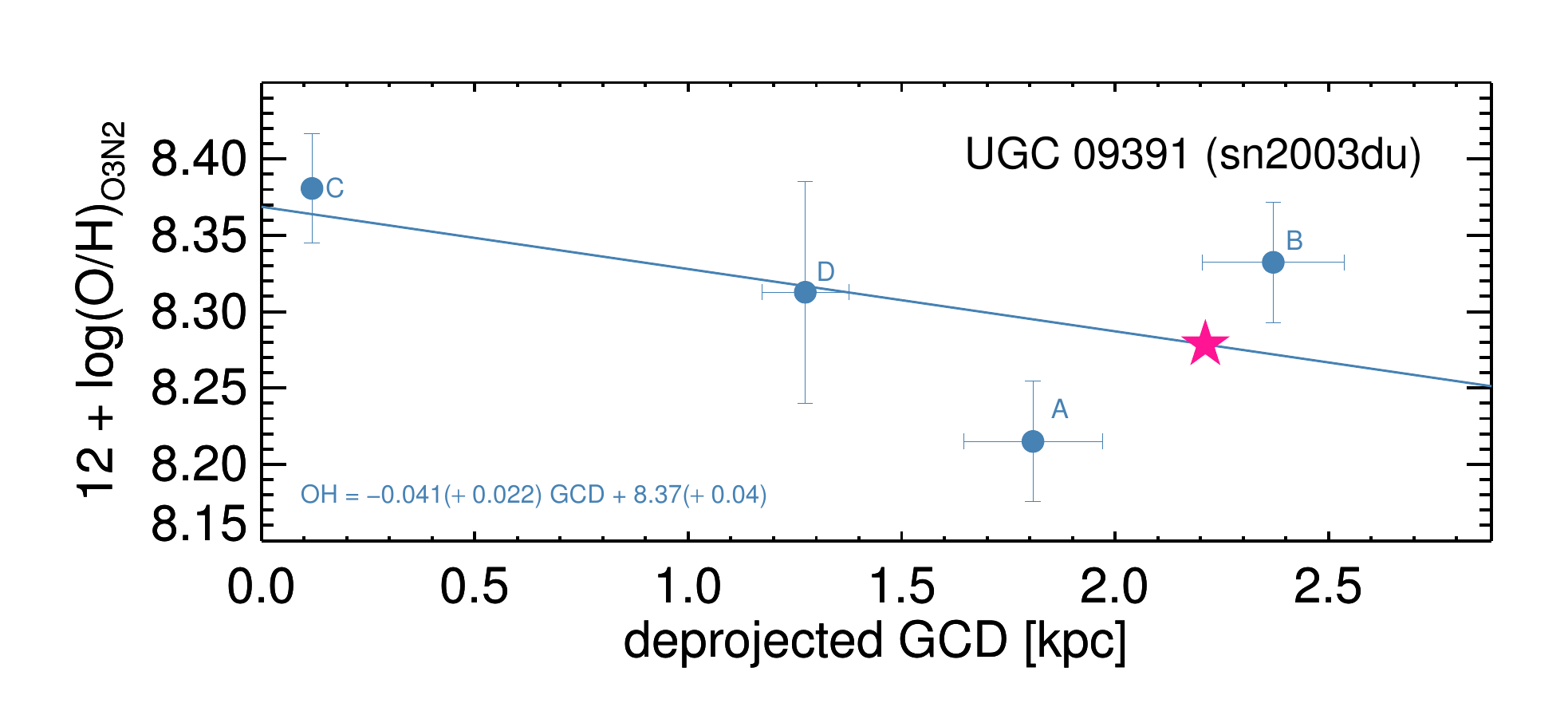}
\includegraphics[width=0.33\textwidth]{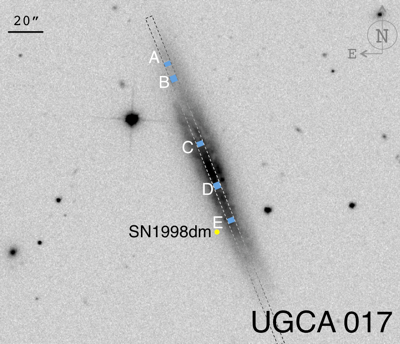}
\includegraphics[width=0.66\textwidth]{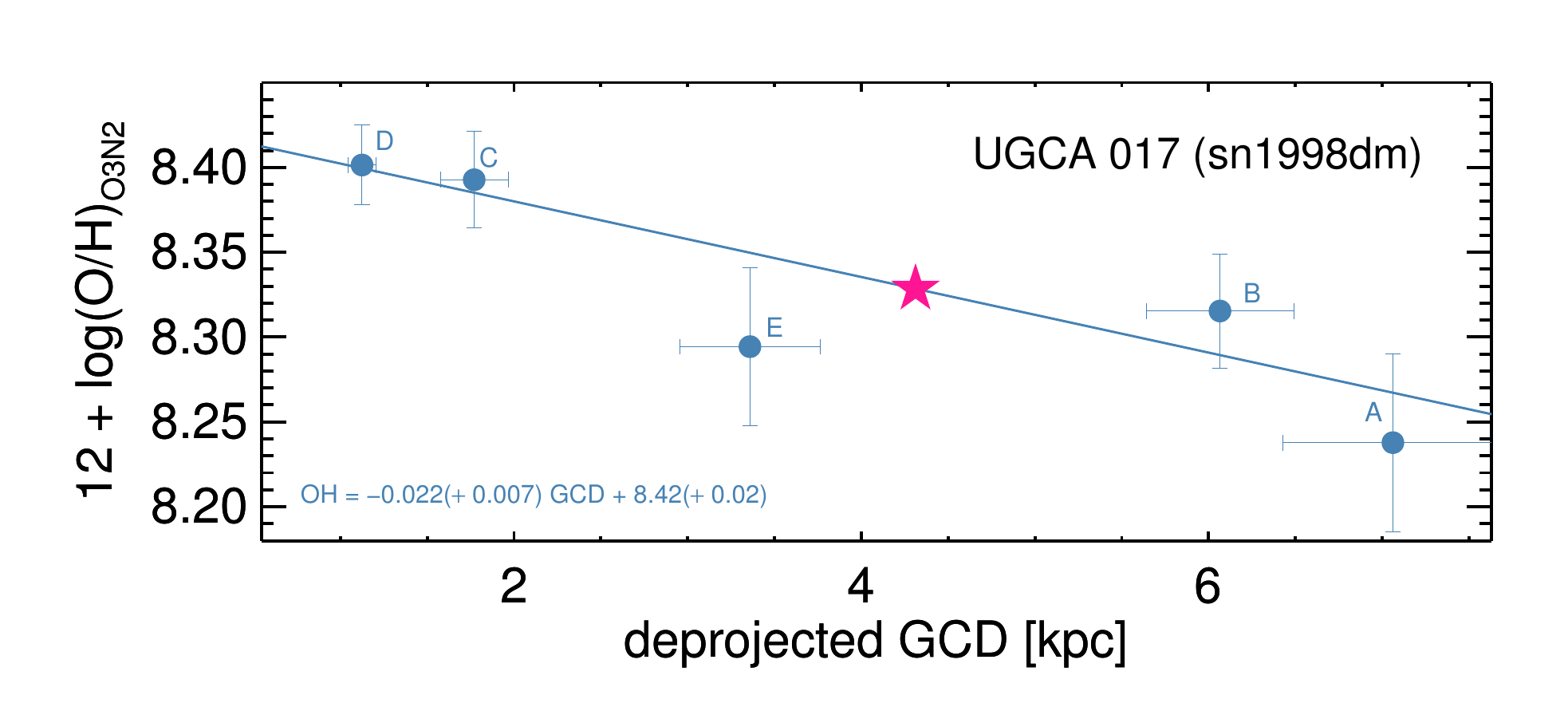}
\caption{Set of galaxies with extracted regions and derived gradients (or closest regions).}
\label{fig:gradientes_todos7}
\end{figure*}

\end{document}